\documentclass[12pt,english,longbibliography,aps,superscriptaddress,footinbib,12pt,sort&compress]{revtex4}
\usepackage{ae,aecompl}
\usepackage[T1]{fontenc}
\usepackage[latin9]{inputenc}
\usepackage{xcolor}
\usepackage{amsmath}
\usepackage{graphicx}
\PassOptionsToPackage{normalem}{ulem}
\usepackage{ulem}

\makeatletter

\newcommand{\lyxmathsym}[1]{\ifmmode\begingroup\def\b@ld{bold}
  \text{\ifx\math@version\b@ld\bfseries\fi#1}\endgroup\else#1\fi}

\providecommand{\tabularnewline}{\\}
\newcommand{\lyxdot}{.}

\providecolor{lyxadded}{rgb}{0,0,1}
\providecolor{lyxdeleted}{rgb}{1,0,0}

\DeclareRobustCommand{\lyxsout}[1]{\ifx\\#1\else\sout{#1}\fi}

\@ifundefined{textcolor}{}
{%
 \definecolor{BLACK}{gray}{0}
 \definecolor{WHITE}{gray}{1}
 \definecolor{RED}{rgb}{1,0,0}
 \definecolor{GREEN}{rgb}{0,1,0}
 \definecolor{BLUE}{rgb}{0,0,1}
 \definecolor{CYAN}{cmyk}{1,0,0,0}
 \definecolor{MAGENTA}{cmyk}{0,1,0,0}
 \definecolor{YELLOW}{cmyk}{0,0,1,0}
}

\usepackage{amsfonts}
\usepackage{aecompl}\usepackage{epstopdf}

\usepackage{pbox}

\makeatother

\begin{document}
\title{Physically-informed artificial neural networks for atomistic modeling
of materials}
\author{G. P. Purja Pun}
\affiliation{Department of Physics and Astronomy, MSN 3F3, George Mason University,
Fairfax, Virginia 22030, USA}
\author{R. Batra}
\affiliation{Department of Materials Science and Engineering, University of Connecticut,
Storrs, CT 06269, USA}
\author{R. Ramprasad}
\affiliation{School of Materials Science and Engineering, Georgia Institute of
Technology, Atlanta, GA 30332, USA}
\author{Y. Mishin}
\affiliation{Department of Physics and Astronomy, MSN 3F3, George Mason University,
Fairfax, Virginia 22030, USA}
\begin{abstract}
Large-scale atomistic computer simulations of materials heavily rely
on interatomic potentials predicting the potential energy and Newtonian
forces on atoms. Traditional interatomic potentials are based on physical
intuition but contain few adjustable parameters and are usually not
accurate. The emerging machine-learning (ML) potentials achieve highly
accurate interpolation between the energies in a large DFT database
but, being purely mathematical constructions, suffer from poor transferability
to unknown structures. We propose a new approach that can drastically
improve the transferability of ML potentials by informing them of
the physical nature of interatomic bonding. This is achieved by combining
a rather general physics-based model (analytical bond-order potential)
with a neural-network regression. The network adjusts the parameters
of the physics-based model on the fly during the simulations according
to the local environments of individual atoms. This approach, called
the physically-informed neural network (PINN) potential, is demonstrated
by developing a general-purpose PINN potential for Al. The potential
provides a DFT-level accuracy of energy predictions and excellent
agreement with experimental and DFT data for a wide range of physical
properties. We suggest that the development of physics-based ML potentials
is the most effective way forward in the field of atomistic simulations.
\end{abstract}
\maketitle

\section{Introduction}

Large-scale molecular dynamics (MD) and Monte Carlo (MC) simulations
of materials are traditionally implemented using classical interatomic
potentials predicting the potential energy and Newtonian forces acting
on atoms. Computations with such potentials are very fast and afford
access to systems with millions of atoms and MD simulation times up
to hundreds of nanoseconds. Such simulations span a wide range of
time and length scales and constitute a critical component of the
multiscale approach in materials modeling and computational design.

Several functional forms of interatomic potentials have been developed
over the years, including the embedded-atom method (EAM) \citep{Daw84,Daw83,Mishin.HMM},
the modified EAM (MEAM) \citep{Baskes87}, the angular-dependent potentials
\citep{Mishin05a}, the charge-optimized many-body potentials \citep{Liang:2012aa},
reactive bond-order potentials \citep{Brenner90,Brenner00,Stuart:2000aa},
and reactive force fields \citep{van-Duin:2001aa} to name a few.
These potentials address particular classes of materials or particular
types of applications. Their functional forms depend on the physical
and chemical models chosen to describe interatomic bonding in the
respective class of materials.

A common feature of all traditional potentials is that they express
the potential energy surface (PES) of the system, $E=E(\mathbf{r}_{1},...,\mathbf{r}_{N},\mathbf{p})$,
as a relatively simple function of atomic coordinates $(\mathbf{r}_{1},...,\mathbf{r}_{N})$,
$N$ being the number of atoms (Fig.~\ref{fig:Flowcharts_1}a). Knowing
the PES, the forces acting on the atoms can be computed by differentiation
and used in MD simulations. The potential functions depend on a relatively
small number of fitting parameters $\mathbf{p}=(p_{1},...,p_{m})$
(typically, $m=10-20$) and are optimized (trained) on a relatively
small database of experimental data and first-principles density functional
theory (DFT) calculations. The traditional potentials are, of course,
much less accurate than DFT calculations. Nevertheless, many of them
demonstrate a reasonably good transferability to atomic configurations
lying well outside the training dataset. This important feature owes
its origin to the incorporation of at least some basic physics in
the potential form. As long as the nature of chemical bonding remains
the same as assumed during the potential development, the potential
can predict the system energy adequately even for new configurations
not seen during the training process. Unfortunately, the construction
of good quality potentials is a long and painful process requiring
personal experience and intuition and is more art than science \citep{Brenner00,Mishin2010a}.
In addition, the traditional potentials are specific to a particular
class of materials and cannot be easily extended to other materials
or improved in a systematic manner.

During the past decade, a new direction has emerged wherein interatomic
potentials are developed by employing machine-learning (ML) methods
\citep{Mueller:2016aa,Behler:2007aa,Behler:2008aa,Behler:2011aa,Behler:2011ab,Behler:2015aa,Behler:2016aa,Bartok:2010aa,Botu:2015bb,Botu:2015aa,Wood:2018aa}.
The idea was originally conceived in the chemistry community in the
1990s in the effort to improve the accuracy of inter-molecular force
fields \citep{Raff:2012aa,Blank:1995aa}, an approach that was later
adopted by the physics and materials science communities. The general
idea is to forego the physical insights and reproduce the PES by interpolating
between DFT data points using high-dimensional nonlinear regression
methods such as the Gaussian process regression \citep{Payne.HMM,Bartok:2010aa,Li:2015aa,Glielmo:2017aa},
interpolating moving least squares \citep{Dawes:2008aa}, kernel ridge
regression \citep{Botu:2015bb,Botu:2015aa,Mueller:2016aa}, compressed
sensing \citep{Seko:2015aa,Mizukami:2015aa}, gradient-domain machine
learning model \citep{Chmiela:2018aa}, or the artificial neural network
(NN) approach \citep{Behler07,Bholoa:2007aa,Behler:2008aa,Sanville08,Eshet2010,Handley:2010aa,Behler:2011aa,Behler:2011ab,Sosso2012,Behler:2015aa,Behler:2016aa,Schutt:148aa,Imbalzano:2018aa}.
If properly trained, a ML potential can predict the system energy
with a nearly DFT accuracy (a few meV/atom). ML potentials are not
specific to a particular class of materials or type of chemical bonding.
They can be improved systematically if weaknesses are discovered or
new DFT data becomes available. The training process can be implemented
on-the-fly by running ab initio MD simulations \citep{Li:2015aa}.

A major weakness of ML potentials is their poor transferability. Being
purely mathematical constructions devoid of any physical meaning,
they can accurately \emph{interpolate} the energy between the training
configurations but are generally incapable of properly \emph{extrapolating}
the energy to unknown atomic environments. As a result, the performance
of ML potentials outside the training domain can be very poor. There
is no reason why a purely mathematical extrapolation scheme would
deliver physically meaningful results outside the training database.
This explains why the existing ML potentials are usually (with rare
exceptions \citep{Bartok_2018}) narrowly focused on, and only tested
for, a particular type of physical properties. This distinguishes
them from the traditional potentials which, although less accurate,
are designed for a much wider range of applications and diverse properties.

In this work we propose a new approach that can drastically improve
the transferability of ML potentials by informing them of the physical
nature of interatomic bonding. We focus on NN potentials as an example,
but the approach is general and can be readily extended to other methods
of nonlinear regression. Like all ML potentials, the proposed physically-informed
NN (PINN) potentials are trained using a large DFT dataset. However,
by contrast to the existing, mathematical NN potentials, the PINN
potentials incorporate the basic physics and chemistry of atomic interactions
leveraged by the extraordinary adaptivity and trainability of NNs.
The PINN potentials thus strike a golden compromise between the two
``extremes'' represented by the traditional, physics-guided interatomic
potentials, and the mathematical NN potentials.

The general idea of combining traditional interatomic potentials with
NNs was previously discussed by Malshe et al.~\citep{Malshe:2008aa},
who constructed an adjustable Tersoff potential \citep{Tersoff88,Tersoff:1988dn,Tersoff:1989wj}
for a Si$_{5}$ cluster. Other authors have also applied machine-learning
methods to parameterize physics-based models of molecular interactions,
primarily in the context of broad exploration of the compositional
space of molecular (mostly organic) matter \citep{Bereau:2015,Bereau:2018aa,Kranz:2018aa}.
Glielmo et al.~\citep{Glielmo:2018aa} recently proposed to construct
$n$-body Gaussian process kernels to capture the $n$-body nature
of atomic interactions in physical systems. The PINN potentials proposed
in this paper are inspired by such approaches but extend them to (1)
more advanced physical models with a broad applicability, and (2)
large-scale systems by introducing \emph{local} energies $E_{i}$
linked to \emph{local} structural parameters $G_{i}^{l}$. The focus
is placed on the exploration of the configurational space of defected
solids and liquids in single-component and, in the future, binary
or multicomponent systems. The main goal is to improve the transferability
of interatomic potentials to unknown atomic environments while keeping
the same high accuracy of training as normally achieved with mathematical
machine-learning potentials.

\section{Physically-informed neural network potentials}

The currently existing, mathematical NN potentials \citep{Behler07,Bholoa:2007aa,Behler:2008aa,Sanville08,Eshet2010,Handley:2010aa,Behler:2011aa,Behler:2011ab,Sosso2012,Behler:2015aa,Behler:2016aa}
partition the total energy $E$ into a sum of atomic energies, $E=\sum_{i}E_{i}$.
A single NN is constructed to express each atomic energy $E_{i}$
as a function of a set of local fingerprint parameters (also called
symmetry parameters \citep{Behler07}) $(G_{i}^{1},G_{i}^{2},...,G_{i}^{k})$.
These parameters encode the local environments of the atoms. The network
is trained by minimizing the error between the energies predicted
by the NN and the respective DFT total energies for a large set of
atomic configurations. The flowchart of the method is depicted in
Fig.~\ref{fig:Flowcharts_1}b.

The proposed PINN model is based on the following considerations.
A traditional, physics-based potential can always be trained to reproduce
the energy of any given atomic configuration with any desired accuracy.
Of course, this potential will not work well for other configurations.
Imagine, however, that the potential parameters have been trained
for a large set of reference structures, one structure at a time,
each time producing a different parameter set $\mathbf{p}$. Suppose
then that, during the subsequent simulations, we have a way of identifying,
on the fly, a reference structure closest to any current atomic configuration.
Then the accuracy of the simulation can be drastically improved by
dynamically choosing the best set of potential parameters for every
atomic configuration accoutered during the simulation. Now, since
the atomic energy $E_{i}$ only depends on the local environment of
atom $i$, the best parameter set for computing $E_{i}$ can be chosen
by only examining the local environment of this atom. The energies
of different atoms are then computed by using different, environment-dependent,
parameter sets while keeping the same, physics-motivated functional
form of the potential.

Instead of generating and storing a large set of discrete reference
structures, we can construct a continuous NN-based function mapping
the local environment of every atom on a parameter set of the interatomic
potential optimized for that particular environment. Specifically,
the local structural parameters (fingerprints) $G_{i}^{l}$ ($l=1,...,k$)
of every atom $i$ are fed into the network, which then maps them
on the optimized parameter set $\mathbf{p}_{i}$ appropriate for atom
$i$. Mathematically, the local energy takes the functional form
\begin{equation}
E_{i}=E_{i}\left(\mathbf{r}_{i1},...,\mathbf{r}_{in},\mathbf{p}_{i}\left(G_{i}^{l}(\mathbf{r}_{i1},...,\mathbf{r}_{in})\right)\right),\label{eq:14-1}
\end{equation}
where $(\mathbf{r}_{i1},...,\mathbf{r}_{in})$ are atomic positions
in the vicinity of atom $i$.

In comparison with the direct mapping $G_{i}^{l}\mapsto E_{i}$ implemented
by the mathematical NN potentials, we have added an intermediate step:
$G_{i}^{l}\mapsto\mathbf{p}_{i}\mapsto E_{i}$. The first step is
executed by the NN and the second by a physics-based interatomic potential.
A flowchart of the two-step mapping is shown in Fig.~\ref{fig:Flowcharts_1}c.
It is important to emphasize that this intermediate step does not
degrade the accuracy relative to the direct mapping, because a feedforward
NN can always be trained to execute \emph{any} real-valued function
\citep{Hornik:1989aa,Pinkus:1999aa}. Thus, for any functional form
of the potential, the NN can always adjust its architecture, weights
and biases to achieve the same mapping as in the direct method. However,
since the chosen potential form captures the essential physics of
atomic interactions, the proposed PINN potential will display a better
transferability to new atomic environments. Even if the potential
parameters predicted by the NN for an unknown environment are not
very accurate, the physics-motivated functional form will ensure that
the results remain at least physically meaningful. This physics-guided
extrapolation is likely to be more reliable than the purely mathematical
extrapolation inherent in the existing NN potentials. Obviously, the
same reasoning applies to the interpolation process as well, which
can also be more accurate.

The functional form of the PINN potential must be general enough to
be applicable across different classes of materials. In this paper
we chose a simple analytical bond-order potential (BOP) \citep{Oloriegbe_PhD:2008aa,Gillespie:2007vl,Drautz07a}
that must work equally well for both covalent and metallic materials.
For a single-component system, the BOP functions are specified in
the Methods section. They capture the physical and chemical effects
such as the pairwise repulsion between atoms, the angular dependence
of the chemical bond strength, the bond-order effect (the more neighbors,
the weaker the bond), and the screening of chemical bonds by surrounding
atoms. In addition to being appropriate for covalent bonding, the
proposed BOP form reduces to the EAM formalism in the limit of metallic
bonding.

\section{Example: PINN potential for Al}

To demonstrate the PINN method, we have constructed a general-purpose
potential for aluminum. The training and validation datasets were
randomly selected from a pre-existing DFT database \citep{Botu:2015aa,Botu:2015bb}.
Some additional DFT calculations have also been performed using the
same methodology as in \citep{Botu:2015aa,Botu:2015bb}. The selected
DFT supercells represent 7 crystal structures for a large set of atomic
volumes under isotropic tension and compression, several slabs with
different surface orientations, including surfaces with adatoms, a
supercell with a single vacancy, five different symmetrical tilt grain
boundaries, and an unrelaxed intrinsic stacking fault on the (111)
plane with different translational states along the {[}211{]} direction.
The database also includes several isolated clusters with the number
of atoms ranging from 2 (dimer) to 79. The ground-state face centered
cubic (FCC) structure was additionally subject to uniaxial tension
and compression in the {[}100{]} and {[}111{]} directions at 0 K temperature.
Most of the atomic configurations were snapshots of DFT MD simulations
in the microcanonical (NVE) or canonical (NVT or NPT) ensembles for
several atomic volumes at several temperatures. Some of the high-temperature
configurations were part-liquid, part crystalline. In total, the database
contains 3649 (127592 atoms). More detailed information about the
database can be found in the Supplementary Tables \ref{tab:Al_database}
and \ref{tab:Al_database_contd}. To avoid overfitting or selection
bias, the 10-fold cross-validation method was during the training.
The database was randomly partitioned in 10 subsets. One of them was
set aside for validation and the remaining data was used for training.
The process repeated 10 times form different choices of the validation
subset.

The local structural parameters $G_{i}^{l}$ chosen for Al are specified
in the Methods section. The NN contained two hidden layers with the
same number of nodes in each. This number was increased until the
training process produced a PINN potential with the root-mean-square
error (RMSE) of training and validation close to 3 to 4 meV/atom,
which was set as our goal. This is the level of accuracy of the DFT
energies included in the database. For comparison, a mathematical
NN potential was constructed using the same methodology. The number
of hidden nodes of the NN was adjusted to give about the same number
of fitted parameters and to achieve approximately the same RMSE of
training and validation as for the PINN potential. Table \ref{tab:fit_errors}
summarizes the training and validation errors averaged over the 10
cross-validation runs. One PINN and one NN potential were selected
for a more detailed examination reported below.

\begin{figure}
\noindent \begin{centering}
\textbf{(a)} \includegraphics[width=0.43\textwidth]{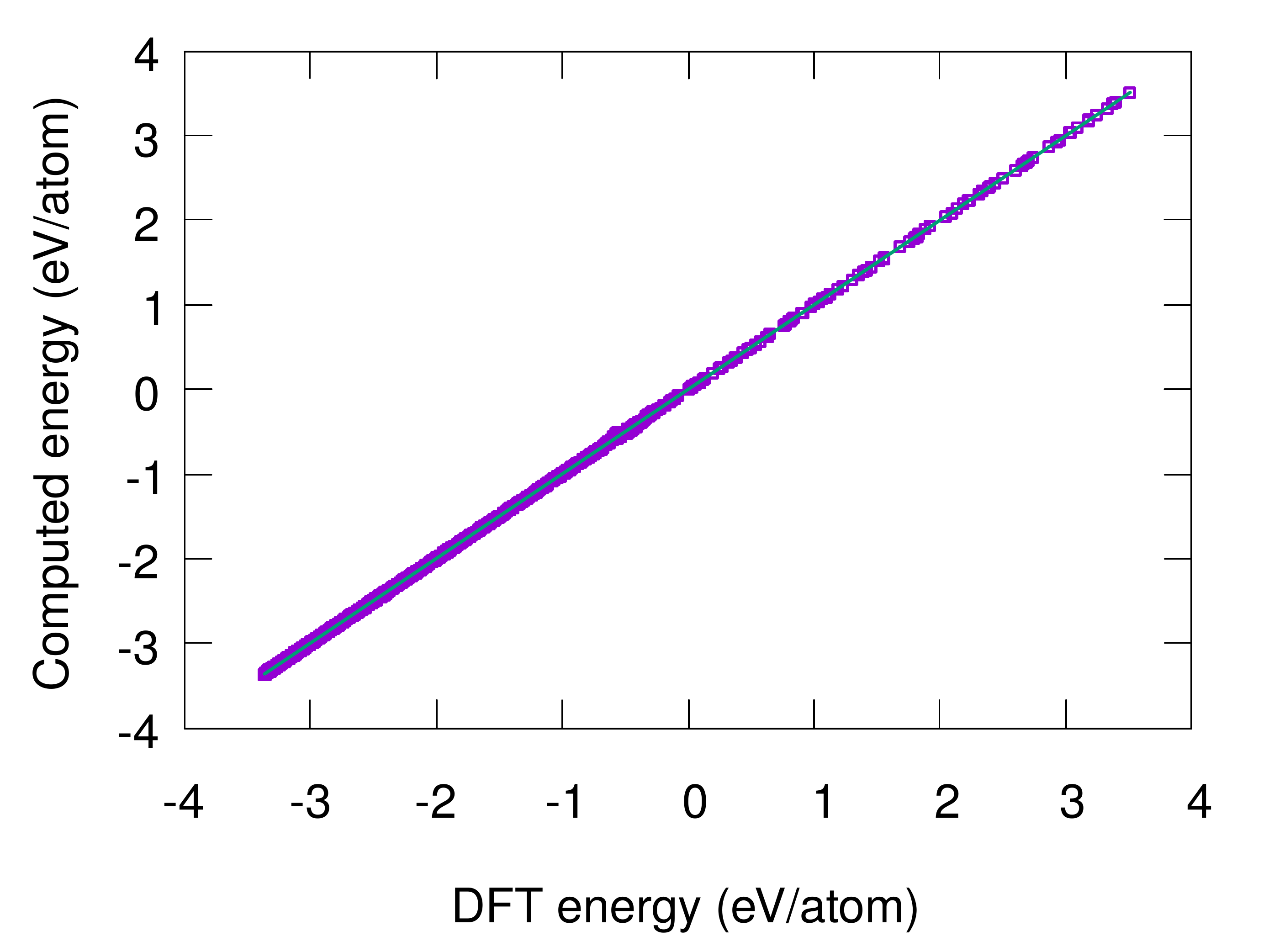}\textbf{(b)}
\includegraphics[width=0.44\textwidth]{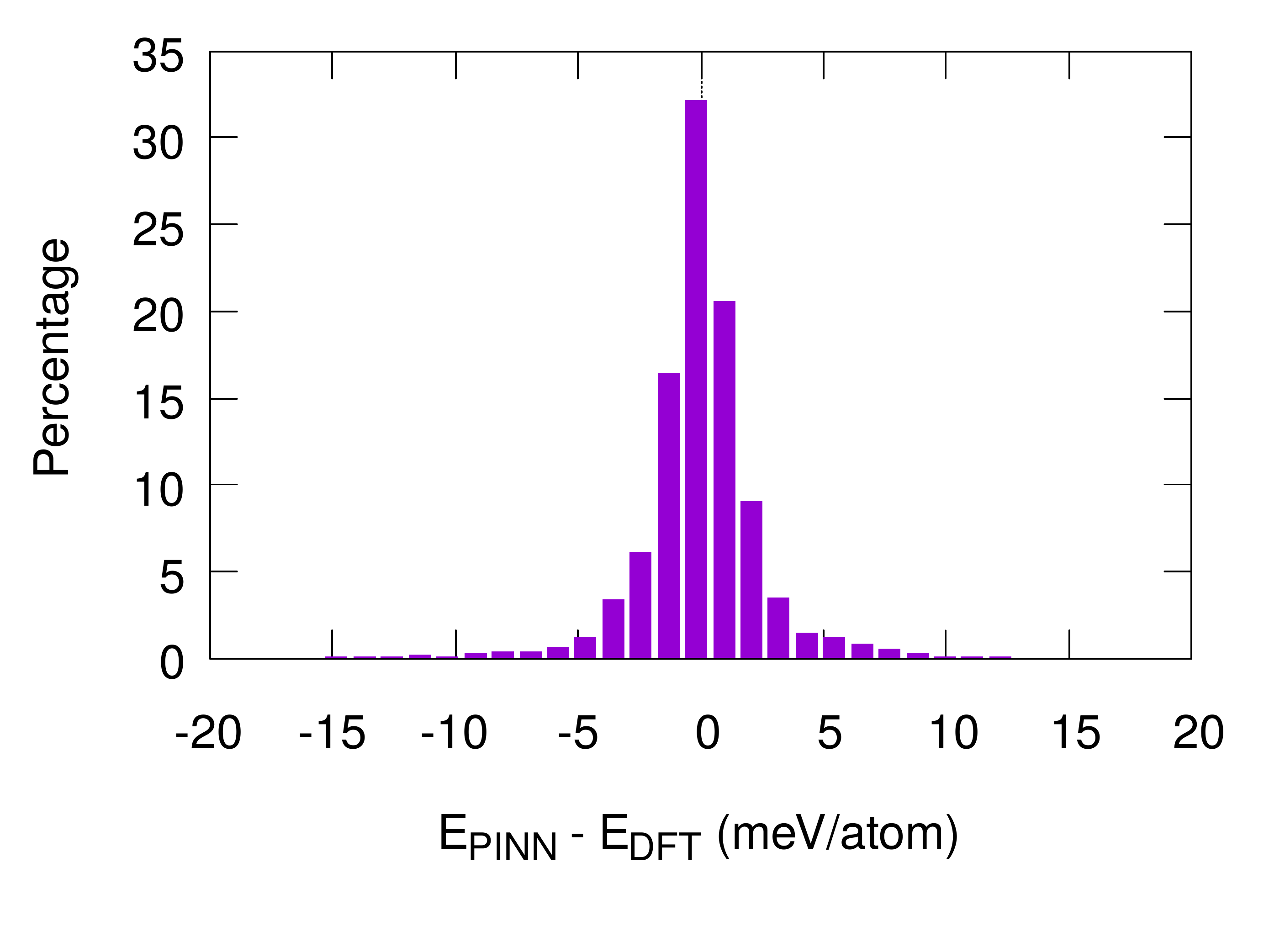}
\par\end{centering}
\bigskip{}

\noindent \begin{centering}
\textbf{(c)} \includegraphics[width=0.435\textwidth]{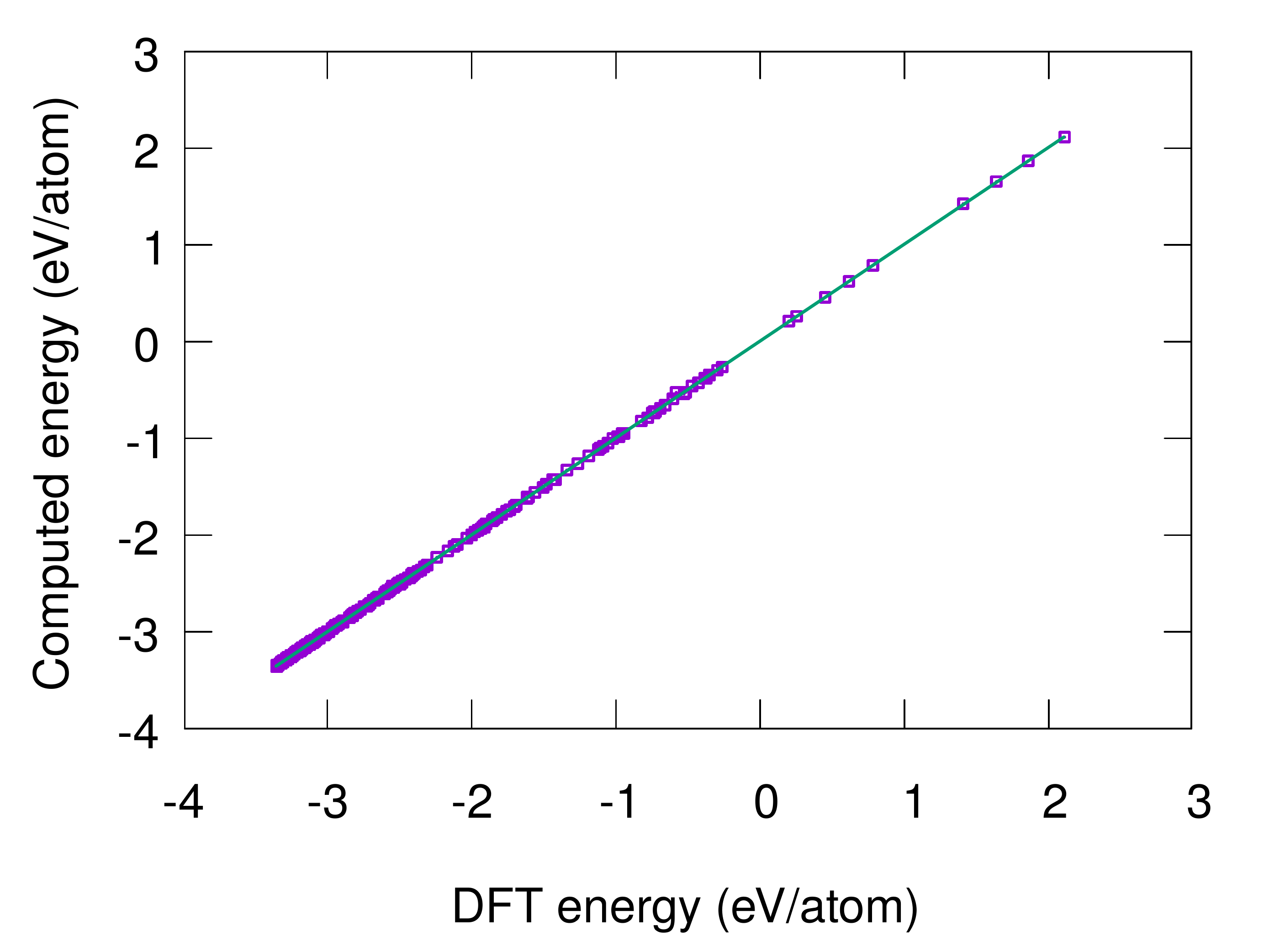}\textbf{(d)}
\includegraphics[width=0.44\textwidth]{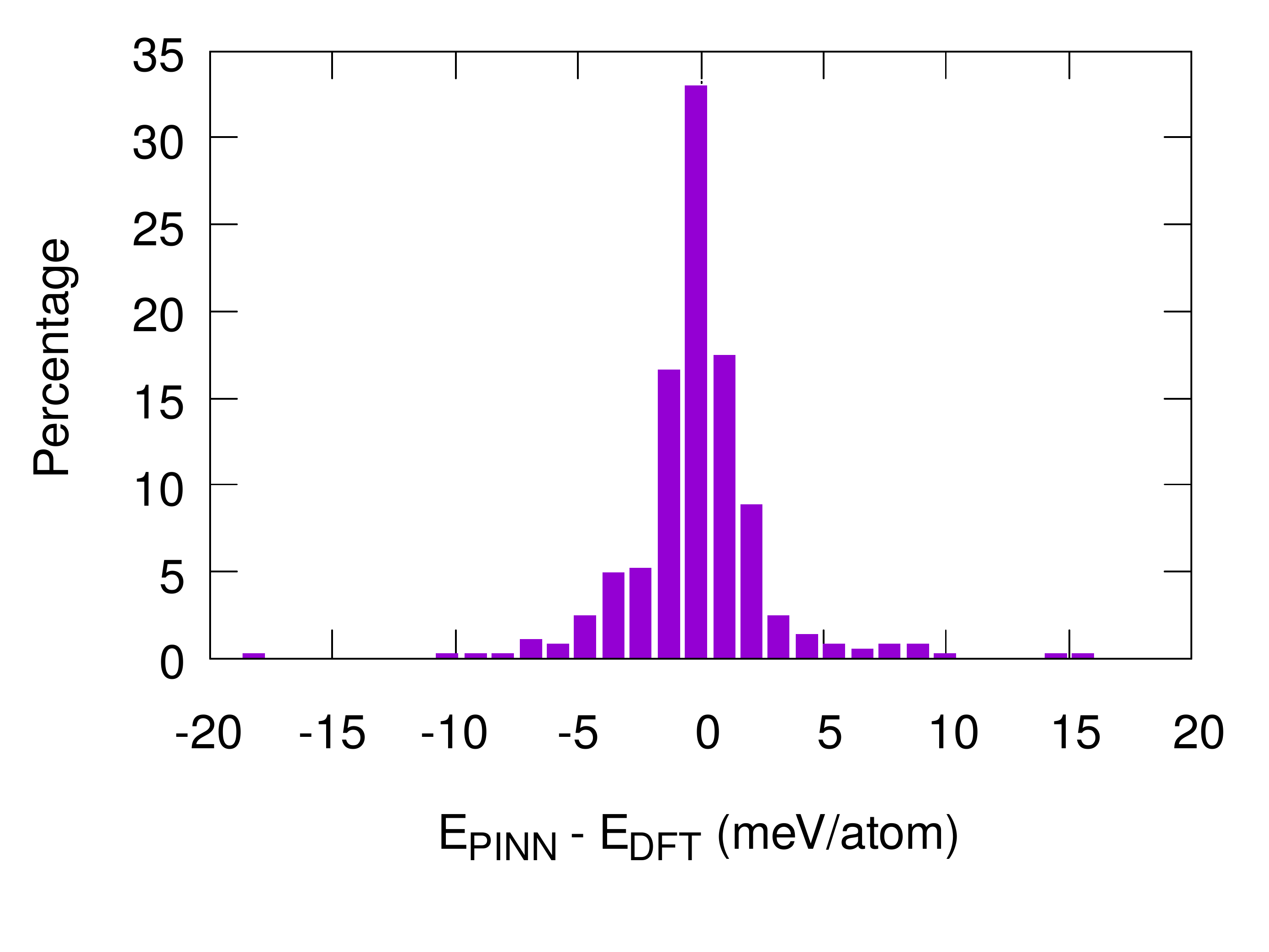}
\par\end{centering}
\caption{(a,c) Energies of atomic configurations in the (a) training and (c)
validation datasets computed with the PINN potentials versus DFT energies.
The straight line represents the perfect fit. (b,d) Error distributions
in the (b) training and (d) validation datasets. \label{fig:PINN_Correlation}}
\end{figure}

Figures \ref{fig:PINN_Correlation} and \ref{fig:NN_Correlation}
demonstrate excellent correlation between the predicted and DFT energies
over a 7 eV/atom wide energy range for both potentials. The error
distribution has a near-Gaussian shape centered at zero. Examination
of errors in individual groups of structures (Fig.~\ref{fig:Error_by_group})
shows that the largest errors originate from the crystal structures
(especially FCC, HCP and simple hexagonal) subjected to large expansion.

\begin{figure}
\noindent \begin{centering}
\includegraphics[width=0.6\textwidth]{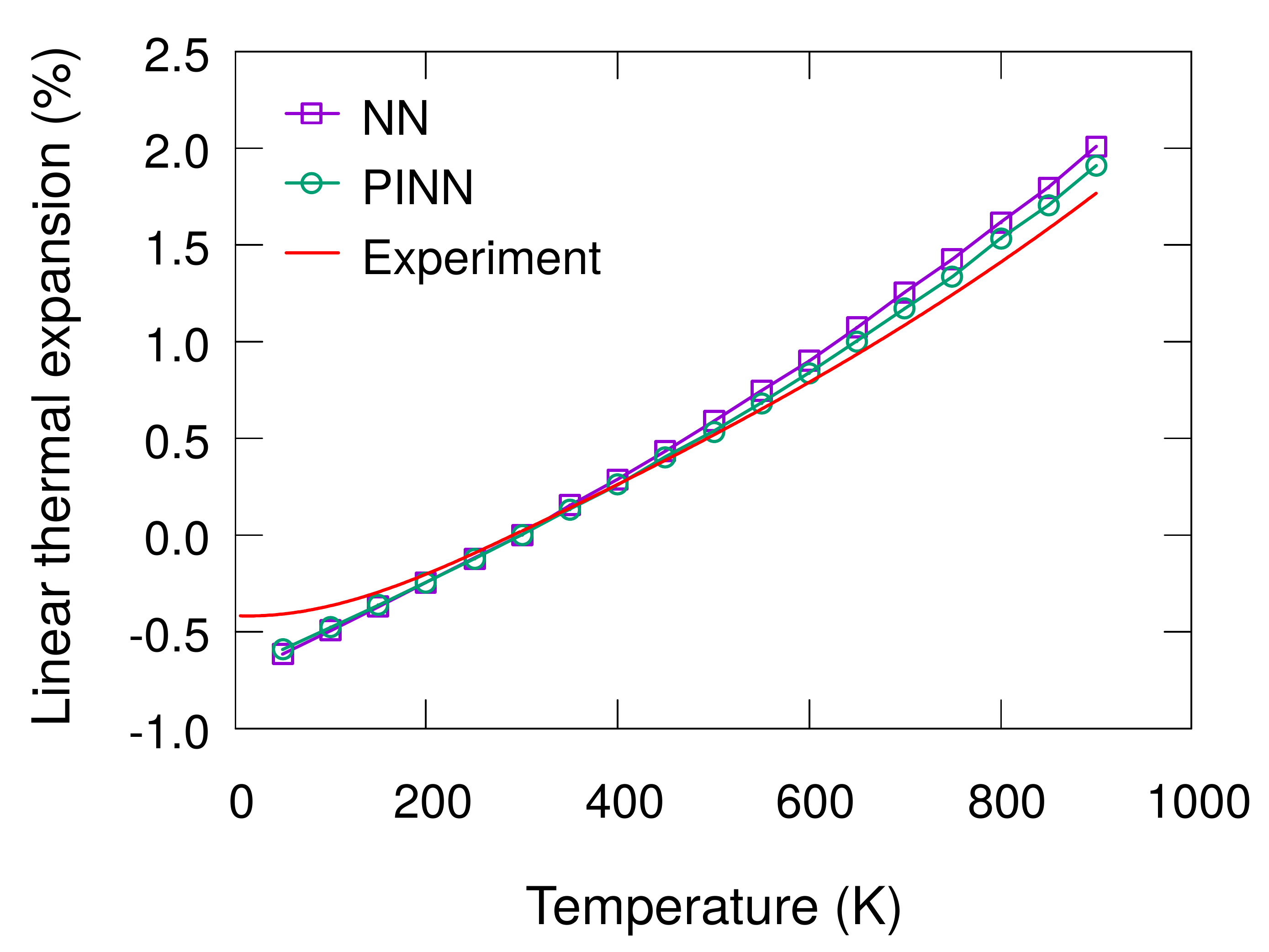}
\par\end{centering}
\caption{Linear thermal expansion of Al relative to room temperature (295 K)
predicted by the PINN and NN potentials in comparison with experiment
\citep{Expansion}. The plots stop near the experimental melting point.
\label{fig:Thermal-expansion}}
\end{figure}

Table \ref{table:al_prop} summarizes some of the physical properties
of Al predicted by the potentials in comparison with DFT data from
the literature. There was no direct fit to any of these properties,
although atomic configurations most relevant to some of the properties
were represented in the training dataset. While both potentials agree
with the DFT data well, the PINN potential tends to be more accurate
for most properties. For the {[}110{]} self-interstitial dumbbell,
the NN potential predicts an unstable configuration that spontaneously
rotate to the {[}100{]} orientation, whereas the PINN correctly predicts
such configurations to be metastable. Figure \ref{fig:Thermal-expansion}
shows the linear thermal expansion factor as a function of temperature
predicted by the potentials in comparison with experimental data.
The PINN potential displays good agreement with experiment without
direct fit, whereas the NN potentials overestimates the thermal expansion
at high temperatures. (The discrepancies at low temperatures are due
to the quantum effects that are not captured by classical simulations.)
As another test, the radial distribution function and the bond angle
distribution in liquid Al were computed at several temperatures for
which experimental and/or DFT data is available (Figs.~\ref{fig:RDF}
and \ref{fig:Bond-angle-distribution}). In this case, both potentials
were found to perform equally well. Any small deviations from the
published DFT calculations are within the uncertainty of the different
DFT flavors (exchange-correlation functionals).

\begin{figure}
\textbf{(a)}\includegraphics[width=0.45\textwidth]{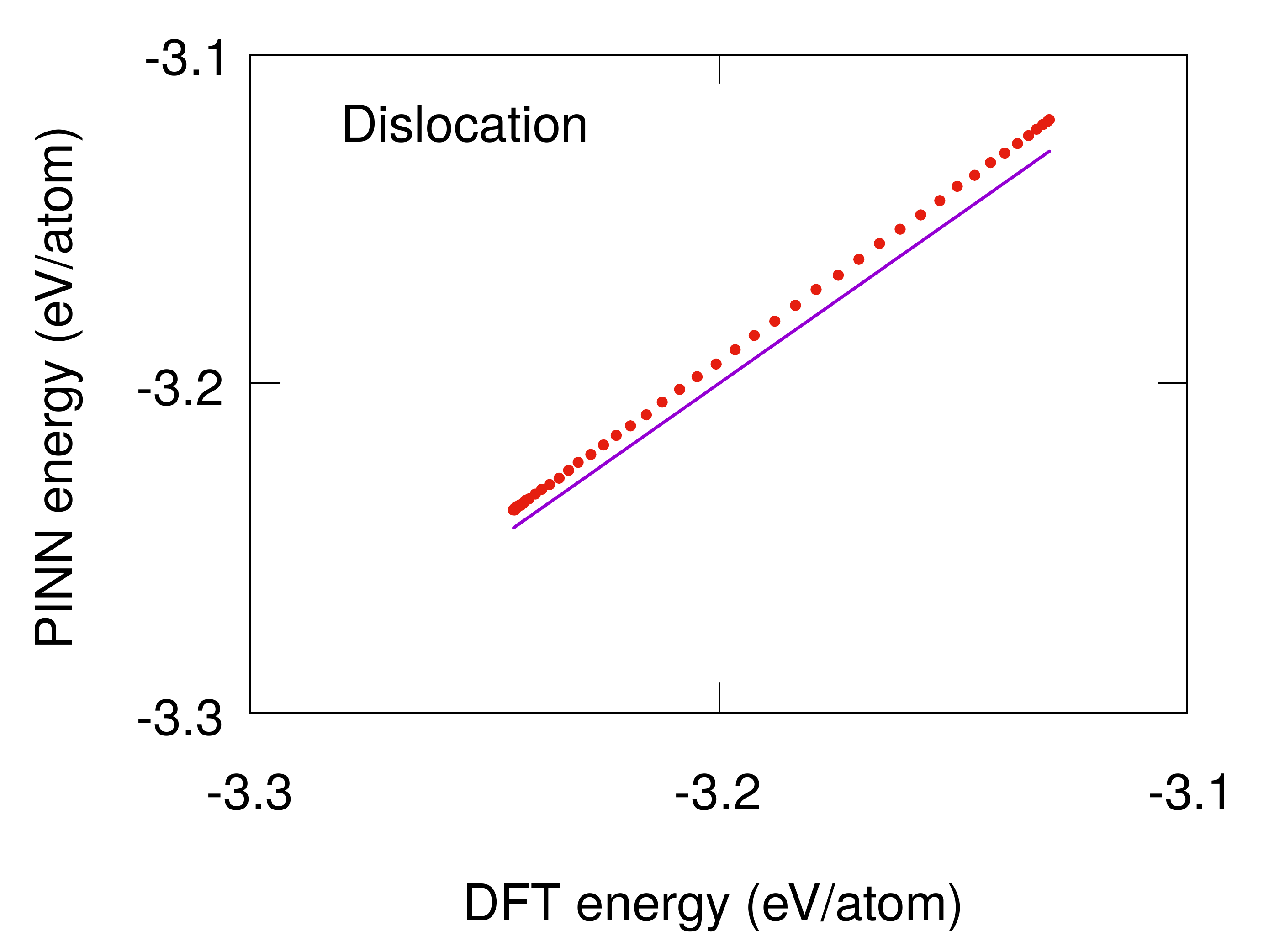}\enskip{}\textbf{(b)}\includegraphics[width=0.45\textwidth]{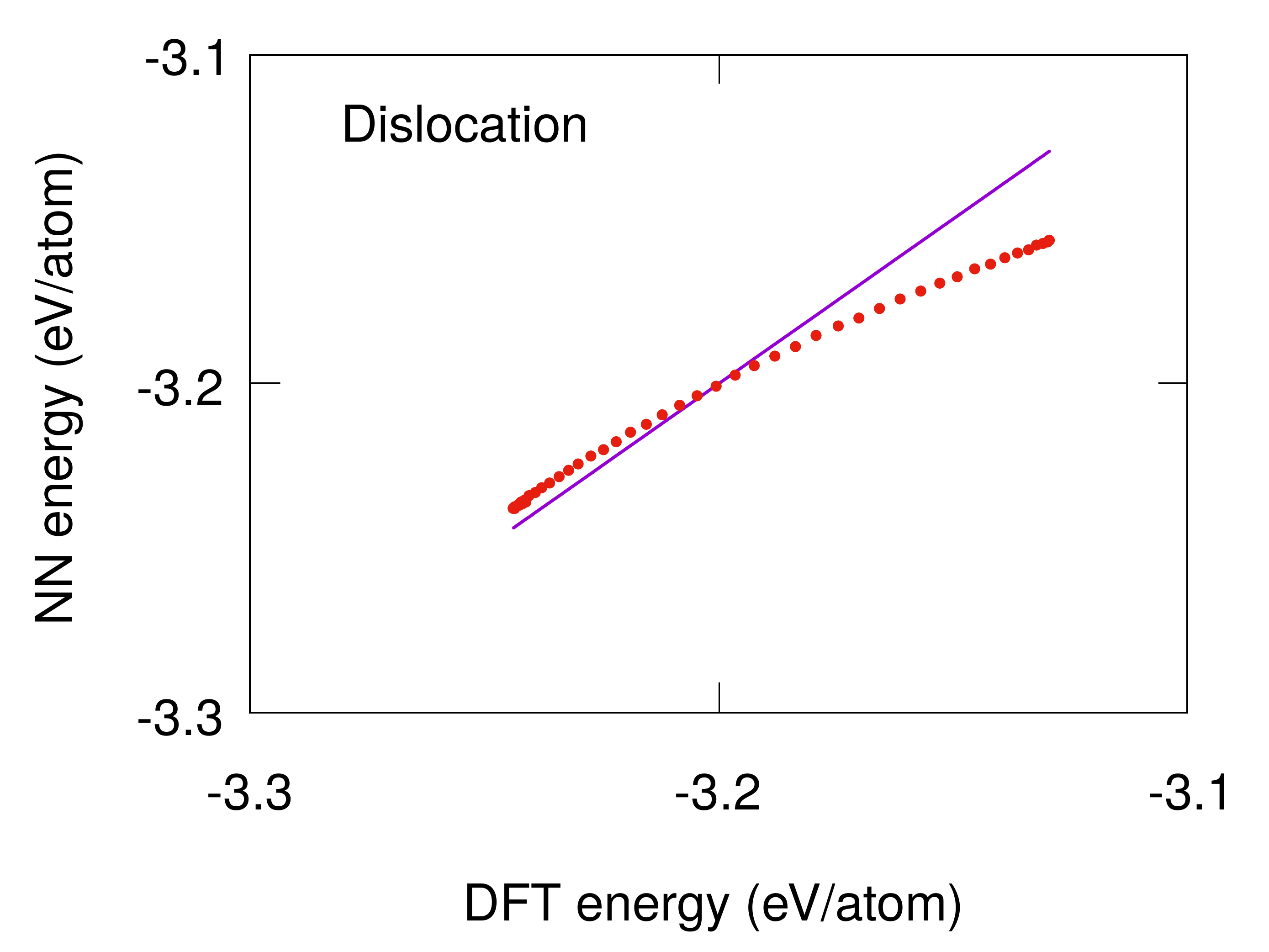}

\bigskip{}
\bigskip{}

\textbf{(c)}\includegraphics[width=0.45\textwidth]{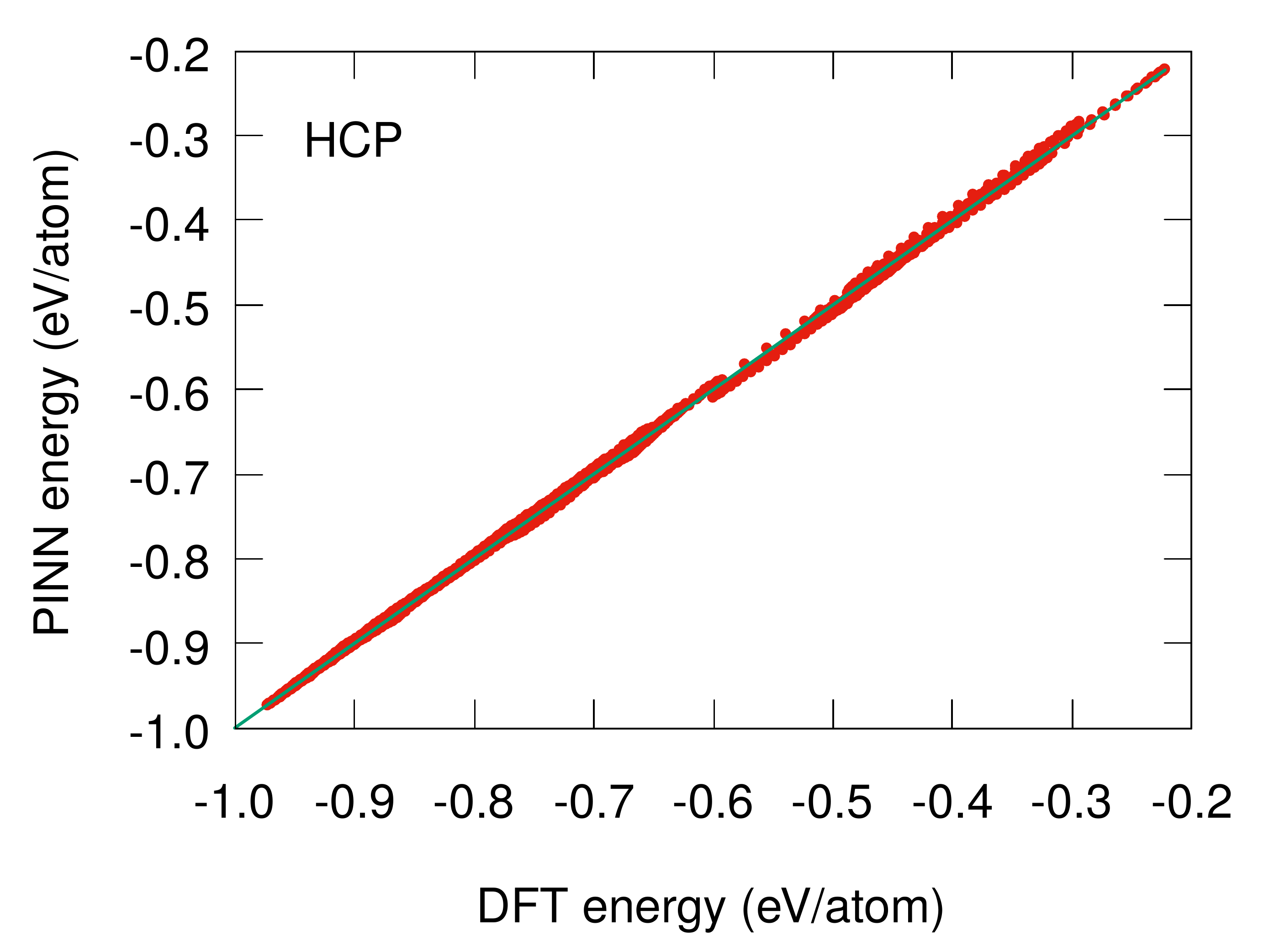}\enskip{}\textbf{(d)}\includegraphics[width=0.45\textwidth]{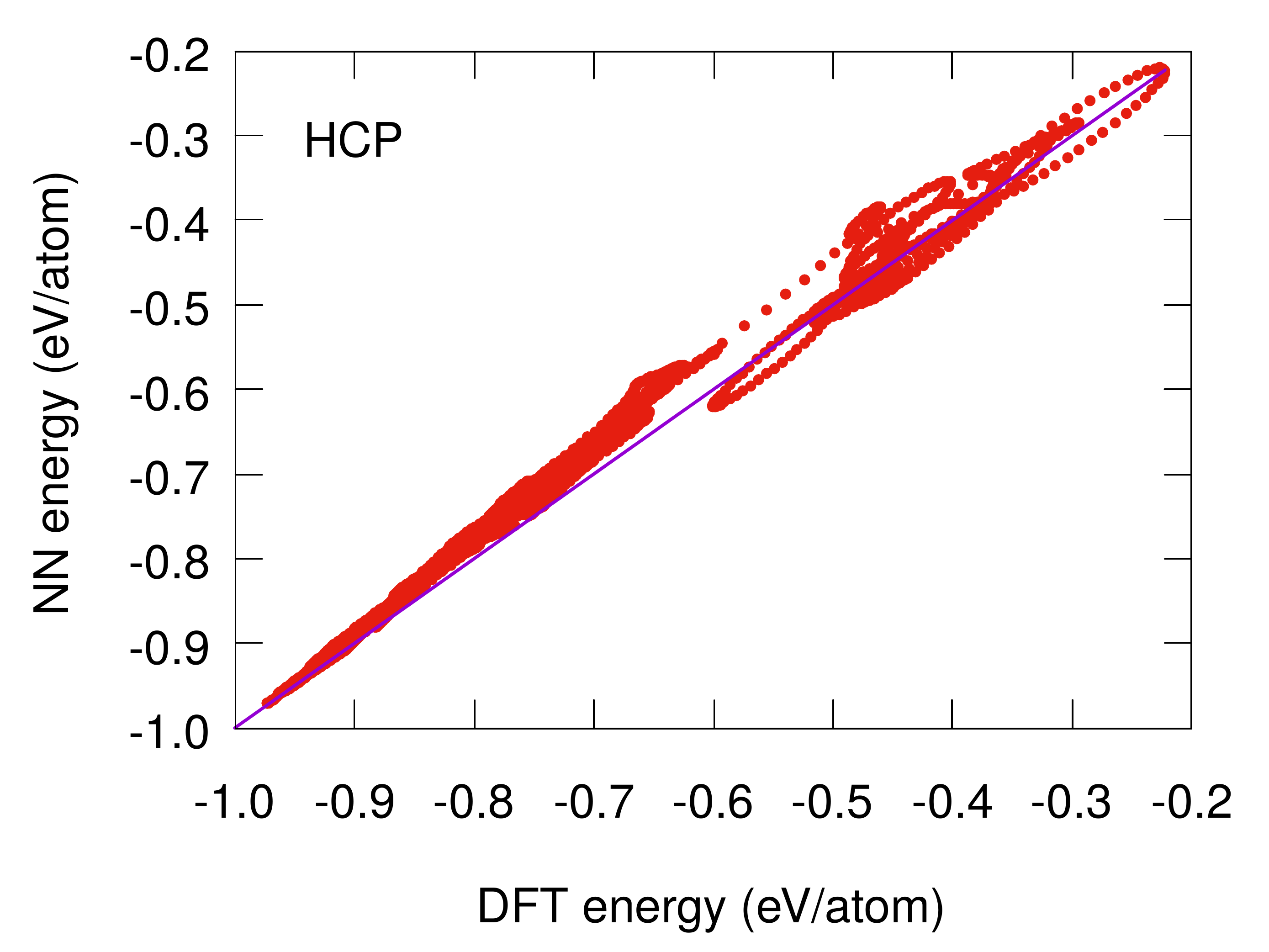}

\caption{(a,b) Energy of an edge dislocation in Al in NVE MD simulations starting
at 700 K. (c,d) Energy of HCP Al in NVT MD simulations at 1000\,K,
1500\,K, 2000\,K and 4000\,K. The energies predicted by the PINN
(a,c) and NN (b,d) potentials are compared with DFT calculations from
\citep{Botu:2015aa,Botu:2015bb}. The straight lines represent the
perfect fit.\label{Fig:validation_1}}
\end{figure}

For testing purposes, we computed the energies of the remaining groups
of structures that were part of the original DFT database \citep{Botu:2015aa,Botu:2015bb}
but were not used here for training or validation. The full information
about the testing dataset (26425 supercells containing a total of
2376388 atoms) can be found in the Supplementary Table \ref{tab:Al_database_validation}.
For example, Fig.~\ref{Fig:validation_1} compares the energies predicted
by the potentials with DFT energies from high-temperature MD simulations
for a supercell containing an edge dislocation or HCP Al. In both
cases, the PINN potential is obviously more accurate. The remaining
testing cases are presented in the Supplementary Information file
(Supplementary Figures \ref{Fig:validation_2}-\ref{Fig:validation_6}).
Although there are cases where both potentials perform equally well,
in most cases the PINN potential predicts the energies of unknown
atomic configurations more accurately than the NN potential.

\begin{figure}
\noindent \begin{centering}
\textbf{(a)}\includegraphics[width=0.45\textwidth]{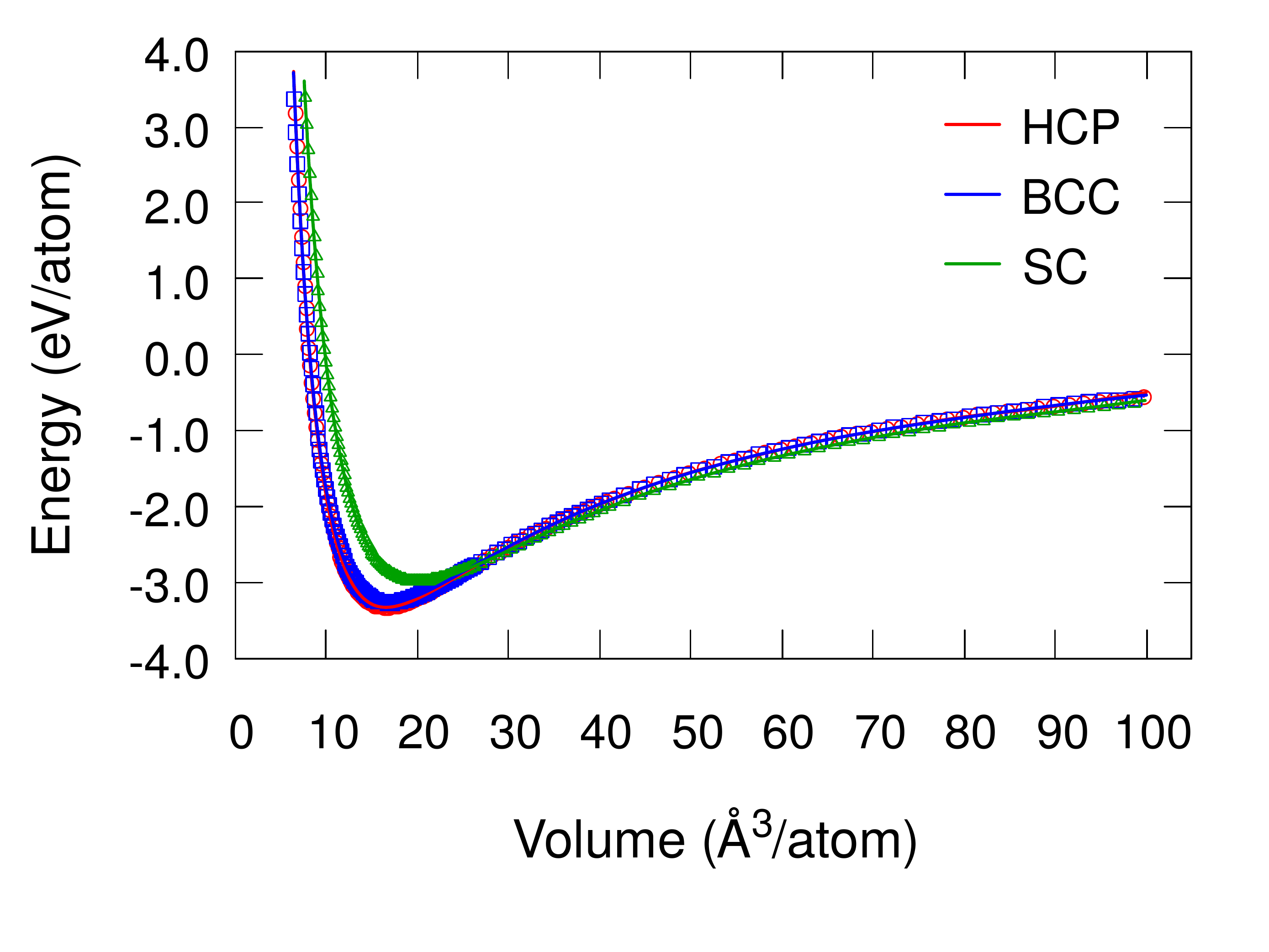}\quad{}\textbf{(b)}\includegraphics[width=0.45\textwidth]{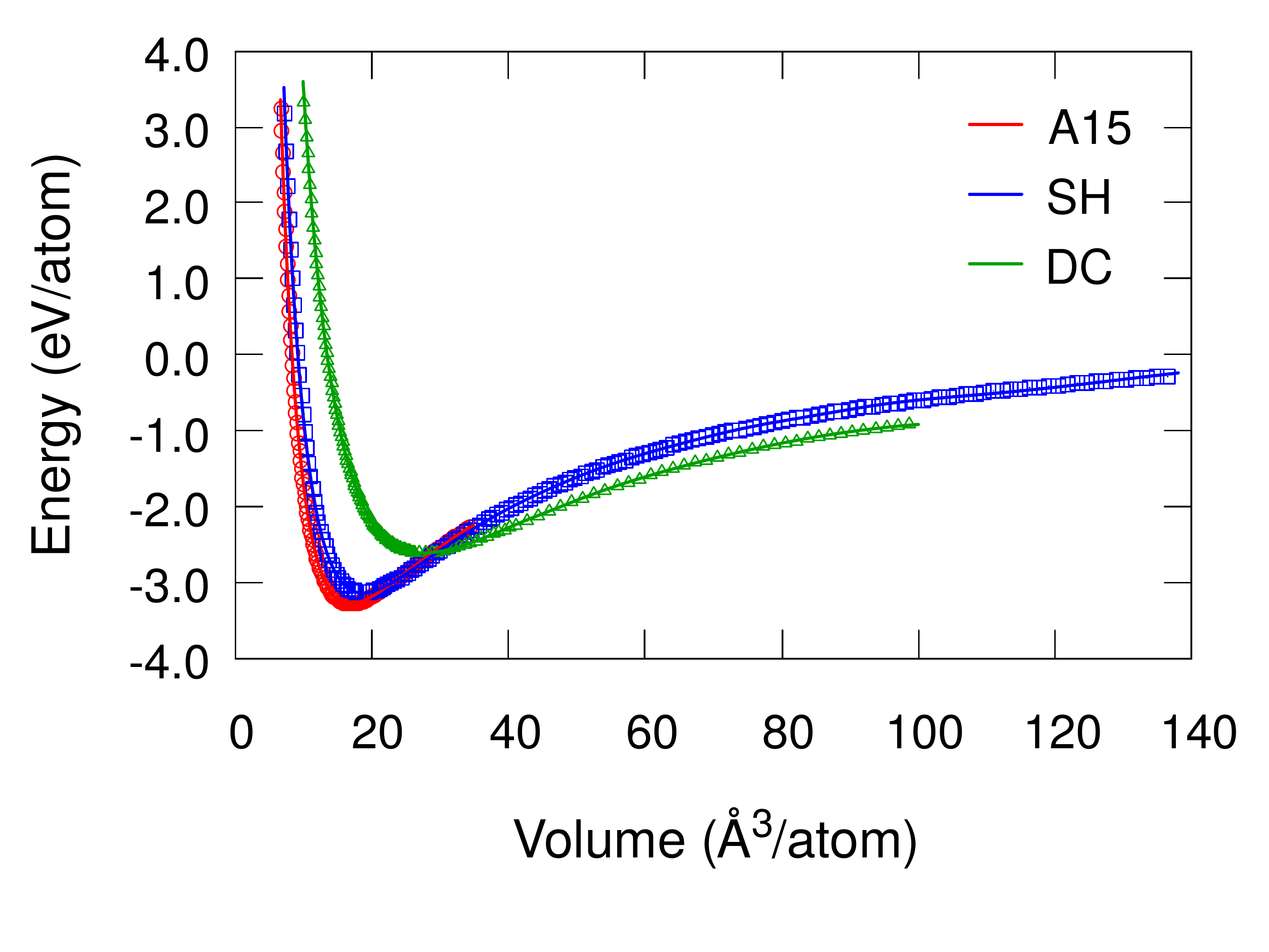}
\par\end{centering}
\caption{Energy-volume relations for Al crystal structures predicted by the
PINN potential (lines) and by DFT calculations (points). (a) Hexagonal
close-packed (HCP), body-centered cubic (BCC), and simple cubic (SC)
structures. (b) A15 (Cr$_{3}$Si prototype), simple hexagonal (SH),
and diamond cubic (DC) structures.\label{fig:EOS_PINN}}
\end{figure}

\begin{figure}
\noindent \begin{centering}
\textbf{(a)}\includegraphics[width=0.47\textwidth]{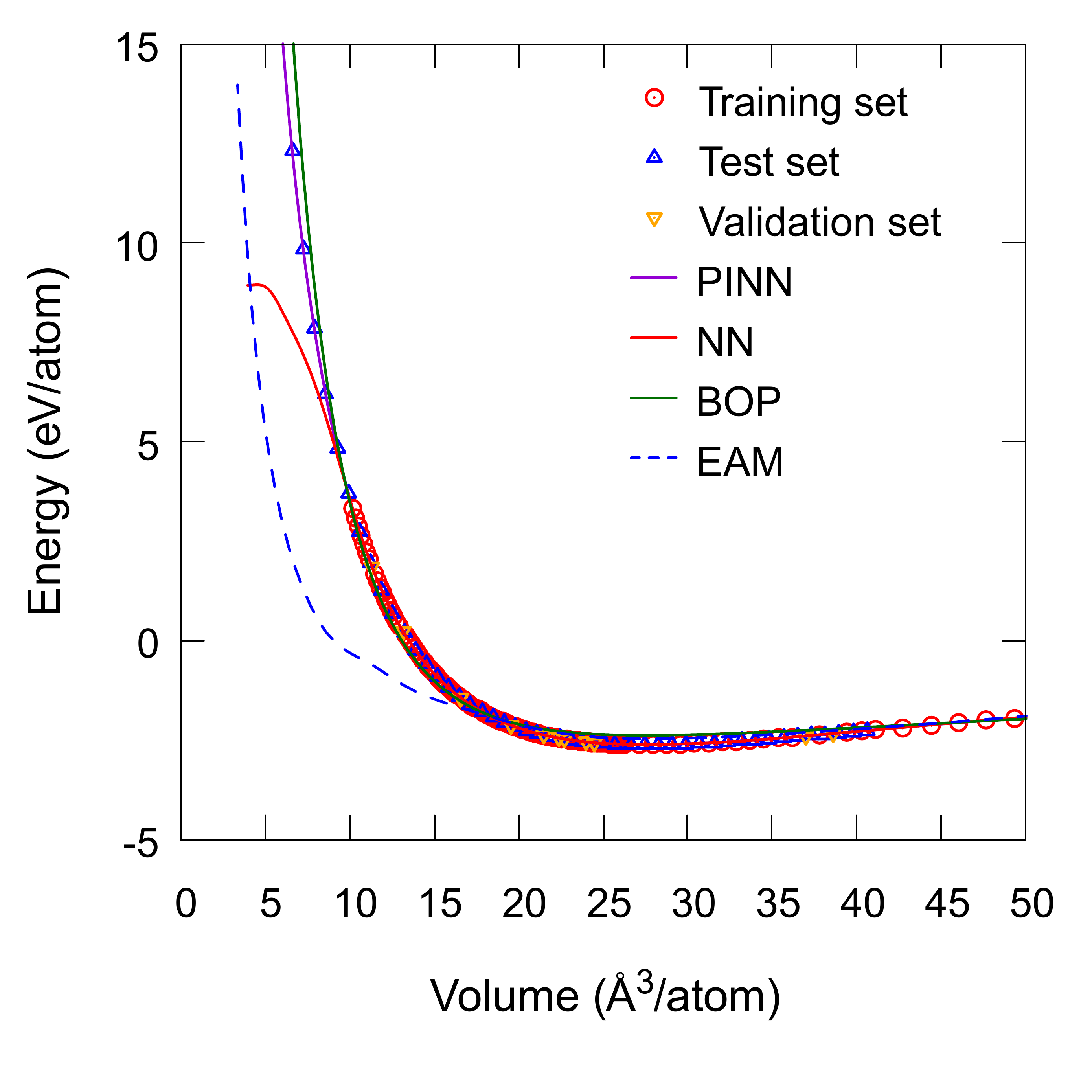}
\par\end{centering}
\bigskip{}
\bigskip{}

\noindent \begin{centering}
\textbf{(b)}\includegraphics[width=0.47\textwidth]{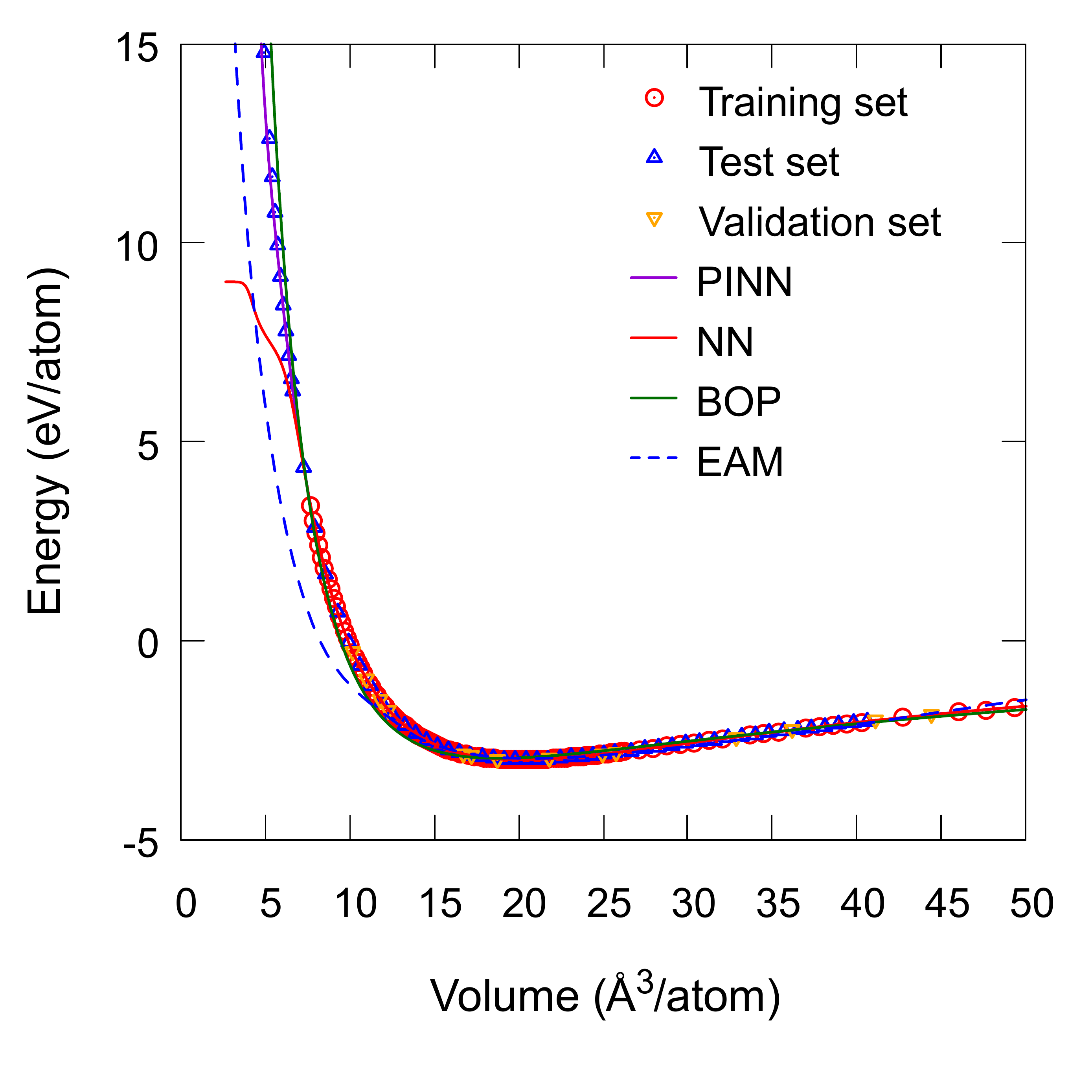}
\par\end{centering}
\caption{Zoom into the repulsive part of the energy-volume relations for the
Al SC and DC structures predicted by the PINN, NN, EAM and BOP potentials
(curves) in comparison with DFT calculations (points) from the training,
testing and validation datasets. \label{fig:EOS-FCC}}
\end{figure}

For further testing, the energies of the crystal structures of Al
were computed for atomic volumes both within and beyond the training
interval. Both potentials accurately reproduce the DFT energy-volume
relations for all volumes spanned by the DFT database (Figs.~\ref{fig:EOS_PINN}
and \ref{fig:EOS_NN}). However, extrapolation to larger or smaller
volumes reveals significant differences. For example, the PINN potential
correctly predicts that the crystal energy continues to rapidly increase
under strong compression (repulsive interaction mode). In fact, the
extrapolated PINN energy goes exactly through the new DFT points that
were not included in the training or validation datasets, see examples
in Fig.~\ref{fig:EOS-FCC}. By contrast, the energy predicted by
the NN model immediately develops wiggles and strongly deviates from
the physically meaningful repulsive behavior. Such artifacts were
found for other structures as well.

\begin{figure}
\textbf{(a)}\includegraphics[width=0.47\textwidth]{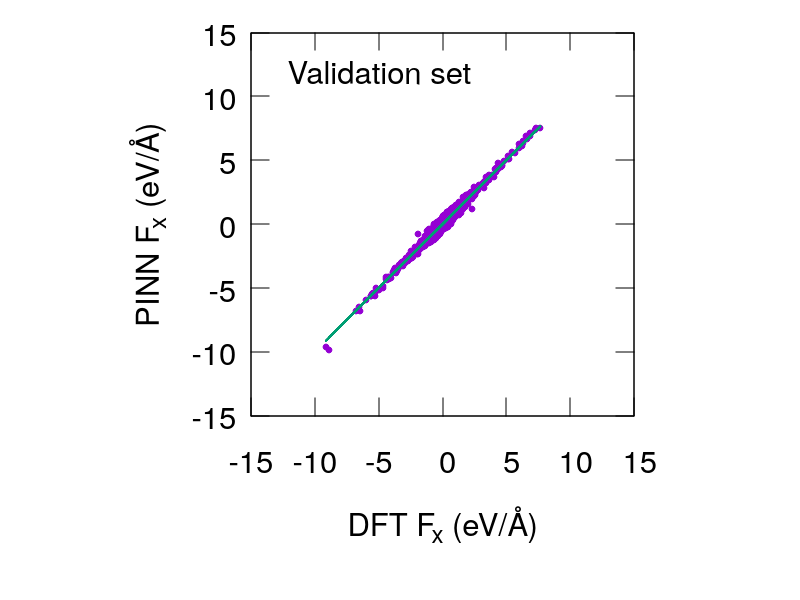}\textbf{(b)}\includegraphics[width=0.47\textwidth]{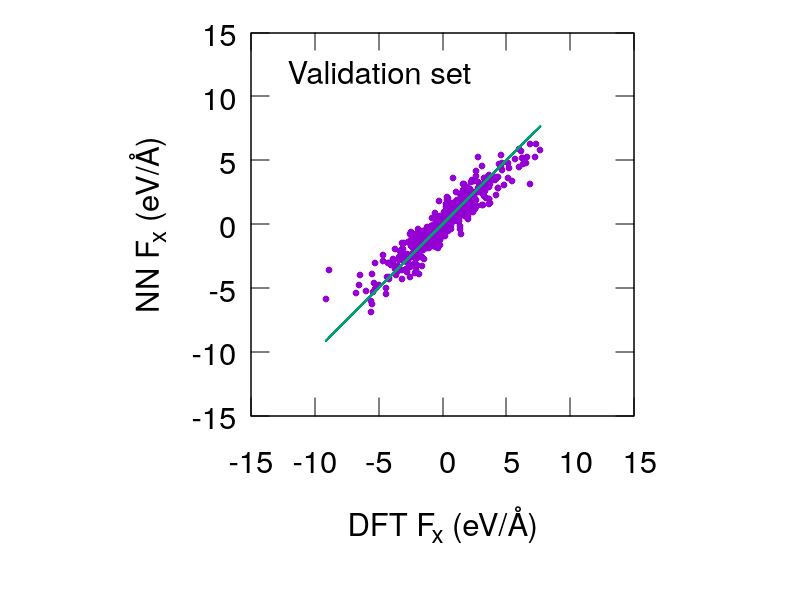}\bigskip{}
\bigskip{}
\textbf{(c)}\includegraphics[width=0.47\textwidth]{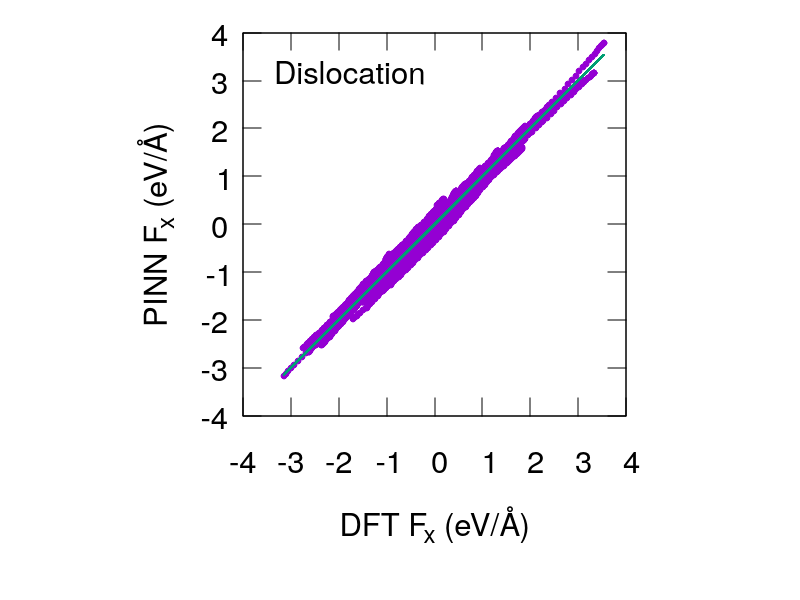}\textbf{(d)}\includegraphics[width=0.47\textwidth]{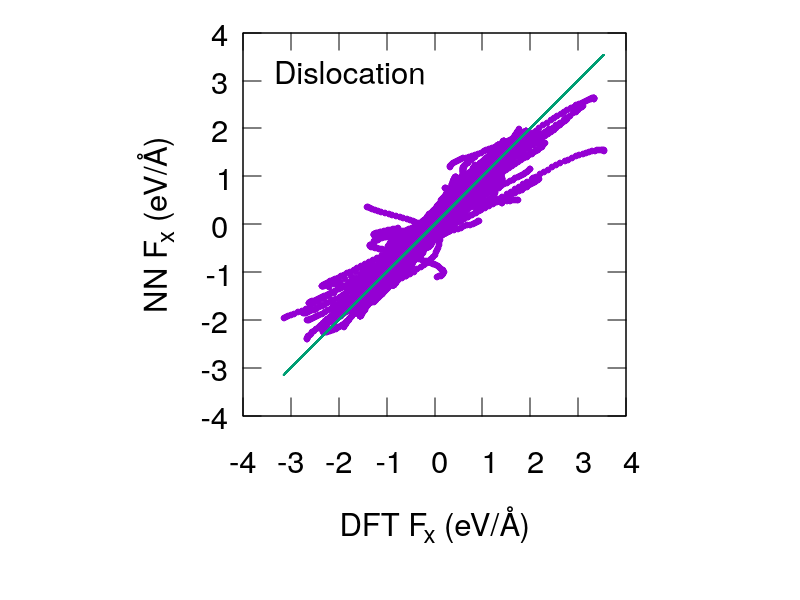}

\bigskip{}
\bigskip{}

\textbf{(e)}\includegraphics[width=0.47\textwidth]{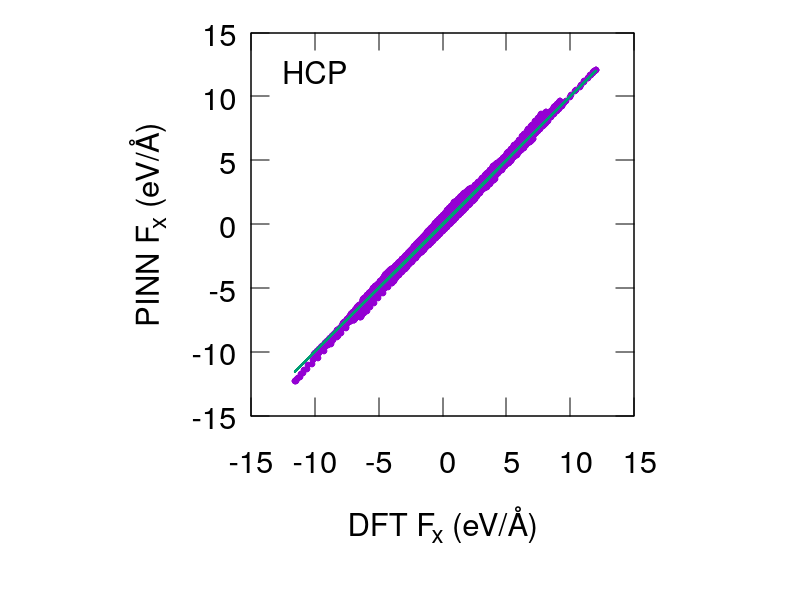}\textbf{(f)}\includegraphics[width=0.47\textwidth]{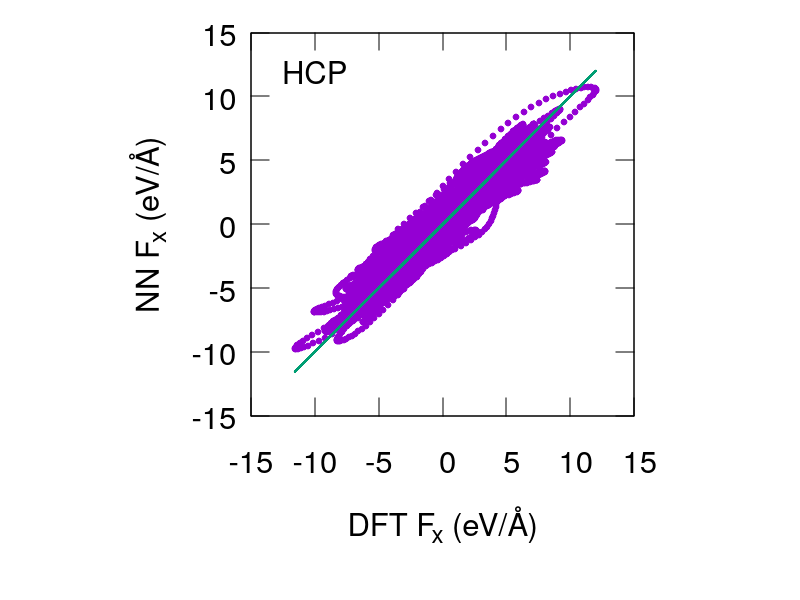}

\caption{The $x$-component of atomic forces for (a,b) validation database,
(c,d) edge dislocation in NVE MD simulations starting at 700 K, and
(e,f) HCP Al in NVT MD simulations at 300\,K, 600\,K, 1000\,K,
1500\,K, 2000\,K and 4000\,K. The forces predicted by the PINN
(a,c,e) and NN (b,d) potentials are compared with DFT calculations
from \citep{Botu:2015aa,Botu:2015bb}. The straight lines represent
the perfect fit. See Figs.~\ref{Fig:validation_forces_S1}-\ref{Fig:validation_forces_HCP}
for the all components of the forces.\label{Fig:validation_forces_1}}
\end{figure}

Furthermore, while the atomic forces were not used for either training
or validation, they were compared with the DFT forces once the training
was complete. For the validation dataset, this comparison probes the
accuracy of interpolation, whereas for the testing dataset the accuracy
of extrapolation. As expected, for the validation dataset the PINN
forces are in better agreement with DFT calculations than the NN forces
(RMSE $\approx0.1$ eV/\AA\   versus $\approx0.2$ eV/\AA)
as illustrated in Fig.~\ref{Fig:validation_forces_1}a,b. For the
testing dataset, the advantage of the PINN model in force predictions
is even more significant. For example, for the dislocation and HCP
cases discussed above, the PINN potential provides more accurate predictions
(RMSE $\approx0.1$ eV/\AA) than the NN potential (RMSE
$\approx0.4$ eV/\AA\  for the dislocation and $0.6$ eV/\AA\ 
for the HCP case) (Fig.~\ref{Fig:validation_forces_1}c-f). This
advantage persists for all other groups of the structures from the
testing database.

\begin{figure}
\noindent \begin{centering}
\par\end{centering}
\noindent \begin{centering}
\textbf{(a)}\includegraphics[width=0.47\textwidth]{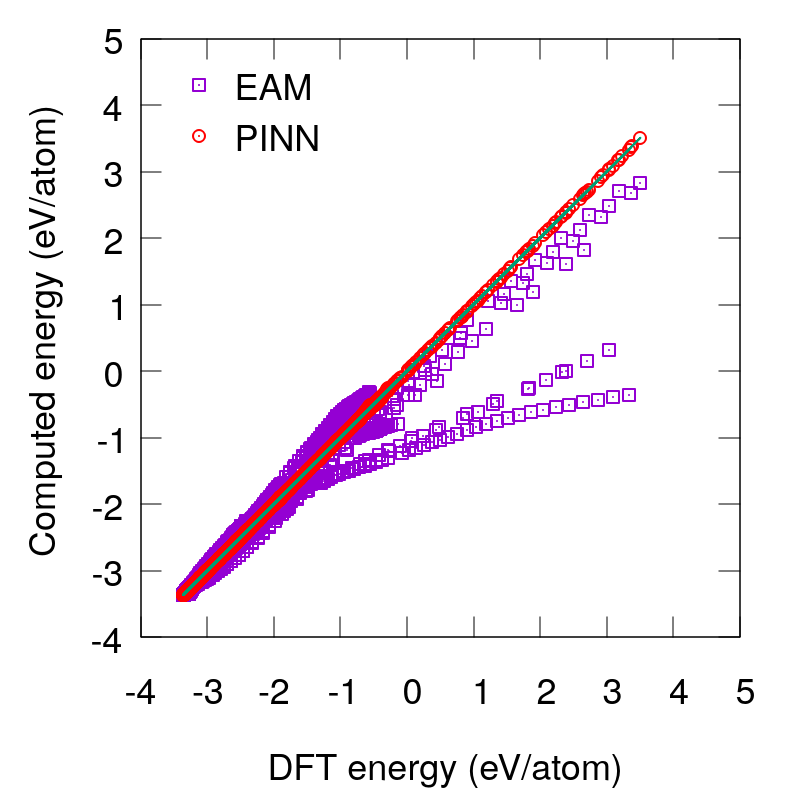}
\par\end{centering}
\noindent \begin{centering}
\bigskip{}
\bigskip{}
\par\end{centering}
\noindent \begin{centering}
\textbf{(b)}\includegraphics[width=0.47\textwidth]{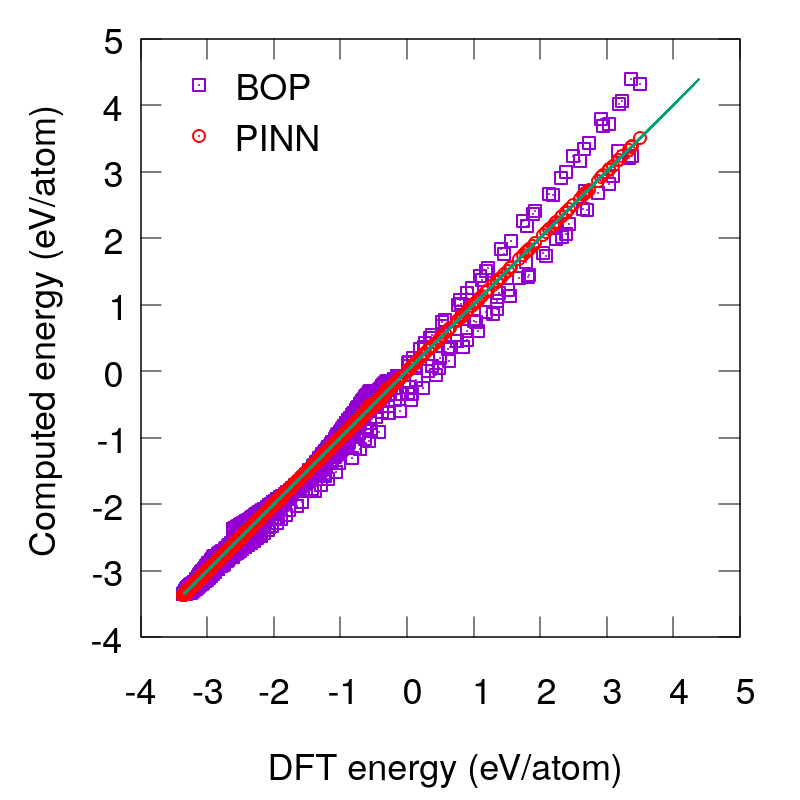}
\par\end{centering}
\caption{Energies of atomic configurations in the DFT database used for training
and validation compared with predictions of the (a) EAM Al potential
\citep{Mishin99b} and (b) BOP potential. The BOP parameters were
fitted to the DFT database and permanently fixed. The PINN potential
predictions are included for comparison. The straight line represents
the perfect fit. \label{fig:EAM_BOP_DFT_comparison}}

\end{figure}

It was also interesting to compare PINN potential with traditional,
parameter-based potentials for Al. One of them was the widely accepted
EAM Al potential \citep{Mishin99b} that had been fitted to a mix
of experimental and DFT data. The other was a BOP potential of the
same functional form as in the PINN model. Its parameters were fitted
in this work using the DFT database as for the PINN/NN potentials
and then fixed once and for all. Fig.~\ref{fig:EAM_BOP_DFT_comparison}
compares the DFT energies with the energies predicted by the EAM and
BOP models across the entire set of reference configurations. The
PINN predictions are shown for comparison. The plots demonstrate that
the traditional, fixed-parameter models generally follow the correct
trend but become increasingly less accurate as the structures deviate
from the equilibrium, low-energy atomic configurations. The adaptivity
to the local atomic environments built into the PINN potential greatly
improves the accuracy.

\section{Discussion and conclusions}

The proposed PINN potential model is capable of achieving the same
high accuracy in interpolating between DFT energies on the PES as
the currently existing mathematical NN potentials. The construction
of PINN potentials requires the same type of DFT database, is equally
straightforward, and does not heavily rely on human intuition. However,
extrapolation outside the domain of atomic configurations represented
in the training database is now based on a physical model of interatomic
bonding. As a result, the extrapolation becomes more reliable, or
at least more failure-proof, than the purely mathematical extrapolation.
The accuracy of interpolation can also be improved for the same reason.
As an example, the PINN Al potential constructed in this paper demonstrates
better accuracy of interpolation and significantly improved transferability
than a regular NN potential with about the same number of parameters.
The advantage of the PINN potential is especially strong for atomic
forces, which are important for molecular dynamics. The potential
could be used for accurate simulations of mechanical behavior and
other processes in Al. Construction of general-purpose PINN potentials
for Si and Ge is currently in progress.

We believe that the development of physics-based ML potentials is
the best way forward in this field. Such potentials need not be limited
to NNs or the particular BOP model adopted in this paper. Other regression
methods can be employed and the interatomic bonding model can be made
more sophisticated, or the other way round, simpler in the interest
of speed.

\begin{figure}
\noindent \begin{centering}
\includegraphics[width=0.95\textwidth]{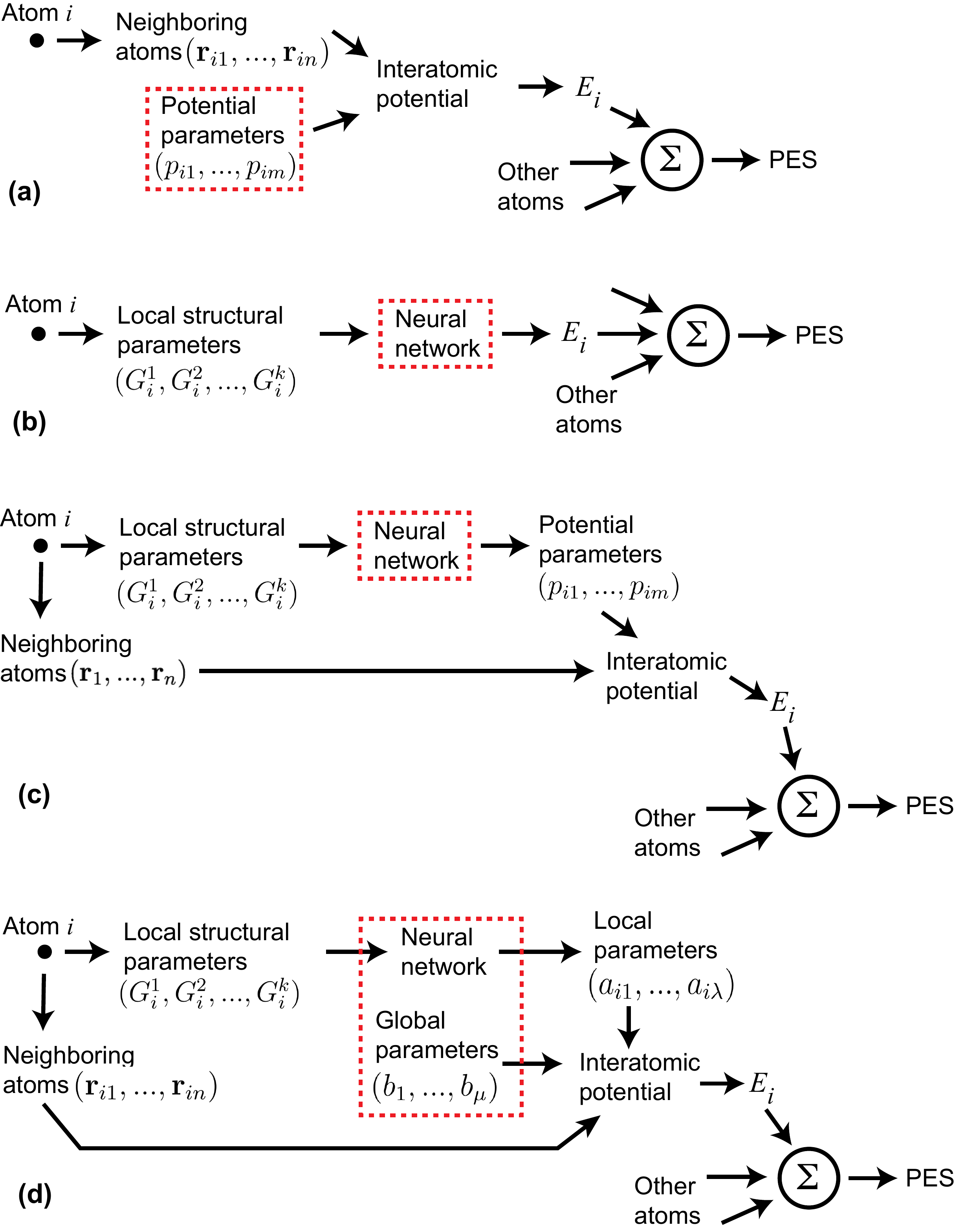}
\par\end{centering}
\caption{Flowcharts of the development of atomistic potentials: (a) Traditional
interatomic potential, (b) Mathematical NN potential, (c) Physically-informed
NN (PINN) potential with all-local parameters, (d) PINN potential
with parameters divided into local and global. The dashed rectangle
outlines the objects requiring parameter optimization. PES is the
potential energy surface of the material.\label{fig:Flowcharts_1}}
\end{figure}

Other modifications are envisioned in the future. For example, not
all potential parameters are equally sensitive to local environments.
To improve the computational efficiency, the parameters can be divided
in two subsets \citep{Malshe:2008aa}: local parameters $\mathbf{a}_{i}=(a_{i1},...,a_{i\lambda})$
adjustable according to the local environments as discussed above,
and global parameters $\mathbf{b}=(b_{1},...,b_{\mu})$ that are fixed
after the optimization and used for all environments (as in the traditional
potentials). The potential format now becomes
\begin{equation}
E_{i}=E_{i}\left(\mathbf{r}_{i1},...,\mathbf{r}_{in},\mathbf{a}_{i}\left(G_{i}^{l}(\mathbf{r}_{i1},...,\mathbf{r}_{in})\right),\mathbf{b}\right).\label{eq:15}
\end{equation}
During the training process, the global parameters $\mathbf{b}$ and
the network weights and biases are optimized \emph{simultaneously},
as shown in Fig.~\ref{fig:Flowcharts_1}d. Extension of PINN potentials
to binary and multicomponent systems is another major task for the
future.

All ML potentials are orders of magnitude faster than straight DFT
calculations but inevitably much slower than the traditional potentials.
Preliminary tests indicate that PINN potentials are about a factor
of two slower than regular NN potentials for the same number of parameters,
the extra overhead being due to the BOP calculation. All computations
reported in this paper utilized in-house software parallelized with
MPI for training and with OpenMP for MD and MC simulations (see example
in Fig.~\ref{fig:S_MD_example}). Collaborative work is underway
to develop highly scalable HPC software packages for physically-informed
ML potential training and MD/MC simulations using multiple CPUs or
GPUs, or both.

\newpage{}
\noindent \begin{center}
\textbf{METHODS}
\par\end{center}

The main ingredients of the proposed PINN method are the local structural
parameters $G_{i}^{l}$, the BOP potential, and the NN.

There are many possible ways of choosing local structural parameters
\citep{Behler07,Behler:2008aa,Eshet2010,Behler:2011aa,Behler:2011ab,Sosso2012,Behler:2015aa,Behler:2016aa}.
After trying several options, the following set of $G_{i}^{l}$'s
was selected. For an atom $i$, we define
\begin{equation}
g_{i}^{(m)}=\sum_{j,k}P_{m}\left(\cos\theta_{ijk}\right)f(r_{ij})f(r_{ik}),\enskip\enskip m=0,1,2,...,\label{eq:5-1-1}
\end{equation}
where $r_{ij}$ and $r_{ik}$ are distances to atoms $j$ and $k$,
respectively, and $\theta_{ijk}$ is the angle between the bonds $ij$
and $ik$. In Eq.(\ref{eq:5-1-1}), $P_{m}(x)$ is the Legendre polynomial
of order $m$ and 
\begin{equation}
f(r)=\dfrac{1}{\sigma^{3}}e^{-(r-r_{0})^{2}/\sigma^{2}}f_{c}(r)\label{eq:5-1}
\end{equation}
is a truncated Gaussian of width $\sigma$ centered at point $r_{0}$.
The truncation function $f_{c}(r)$ is defined by
\begin{equation}
f_{c}(r)=\begin{cases}
\dfrac{(r-r_{c})^{4}}{d^{4}+(r-r_{c})^{4}}\enskip & r\leq r_{c}\\
0,\enskip & r\geq r_{c}.
\end{cases}\label{eq:6-1}
\end{equation}
This function and its derivatives up to the third go to zero at a
cutoff distance $r_{c}$. The parameter $d$ controls the truncation
range.

For example, $P_{0}(x)=1$ and $g_{i}^{(0)}$ characterizes the local
atomic density near atom $i$. Likewise, $P_{1}(x)=x$ and $g_{i}^{(1)}$
can be interpreted as the dipole moment of a set of unit charges placed
at the atomic positions $j$ and $k$. As such, this parameter measures
the degree of local deviation from spherical symmetry in the environment
($g_{i}^{(1)}=0$ for spherical symmetry). For $m=2$, we have $P_{2}(x)=(3x^{2}-1)/2$
and $g_{i}^{(2)}$ is related to the quadrupole moment of a set of
unit charges placed at the atomic positions around atom $i$. We found
that polynomials up to degree $m=6$ should be included to accurately
represent the diverse atomic environment. Each $g_{i}^{(l)}$ is computed
for several values of $\sigma$ and $r_{0}$ spanning a range of interatomic
distances. For each atom, the set of $n$ $g_{i}^{(m)}$'s obtained
is arranged in a one-dimensional array $(G_{i}^{1},G_{i}^{2},...,G_{i}^{k})$.
In this work we chose $\sigma=1.0$ and used polynomials with $m=0,1,2,4,6$
for 12 $r_{0}$ values, giving a total of $k=60$ $G_{i}^{l}$'s.

In the BOP model adopted in this work, the energy of an atom $i$
is postulated in the form
\begin{equation}
E_{i}=\dfrac{1}{2}\sum_{j\neq i}\left[e^{A_{i}-\alpha_{i}r_{ij}}-S_{ij}b_{ij}e^{B_{i}-\beta_{i}r_{ij}}\right]f_{c}(r_{ij})+E_{i}^{(p)},\label{eq:16}
\end{equation}
where $r_{ij}$ is the distance between atoms $i$ and $j$ and the
summation is over all atom $j$ other than $i$ within the cutoff
radius $r_{c}$. The bond-order parameter $b_{ij}$ is taken in the
form 
\begin{equation}
b_{ij}=(1+z_{ij})^{-1/2},\label{eq:17}
\end{equation}
where 
\begin{equation}
z_{ij}=a_{i}^{2}\sum_{k\neq i,j}S_{ik}(\cos\theta_{ijk}+h_{i})^{2}f_{c}(r_{ik})\label{eq:18}
\end{equation}
represents the number of chemical bonds (other than $ij)$ formed
by atom $i$. Larger $z_{ij}$ values (more bonds) lead to a smaller
$b_{ij}$ and thus weaker $ij$ bond.

The screening factor $S_{ij}$ reduces the strength of bonds by surrounding
atoms. For example, when counting the bonds in (\ref{eq:18}), we
screen them by $S_{ik}$, so that strongly screened bonds contribute
less to $z_{ij}$. The screening factor $S_{ij}$ is given by 
\[
S_{ij}=\prod_{k\neq i,j}S_{ijk},
\]
where the partial screening factor $S_{ijk}$ represents the contribution
of a neighboring atom $k$ (different from $i$ and $j$) to the screening
of the bond $ij$. $S_{ijk}$ is given by
\begin{equation}
S_{ijk}=1-f_{c}(r_{ik}+r_{jk}-r_{ij})e{}^{-\lambda_{i}^{2}(r_{ik}+r_{jk}-r_{ij})}.\label{eq:19}
\end{equation}
It has the same value for all atoms $k$ located on the surface of
an imaginary spheroid whose poles coincide with the atoms $i$ and
$j$. For all atoms $k$ outside this ``cutoff spheroid'', on which
$r_{ik}+r_{jk}-r_{ij}=r_{c}$, we have $S_{ijk}=1$ -- such atoms
are too far away to screen the bond. If an atom $k$ is placed on
the line between the atoms $i$ and $j$, we have $r_{ik}+r_{jk}-r_{ij}=0$
and $S_{ijk}$ is small -- the bond $ij$ is strongly screened (almost
broken) by the atom $k$. This behavior reasonably reflects the nature
of chemical bonding.

Finally, the promotion energy $E_{i}^{(p)}$ is taken in the form
\begin{equation}
E_{i}^{(p)}=-\sigma_{i}\left({\displaystyle \sum_{j\neq i}S_{ij}b_{ij}}f_{c}(r_{ij})\right)^{1/2}.\label{eq:20}
\end{equation}
For a covalent material, $E_{i}^{(p)}$ accounts for the energy cost
of changing the electronic structure of a free atoms before it forms
chemical bonds. For example, for group IV elements, this is the cost
of the s$^{2}$p$^{2}$ $\rightarrow$ sp$^{3}$ hybridization. On
the other hand, $E_{i}^{(p)}$ can be interpreted as the embedding
energy 
\begin{equation}
F(\bar{\rho}_{i})=-\sigma_{i}\left(\bar{\rho}_{i}\right)^{1/2}\label{eq:21-1}
\end{equation}
appearing in the EAM formalism \citep{Daw83,Daw84}. Here, the host
electron density on atom $i$ is given by $\bar{\rho}_{i}=\sum_{j\neq i}S_{ij}b_{ij}f_{c}(r_{ij})$.
Due to this feature, this BOP model can be applied to both covalent
and metallic systems.

The BOP functions depend on 8 parameters $A_{i}$, $B_{i}$, $\alpha_{i}$,
$\beta_{i}$, $a_{i}$, $h_{i}$, $\sigma_{i}$ and $\lambda_{i}$,
which constitute the parameter set $(p_{1},...,p_{m})$ with $m=8$.
The cutoff parameters were fixed at $r_{c}=6$ \AA\ and $d=1.5$
\AA.

The feedforward NN contained two hidden layers and had the $60\times15\times15\times8$
architecture for the PINN potential and $60\times16\times16\times1$
for the NN potential. The number of nodes in the hidden layers was
chosen to reach the target accuracy of about 4 meV/atom without overfitting.

The training/validation database consisted of DFT total energies for
a set of supercells. The DFT calculations were performed using projector-augmented
wave (PAW) pseudopotentials as implemented in the electronic structure
\textcolor{black}{Vienna Ab initio Simulation Package (VASP)} \citep{Kresse1996,Kresse1999}.
The generalized gradient approximation (GGA) was used in conjunction
with the Perdew, Burke, and Ernzerhof (PBE) density functional \citep{perdew92:gga_apps,PerdewBE96}.
The plane-wave basis functions up to a kinetic energy cutoff of 520
eV were used, with the $k$-point density chosen to achieve convergence
to a few meV/atom level. Further details of the DFT calculations can
be found in \citep{Botu:2015aa,Botu:2015bb}. The energy of a given
supercell $s$, $E^{s}=\sum_{i}E_{i}^{s}$, predicted by the potential
was compared with the DFT energy $E_{\textrm{DFT}}^{s}$. Note that
the original $E_{\textrm{DFT}}^{s}$ values were not corrected to
remove the energy of a free atom. To facilitate comparison with literature
data, prior to the training all DFT energies were uniformly shifted
by 0.38446 eV/atom to match the experimental cohesive energy of Al,
3.36 eV/atom \citep{Kittel}. The NN was trained by adjusting its
weights $w_{\epsilon\kappa}$ and biases $b_{\kappa}$ to minimize
the objective function
\begin{equation}
\mathcal{E}=\sum_{s}\left(E^{s}-E_{\textrm{DFT}}^{s}\right)^{2}+\tau\left(\sum_{\epsilon\kappa}\left|w_{\epsilon\kappa}\right|^{2}+\sum_{\nu}\left|b_{\kappa}\right|^{2}\right)+\gamma\left(\sum_{\eta}\left|p_{\eta}-\overline{p}_{\eta}\right|^{2}\right).\label{eq:123-1}
\end{equation}
The second term was added to avoid overfitting by controlling the
magnitudes of the weights and biases. The parameter $\tau$ controls
the degree of regularization. The third term ensures the variations
of the PINN parameters relative to their database-averaged values
$\overline{p}_{\eta}$ remain small. The minimization of $\mathcal{E}$
was implemented by the Davidson-Fletcher-Powell algorithm of unconstrained
optimization. The optimization was repeated several times starting
from different random states and the solution with the smallest $\mathcal{E}$
was selected as final. The PINN and NN forces were computed by analytical
calculations using chain-rule differentiation.


\noindent \bigskip{}
\bigskip{}

\noindent \textbf{Acknowledgements}

\noindent The authors are grateful to Dr.~James Hickman for performing
some of the additional Al DFT calculations used for this work. The
authors acknowledge support of the Office of Naval Research under
Awards No.~N00014-18-1-2612 (G.~P.\ P.~P. and Y.~M.) and N00014-17-1-2148
(R.\ B. and R.\ R.). This work was also supported in part by a grant
of computer time from the DoD High Performance Computing Modernization
Program at ARL DSRC, ERDC DSRC and Navy DSRC.

\newpage\clearpage{}

\begin{table}[h]
\caption{Fitting and validation errors and related neural network information
for the straight NN and PINN models.}

\medskip{}
\label{tab:fit_errors} %
\begin{tabular}{lccccccccc}
\hline 
Model & NN architecture &  & Number of &  &  & RMSE of training &  &  & RMSE of validation\tabularnewline
 &  &  & parameters &  &  & (meV/atom) &  &  & (meV/atom)\tabularnewline
\hline 
NN & $60\times16\times16\times1$ &  & 1265 &  &  & 3.36 &  &  & 3.85\tabularnewline
PINN & $60\times15\times15\times8$ &  & 1283 &  &  & 3.46 &  &  & 3.59\tabularnewline
\hline 
\end{tabular}
\end{table}

\begin{table}[h]
\caption{Aluminum properties predicted by the PINN and NN potentials in comparison
with DFT calculations from the literature. $E_{0}$ - equilibrium
cohesive energy, $a_{0}$ - equilibrium lattice parameter, $B$ -
bulk modulus, $c_{ij}$ - elastic constants, $\gamma_{s}$ - surface
energy, $E_{v}^{f}$ - vacancy formation energy, $E_{v}^{m}$ - vacancy
migration barrier, $E_{I}^{f}$ - interstitial formation energy for
the tetrahedral ($T_{d}$) and octahedral ($O_{h}$) positions and
split dumbbell configurations with different orientations, $\gamma_{\textrm{SF}}$
- intrinsic stacking fault energy, $\gamma_{\textrm{us}}$ - unstable
stacking fault energy. All defect energies are statically relaxed
unless otherwise indicated. {*}Unstable and flips to the $\left\langle 100\right\rangle $
dumbbell orientation.}
\label{table:al_prop} \centering %
\begin{tabular}{lccc}
\hline 
Property & DFT & NN & PINN\tabularnewline
\hline 
$E_{0}$ (eV/atom) & $-3.7480$$^{a}$ & $-3.3606$ & $-3.3609$\tabularnewline
$a_{0}$ (\AA) & $4.039$$^{a,d}$; $3.9725\lyxmathsym{\textendash}4.0676$$^{c}$ & $4.0409$ & $4.0396$\tabularnewline
$B$ (GPa) & $83$$^{a}$; $81$$^{f}$ & $80$ & $79$\tabularnewline
$c_{11}$ (GPa) & $104$$^{a}$; $103\lyxmathsym{\textendash}106$$^{d}$ & $108$ & $117$\tabularnewline
$c_{12}$ (GPa) & $73$$^{a}$; $57\lyxmathsym{\textendash}66$$^{d}$ & $66$ & $60$\tabularnewline
$c_{44}$ (GPa) & $32$$^{a}$; $28\lyxmathsym{\textendash}33$$^{d}$ & $25$ & $32$\tabularnewline
$\gamma_{s}$(100) (Jm$^{-2}$) & $0.92$$^{b}$ & $0.897$ & $0.899$\tabularnewline
$\gamma_{s}$(110) (Jm$^{-2}$) & $0.98$$^{b}$ & $0.986$ & $0.952$\tabularnewline
$\gamma_{s}$(111) (Jm$^{-2}$) & $0.80$$^{b}$ & $0.837$ & $0.819$\tabularnewline
$E_{v}^{f}$ (eV) & $0.665\lyxmathsym{\textendash}1.346$$^{c}$; $0.7$$^{e}$ & $0.640$ & $0.678$\tabularnewline
$E_{v}^{f}$ (eV) unrelaxed & $0.78$$^{e}$ & $0.71$ & $0.77$\tabularnewline
$E_{v}^{m}$ (eV) & $0.304-0.621$$^{c}$ & $0.627$ & $0.495$\tabularnewline
$E_{I}^{f}$ ($T_{d}$) (eV) & $2.200\lyxmathsym{\textendash}3.294$$^{c}$ & $2.683$ & $2.840$\tabularnewline
$E_{I}^{f}$ ($O_{h}$) (eV) & $2.531\lyxmathsym{\textendash}2.948$$^{c}$ & $1.600$ & $2.367$\tabularnewline
$E_{I}^{f}$ $\left\langle 100\right\rangle $ (eV) & $2.295\lyxmathsym{\textendash}2.607$$^{c}$ & $1.529$ & $2.246$\tabularnewline
$E_{I}^{f}$ $\left\langle 110\right\rangle $ (eV) & $2.543\lyxmathsym{\textendash}2.981$$^{c}$ & $1.529$$^{*}$ & $2.713$\tabularnewline
$E_{I}^{f}$ $\left\langle 111\right\rangle $ (eV) & $2.679\lyxmathsym{\textendash}3.182$$^{c}$ & $2.631$ & $2.815$\tabularnewline
$\gamma_{\textrm{SF}}$ (mJ/m$^{2}$) & $134$$^{i}$ ; $146$$^{g}$; $158$$^{h}$ & $128$ & $121$\tabularnewline
$\gamma_{\textrm{us}}$ (mJ/m$^{2}$) & $162$$^{j}$ ; $169$$^{i}$ ; $175$$^{h}$ & $143$ & $132$\tabularnewline
\hline 
\multicolumn{4}{l}{$^{a}$ Ref.\,\citep{Jong:2015fk}; $^{b}$ Ref.\,\citep{Tran:2016qq};
$^{c}$ Ref.\,\citep{Qiu:2017ve}; $^{d}$ Ref.\,\citep{Zhuang:2016};
$^{e}$ Ref.\,\citep{Iyer:2014}; $^{f}$ Ref.\,\citep{Sjostrom:2016};
$^{g}$ Ref.\,\citep{Devlin:1974qp}}\tabularnewline
\multicolumn{4}{l}{$^{h}$ Ref.\,\citep{OgataLY02}; $^{i}$ Ref.\,\citep{Jahnatek:2009aa};
$^{j}$ Ref.\,\citep{Kibey:2007aa}}\tabularnewline
\end{tabular}
\end{table}

\newpage\clearpage{}

\newpage\clearpage{}
\noindent \begin{center}
\textsf{\textbf{\LARGE{}Supplementary Information}}{\LARGE\par}
\par\end{center}

\setcounter{figure}{0} \makeatletter  \renewcommand{\thefigure}{S\arabic{figure}}

\setcounter{table}{0} \makeatletter  \renewcommand{\thetable}{S\arabic{table}}
\noindent \begin{center}
\textbf{\Large{}Physically-informed artificial neural networks for
atomistic modeling of materials}{\Large\par}
\par\end{center}

\noindent \begin{center}
G. P. Purja Pun
\par\end{center}

\noindent \begin{center}
Department of Physics and Astronomy, MSN 3F3, George Mason University,
Fairfax, Virginia 22030, USA
\par\end{center}

\noindent \begin{center}
R. Batra
\par\end{center}

\noindent \begin{center}
Department of Materials Science and Engineering, University of Connecticut,
Storrs, CT 06269, USA
\par\end{center}

\noindent \begin{center}
R. Ramprasad
\par\end{center}

\noindent \begin{center}
School of Materials Science and Engineering, Georgia Institute of
Technology, Atlanta, GA 30332, USA
\par\end{center}

\noindent \begin{center}
Y. Mishin
\par\end{center}

\noindent \begin{center}
Department of Physics and Astronomy, MSN 3F3, George Mason University,
Fairfax, Virginia 22030, USA
\par\end{center}

\noindent 
\begin{table}
\caption{Al DFT database used in this work. The DFT data indicated by an asterisk
were computed in this work. The remaining data were randomly selected
from the database generated by Botu \textit{et al.}\,\citep{Botu:2015aa,Botu:2015bb}.
The structures are divided into datasets and further into groups according
to the structure type and physical conditions (temperature, deformation).
For NVE simulations, the table indicates the temperature of initial
thermalization with ideal atomic positions.}
\medskip{}
\label{tab:Al_database} %
\begin{tabular}{clclcc}
\hline 
Dataset & Structure & Group & Physical condition & $N_{A}$ & $N_{tv}$\tabularnewline
\hline 
Crystals & FCC{*} & 25 & Isotropic strain at 0\,K & 4 & 174\tabularnewline
 & BCC{*} & 14 & Isotropic strain at 0\,K & 2 & 174\tabularnewline
 & HCP{*} & 34 & Isotropic strain at 0\,K & 4 & 174\tabularnewline
 & SC{*} & 38 & Isotropic strain at 0\,K & 8 & 161\tabularnewline
 & DC{*} & 23 & Isotropic strain at 0\,K & 8 & 152\tabularnewline
 & FCC{*} & 26 & Uniaxial $\langle100\rangle$ at 0\,K & 4 & 81\tabularnewline
 & A15{*} & 13 & Isotropic strain at 0\,K & 8 & 137\tabularnewline
 & SH{*} & 35 & Isotropic strain at 0\,K & 1 & 169\tabularnewline
 & FCC{*} & 27 & Uniaxial $\langle100\rangle$ at 0\,K & 1 & 61\tabularnewline
 & FCC{*} & 28 & Uniaxial $\langle111\rangle$ at 0\,K & 24 & 60\tabularnewline
\hline 
FCC 1 & FCC ($a=4.036$\,\AA) & 24 & NVE-MD (2500\,K) & 32 & 60\tabularnewline
 & FCC ($a=4.036$\,\AA) & 24 & NVE-MD (700\,K) & 32 & 60\tabularnewline
 & FCC ($a=3.302$\,\AA){*} & 37 & NVT-MD (4000\,K) & 32 & 60\tabularnewline
 & FCC ($a=3.530$\,\AA){*} & 36 & NVT-MD (4000\,K) & 32 & 60\tabularnewline
\hline 
FCC 2 & FCC ($a=3.75$\,\AA) & 7 & NVE-MD (1200\,K) & 32 & 60\tabularnewline
 & FCC ($a=3.96$\,\AA) & 8 & NVE-MD(700\,K) & 32 & 60\tabularnewline
 & FCC ($a=4.00$\,\AA) & 12 & NVE-MD(700\,K) & 32 & 60\tabularnewline
 & FCC ($a=4.10$\,\AA) & 10 & NVE-MD(700\,K) & 32 & 60\tabularnewline
 & FCC ($a=4.15$\,\AA) & 9 & NVE-MD(700\,K) & 32 & 60\tabularnewline
 & FCC ($a=4.35$\,\AA) & 11 & NVE-MD(1200\,K) & 32 & 60\tabularnewline
\hline 
Surfaces & Surface (100) & 1 & NVE-MD (700\,K) & 144 & 50\tabularnewline
 & Surface (110) & 2 & NVE-MD (700\,K) & 128 & 60\tabularnewline
 & Surface (111) & 3 & NVE-MD (700\,K) & 16 & 60\tabularnewline
 & Surface (100) & 4 & NVE-MD (1000\,K) & 108 & 60\tabularnewline
 & Surface (311) & 5 & NVE-MD (1000\,K) & 88 & 60\tabularnewline
 & Surface (111) & 6 & NVE-MD (1000\,K) & 108 & 60\tabularnewline
\hline 
Defects & 1 Vacancy & 44 & NVE-MD (700\,K) & 31 & 210\tabularnewline
 & 1 adatom on (100) & 40 & NVE-MD (700\,K) & 76 & 60\tabularnewline
 & 2 adatoms on (111) & 41 & NVE-MD (700\,K) & 66 & 60\tabularnewline
 & Dimer on (111) & 42 & NVE-MD (700,2000\,K) & 66 & 60\tabularnewline
 & Trimer on (111) & 43 & NVE-MD (700,2000\,K) & 103 & 60\tabularnewline
\hline 
\multicolumn{2}{l}{Continued in Table \ref{tab:Al_database_contd}} &  &  &  & \tabularnewline
\end{tabular}
\end{table}

\noindent 
\begin{table}
\noindent \caption{Aluminum DFT database (continued from Table \ref{tab:Al_database}).}
\medskip{}
\label{tab:Al_database_contd} %
\begin{tabular}{llclcc}
\hline 
Dataset & Structure & Group & Physical condition & $N_{A}$ & $N_{tv}$\tabularnewline
\hline 
Clusters & Dimer & 20 & NVE-MD (300\,K) & 2 & 60\tabularnewline
 & 2.5\,$\textrm{\AA}$ cluster & 15 & NVE-MD (300\,K) & 6 & 60\tabularnewline
 & 4\,$\textrm{\AA}$ cluster & 16 & NVE-MD (300\,K) & 13 & 60\tabularnewline
 & 4.5\,$\textrm{\AA}$ cluster & 18 & NVE-MD (300\,K) & 19 & 60\tabularnewline
 & 5\,$\textrm{\AA}$ cluster{*} & 13 & NVE-MD (1200\,K) & 42 & 60\tabularnewline
 & 6.5\,$\textrm{\AA}$ cluster{*} & 19 & NVE-MD (1200\,K) & 79 & 60\tabularnewline
 & Small icosahedron{*} & 21 & NVE-MD (900\,K) & 55 & 60\tabularnewline
 & Wulff cluster{*} & 22 & NVE-MD (1000\,K) & 79 & 60\tabularnewline
 & Wulff cluster{*} & 22 & NVE-MD (2000\,K) & 79 & 60\tabularnewline
\hline 
Interfaces & GB (510) & 23 & NVE-MD (700\,K) & 70 & 60\tabularnewline
 & GB (111) & 19 & NVE-MD (700\,K) & 24 & 60\tabularnewline
 & GB (210) & 20 & NVE-MD (700\,K) & 60 & 60\tabularnewline
 & GB (310) & 21 & NVE-MD (700\,K) & 42 & 60\tabularnewline
 & GB (320) & 22 & NVE-MD (700\,K) & 96 & 60\tabularnewline
 & SF$\langle211\rangle$(111){*} & 1 & Only atomic relaxation & 30 & 60\tabularnewline
\hline 
Total &  &  &  &  & 3649\tabularnewline
\hline 
\multicolumn{6}{l}{$N_{A}$ - number of atoms per supercell}\tabularnewline
\multicolumn{6}{l}{$N_{tv}$ - number of configurations for training and validation}\tabularnewline
\multicolumn{6}{l}{\textcolor{black}{\uline{Notations:}} BCC (body centered cubic),
HCP (hexagonal closed packed)}\tabularnewline
\multicolumn{6}{l}{SC (simple cubic), DC (diamond cubic), SH (simple hexagonal)}\tabularnewline
\multicolumn{6}{l}{GB (grain boundary), SF (stacking fault). $a$ is the cubic lattice
parameter of the FCC structure}\tabularnewline
\end{tabular}
\end{table}

\noindent 
\begin{table}
\caption{Al DFT database used for testing. The data was extracted from the
database generated by Botu \textit{et al.}\,\citep{Botu:2015aa,Botu:2015bb}.
The structures are divided into datasets and further into groups according
to the structure type and physical conditions (temperature, deformation).
For NVE simulations, the table indicates the temperature of initial
thermalization with ideal atomic positions.}
\label{tab:Al_database_validation} \centering %
\begin{tabular}{clclc}
\hline 
Dataset & Structure & Run-type & $N_{A}$ & $N_{t}$\tabularnewline
\hline 
BCC & BCC ($a=2.621$\,\AA) & NVT-MD$^{a}$ & 54 & 2589\tabularnewline
 & BCC ($a=2.802$\,\AA) & NVT-MD$^{a}$ & 54 & 2607\tabularnewline
\hline 
HCP & HCP$^{\dagger}$ ($a=1.847$\,\AA) & NVT-MD$^{a}$ & 32 & 3880\tabularnewline
 & HCP$^{\dagger}$ ($a=1.975$\,\AA) & NVT-MD$^{a}$ & 32 & 3853\tabularnewline
\hline 
FCC 3 & FCC & NPT-MD (300,600,900\,K) & 32 & 6330\tabularnewline
 & FCC (EAM generated) & NPT-MD (300,600,900\,K) & 256 & 30\tabularnewline
\hline 
Defects & 2 Vacancies & NVE-MD (700\,K) & 254 & 578\tabularnewline
 & 6 Vacancies & NVE-MD (700\,K) & 860 & 165\tabularnewline
 & 8 adatoms on (111) & NVE-MD (1500\,K) & 253 & 1420\tabularnewline
 & 15 adatoms on (111) & NVE-MD (1500\,K) & 260 & 1397\tabularnewline
 & Dislocation & NVE-MD (700\,K) & 378 & 50\tabularnewline
\hline 
Clusters & 8\,\AA\  cluster & NVE-MD (1200\,K) & 135 & 1707\tabularnewline
 & 10\,\AA\  cluster & NVE-MD (1200\,K) & 249 & 249\tabularnewline
 & Octahedron cluster & NVE-MD (1000\,K) & 201 & 1570\tabularnewline
\hline 
Total &  &  &  & 26425\tabularnewline
\hline 
\multicolumn{5}{l}{$N_{A}$ - number of atoms per supercell}\tabularnewline
\multicolumn{5}{l}{$N_{t}$ - number of configurations for testing}\tabularnewline
\multicolumn{5}{l}{$^{a}$ 300\,K, 600\,K, 1000\,K, 1500\,K, 2000\,K and 4000\,K}\tabularnewline
\multicolumn{5}{l}{$^{\dagger}$ $c/a=1.648$}\tabularnewline
\end{tabular}
\end{table}

\newpage\clearpage{}

\begin{figure}
\noindent \begin{centering}
\textbf{(a)} \includegraphics[width=0.44\textwidth]{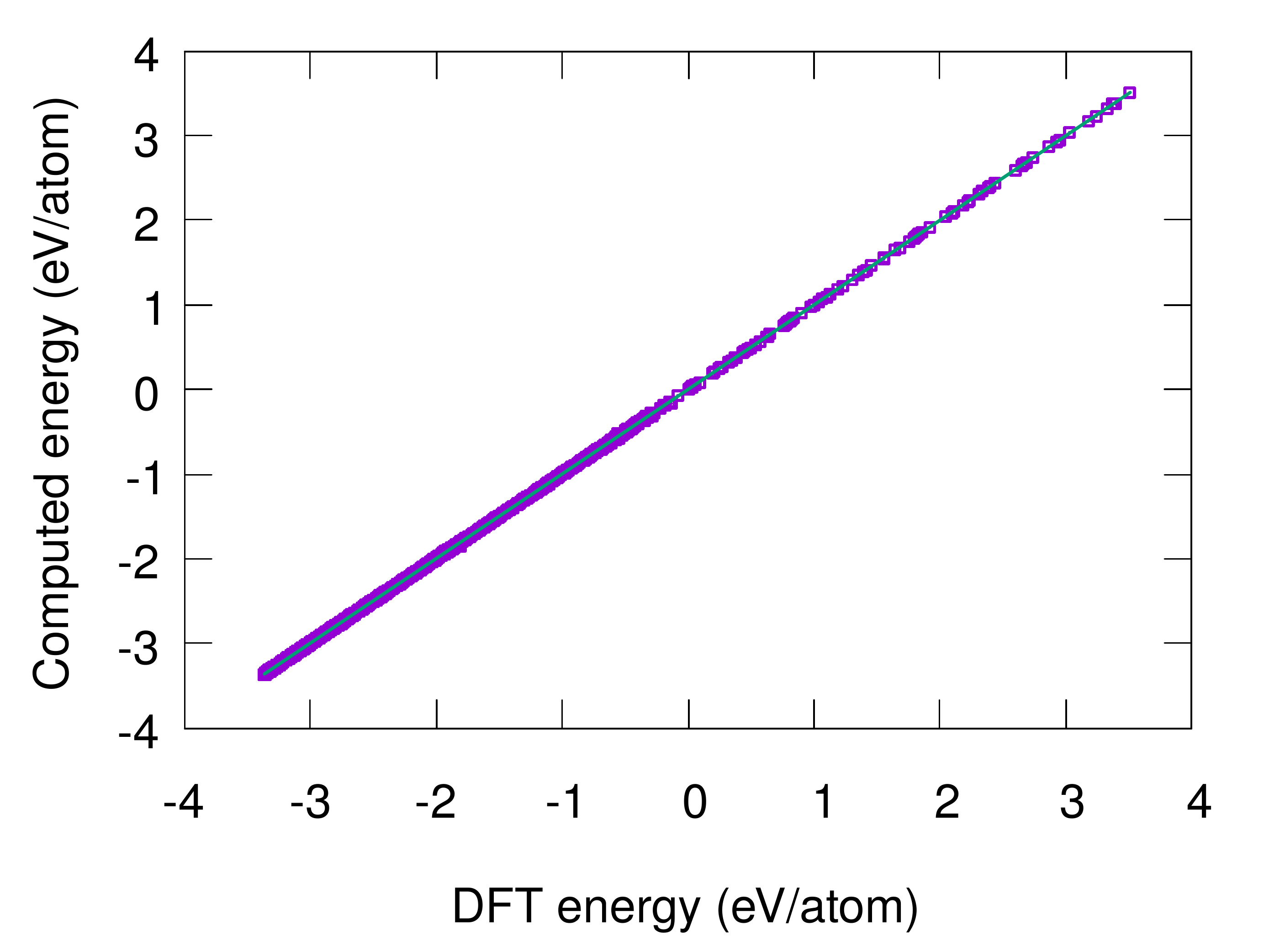}\quad{}\textbf{(b)}
\includegraphics[width=0.44\textwidth]{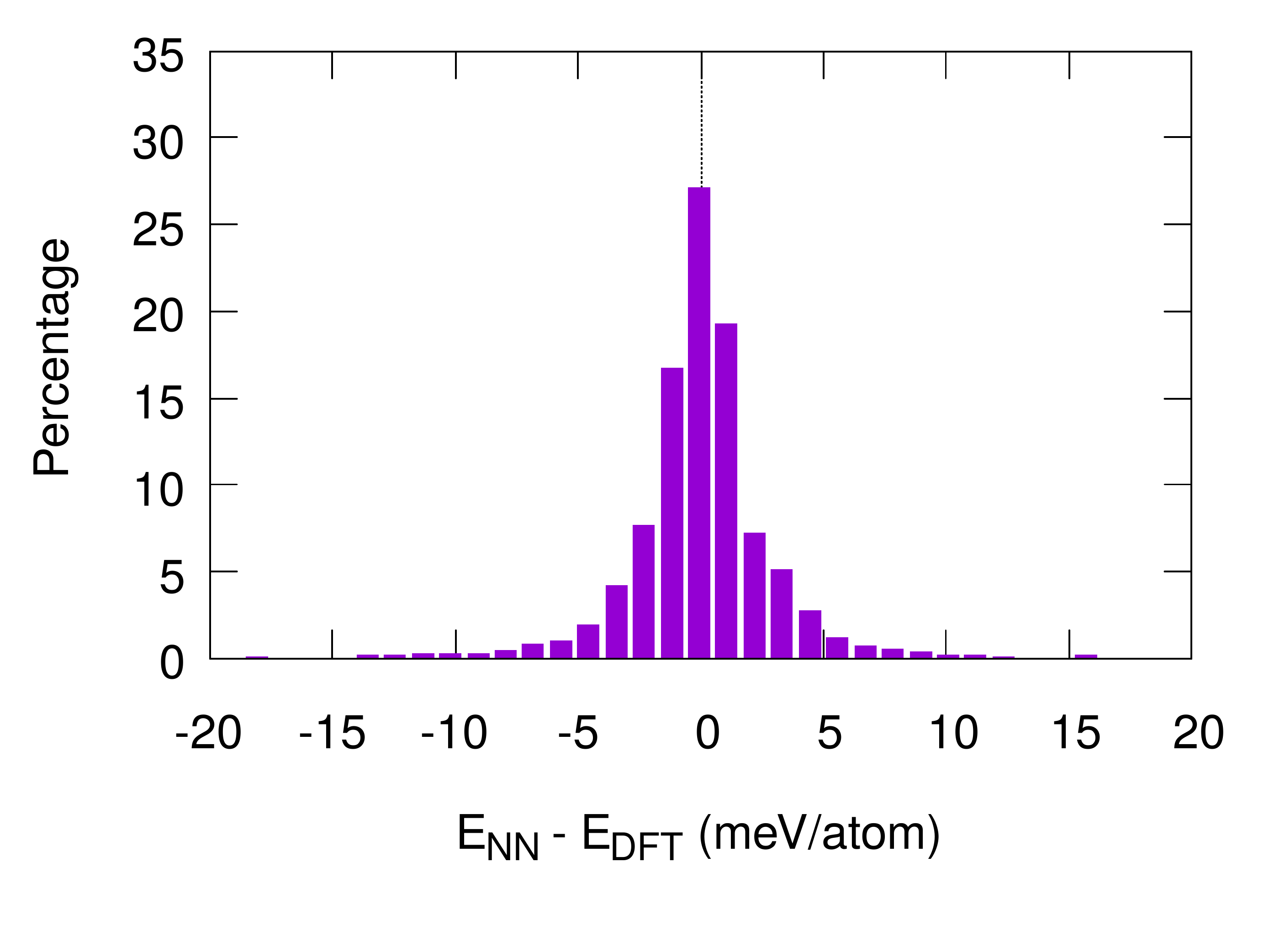}
\par\end{centering}
\caption{(a) Energies of atomic configurations in the training dataset computed
with the mathematical NN potentials versus DFT energies. The straight
line represents the perfect fit. (b) Error distribution in the training
dataset. \label{fig:NN_Correlation}}
\end{figure}

\begin{figure}
\noindent \begin{centering}
\textbf{(a)} \includegraphics[width=0.6\textwidth]{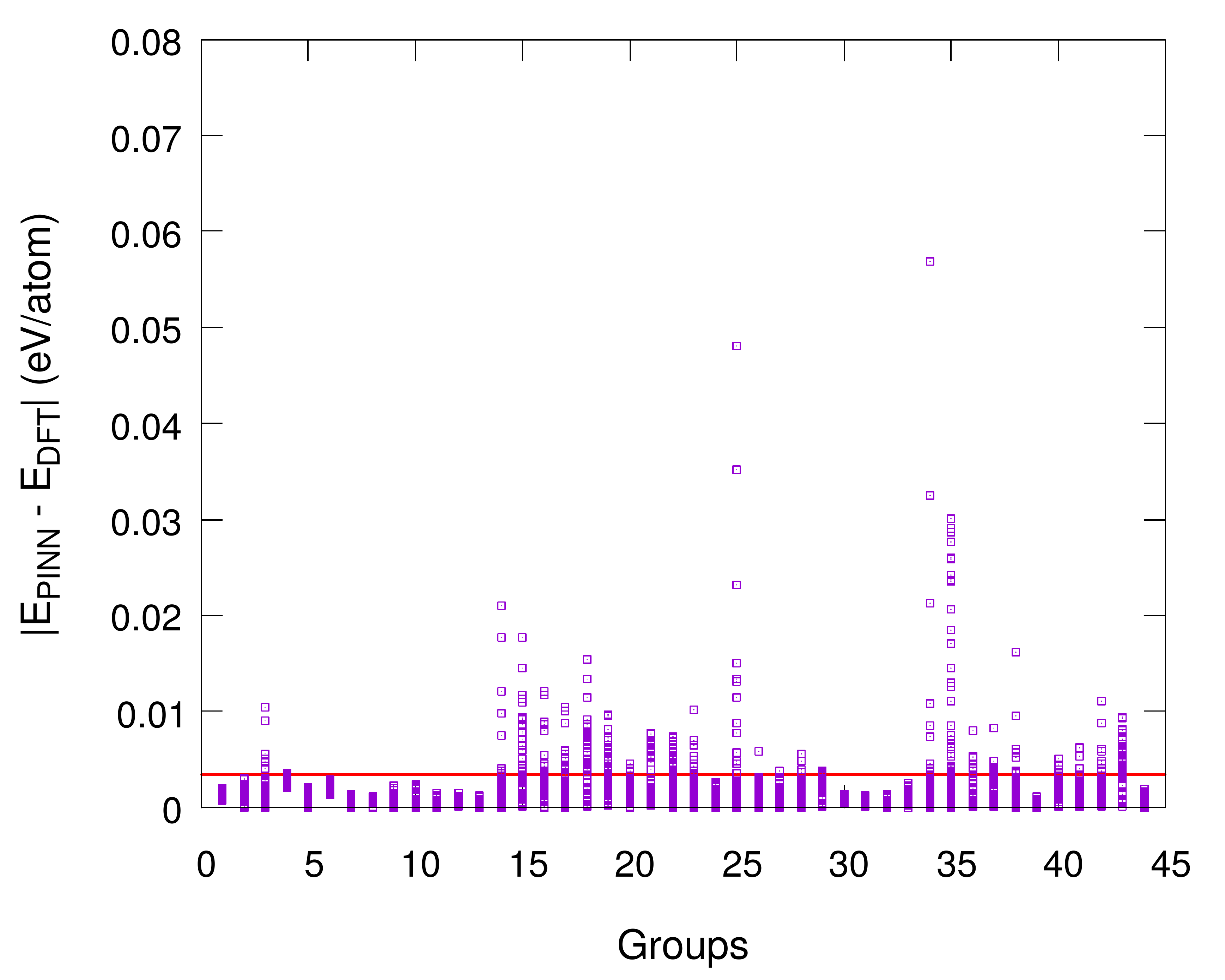}
\par\end{centering}
\bigskip{}

\noindent \begin{centering}
\textbf{(b)} \includegraphics[width=0.6\textwidth]{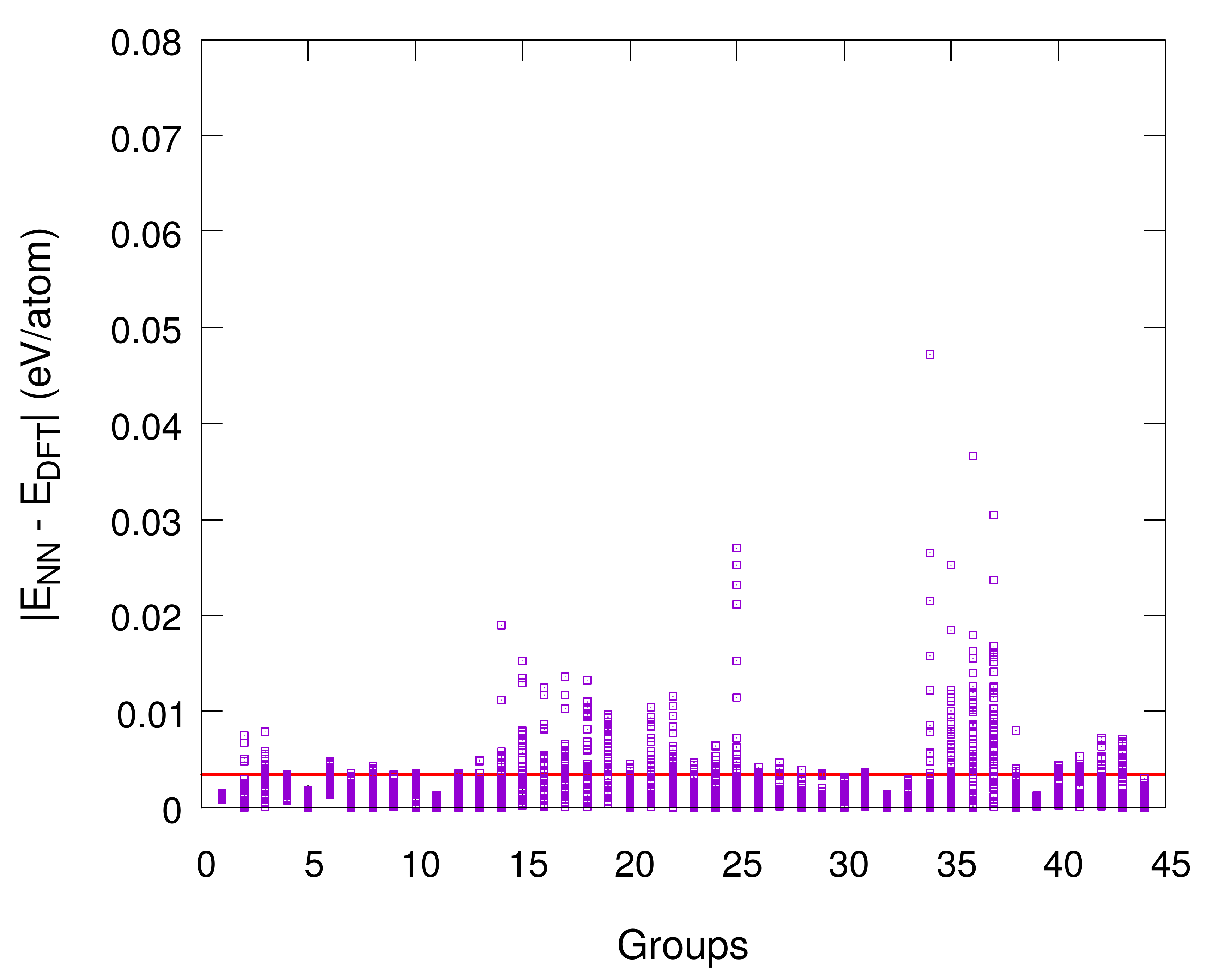}
\par\end{centering}
\caption{Absolute deviations of energies predicted by the PINN (a) and NN (b)
potentials from the DFT energies in individual groups of the training
dataset. (Refer to Tables\,\ref{tab:Al_database} and \ref{tab:Al_database_contd}
for the group numbers). The red line marks the RMSE (Table\,\ref{tab:fit_errors}).\label{fig:Error_by_group}}
\end{figure}

\begin{figure}
\textbf{(a)}\includegraphics[width=0.45\textwidth]{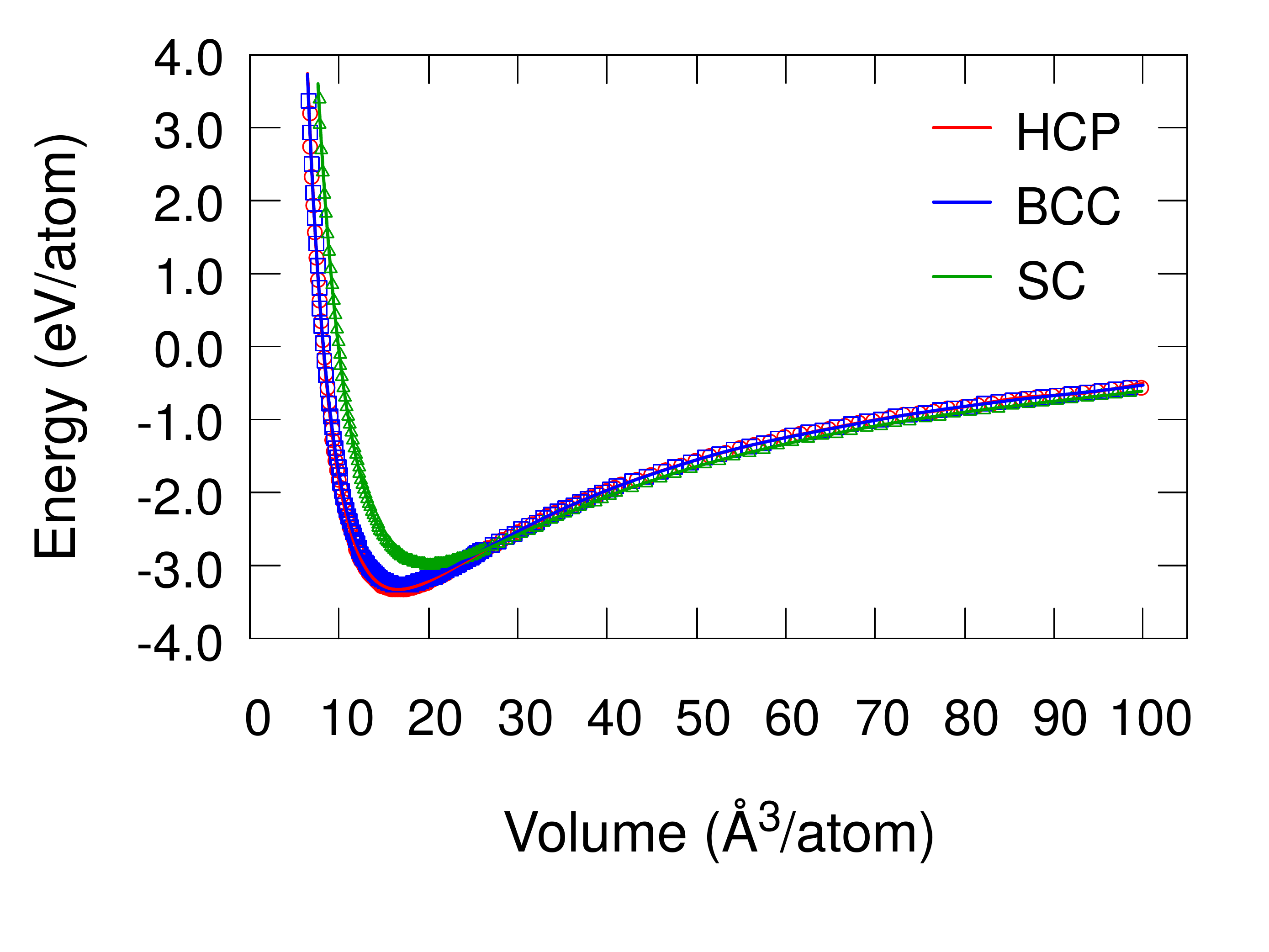}\quad{}\textbf{(b)}\includegraphics[width=0.45\textwidth]{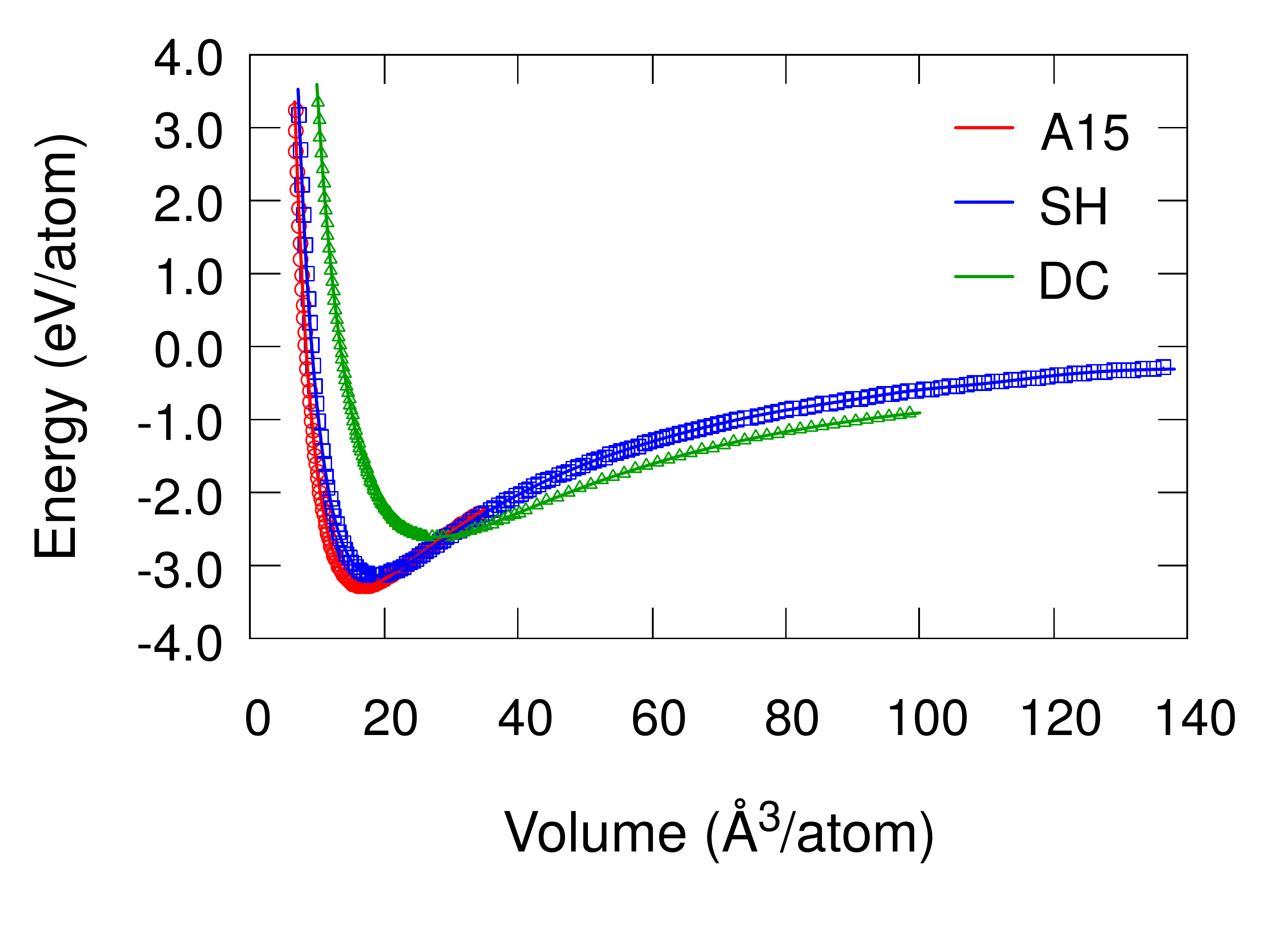}

\caption{Energy-volume relations for Al crystal structures predicted by the
NN potential (lines) and by DFT calculations (points). (a) Hexagonal
close-packed (HCP), body-centered cubic (BCC), and simple cubic (SC)
structures. (b) A15 (Cr$_{3}$Si prototype), simple hexagonal (SH),
and diamond cubic (DC) structures.\label{fig:EOS_NN}}
\end{figure}

\begin{figure}
\textbf{(a)} \includegraphics[width=0.45\textwidth]{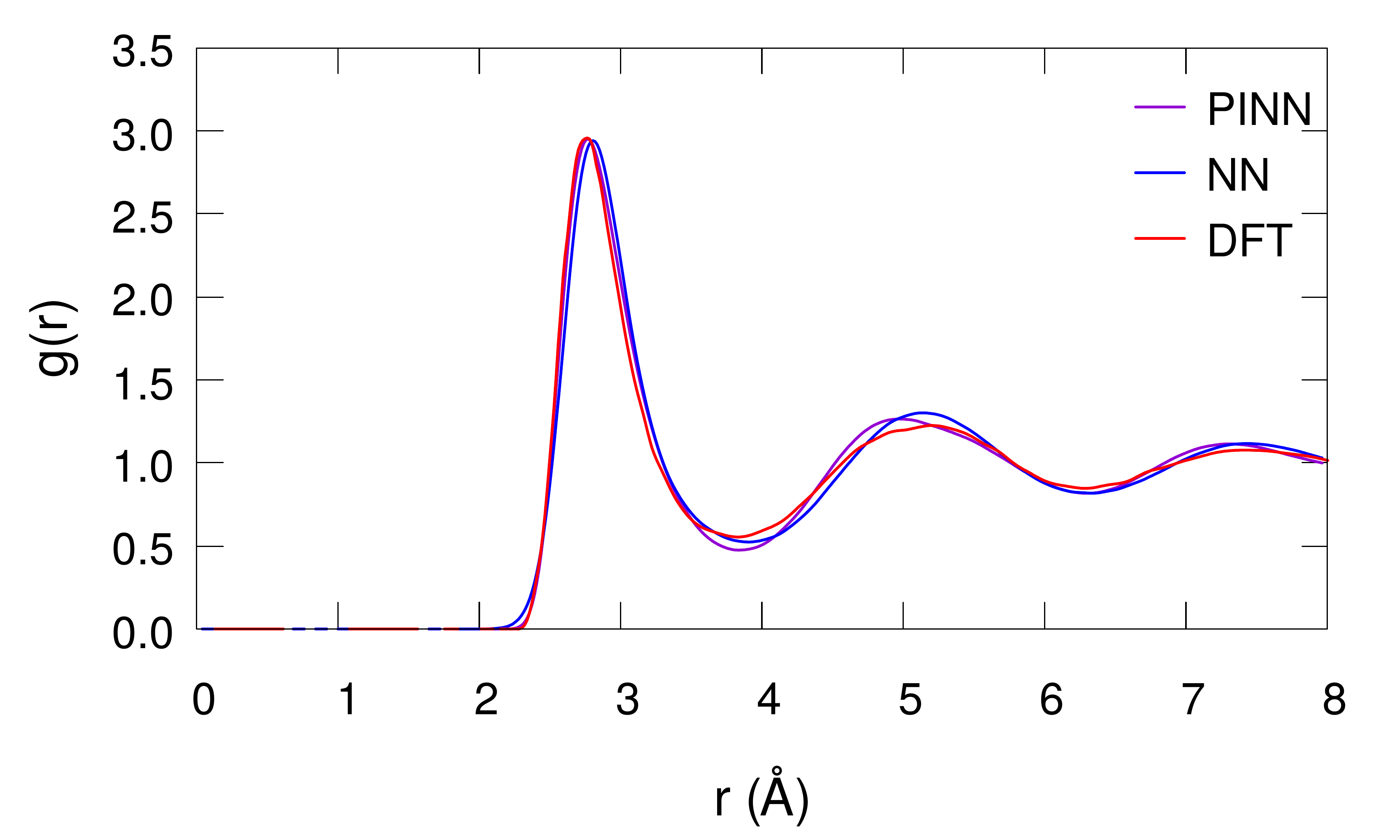}\enskip{}\textbf{(b)}\includegraphics[width=0.45\textwidth]{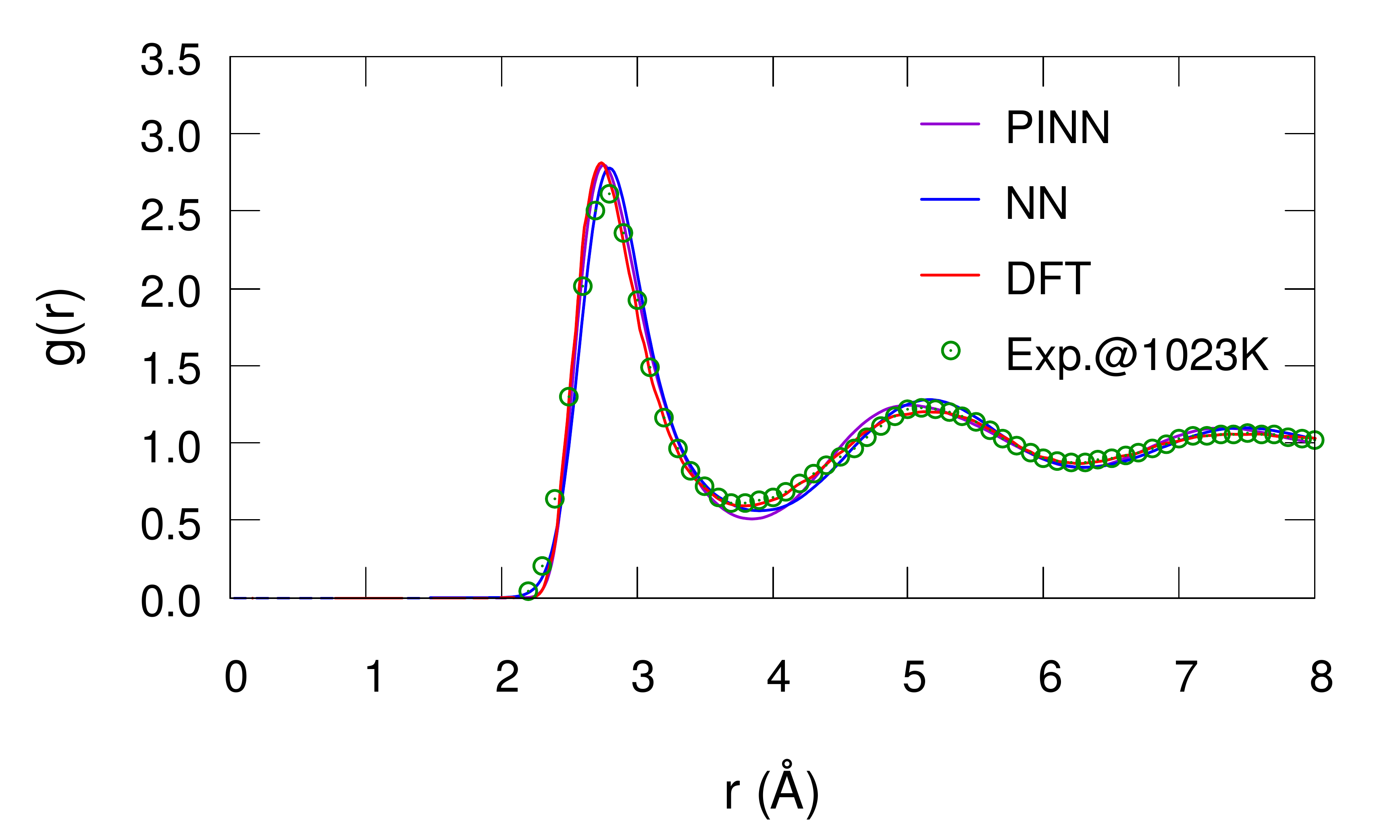}

\bigskip{}
\bigskip{}

\textbf{(c)}\includegraphics[width=0.45\textwidth]{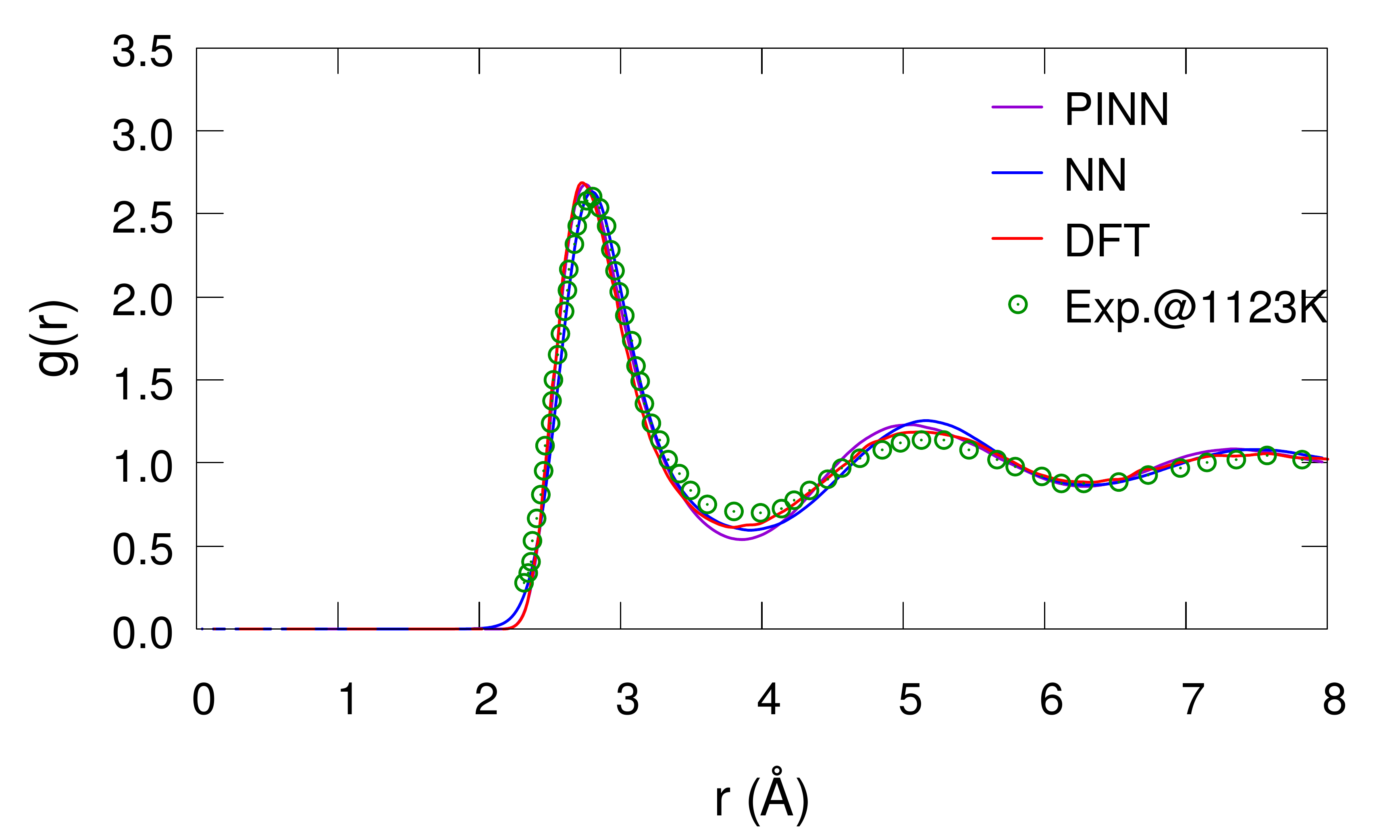}\enskip{}\textbf{(d)}\includegraphics[width=0.45\textwidth]{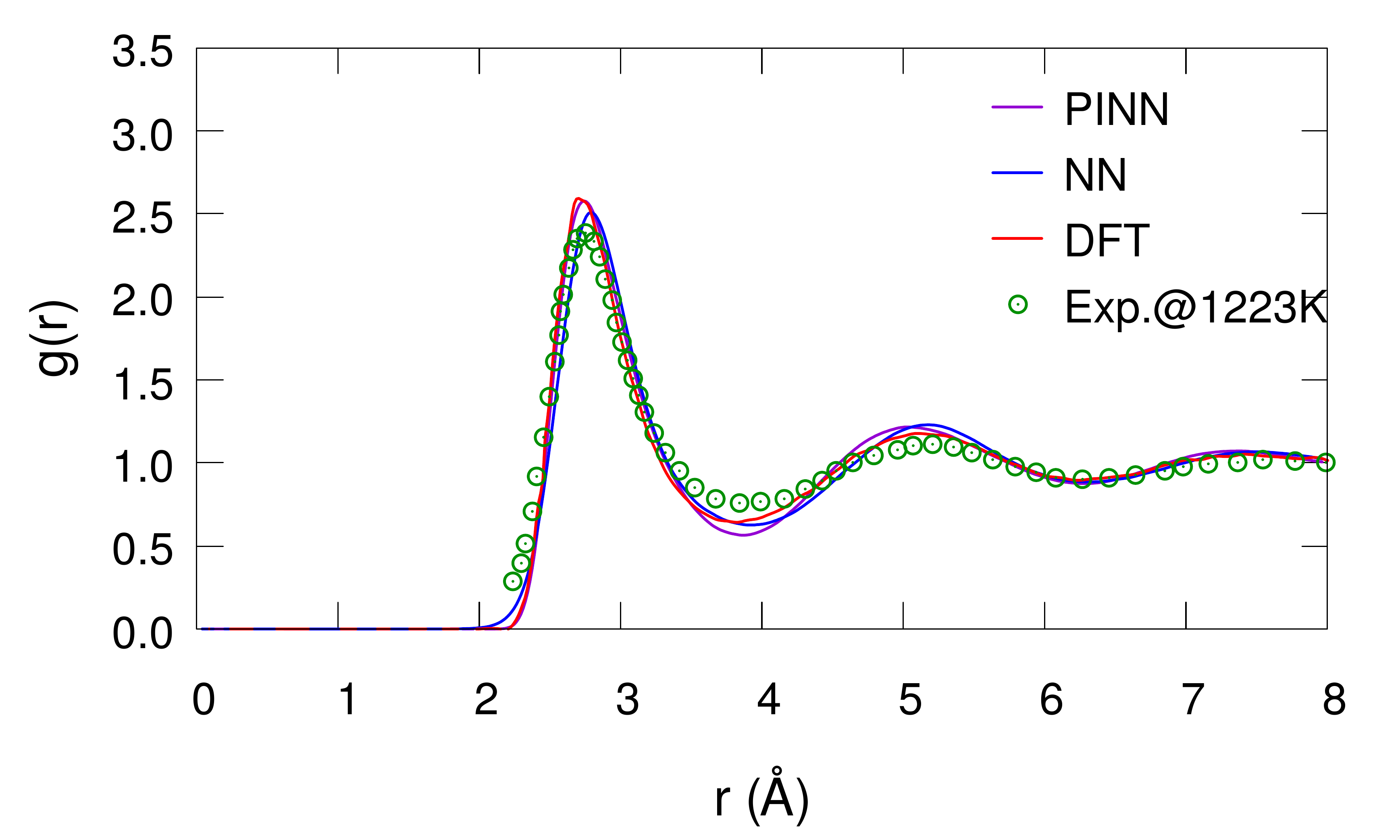}

\caption{Radial distribution functions $g(r)$ in liquid Al at the temperatures
of (a) 875\,K, (b) 1000\,K, (c) 1125\,K and (d) 1250\,K predicted
by the PINN and NN potentials in comparison with experimental data
\citep{Mauro:2011wc} and DFT calculations (Ref.~\citep{Jakse:2013aa}
and therein).\label{fig:RDF}}
\end{figure}

\begin{figure}
\noindent \begin{centering}
\includegraphics[width=0.5\textwidth]{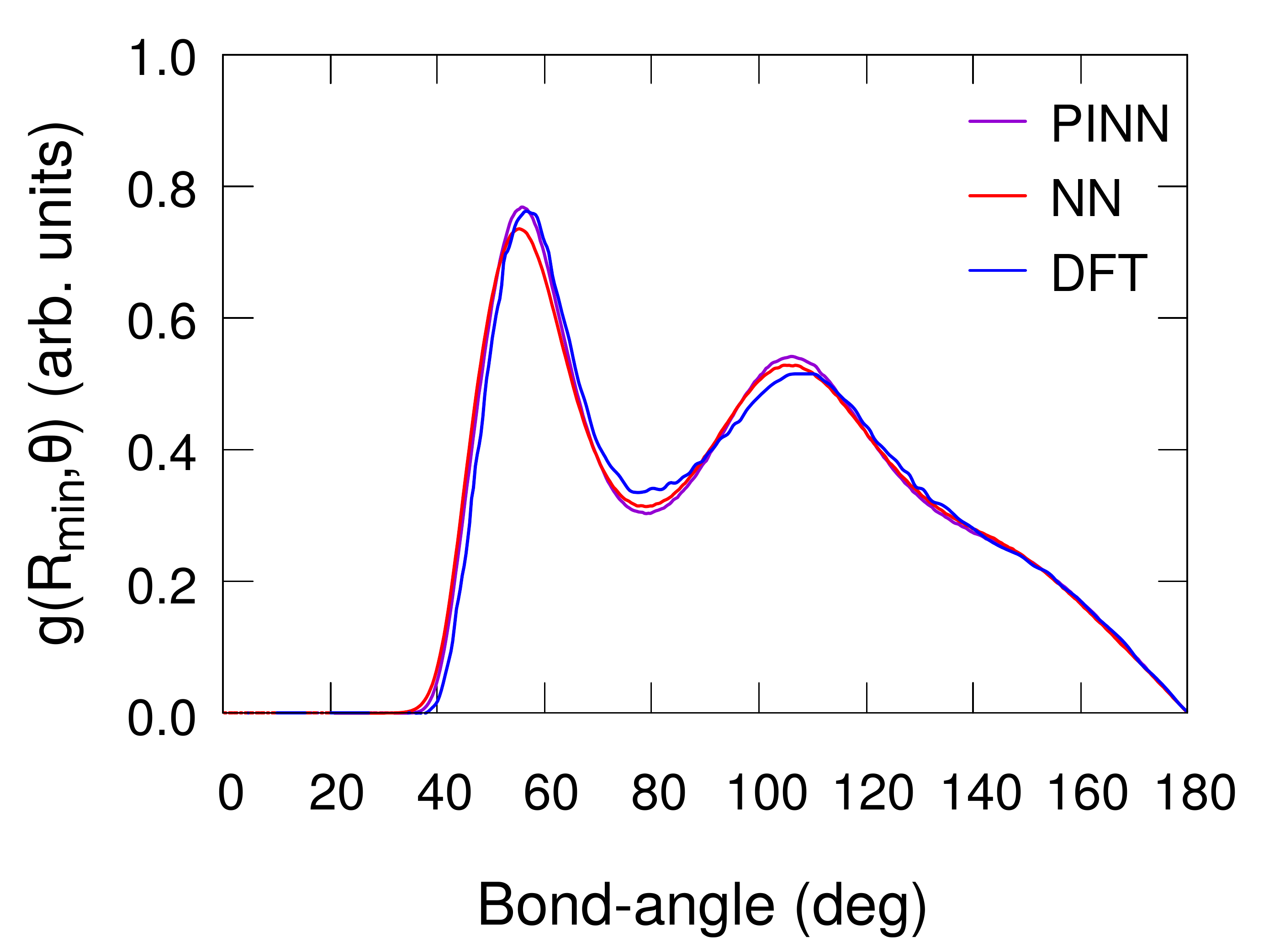}
\par\end{centering}
\caption{Bond angle distribution, $g(R_{min},\theta)$, in liquid aluminum
at 1000\,K in comparison with DFT calculations \citep{Alemany:2004}.
The calculation included the neighbors within the first minimum $R_{min}$
of the radial distribution function (cf.~Fig.~\ref{fig:RDF}). \label{fig:Bond-angle-distribution}}
\end{figure}

\begin{figure}
\textbf{(a)}\includegraphics[width=0.45\textwidth]{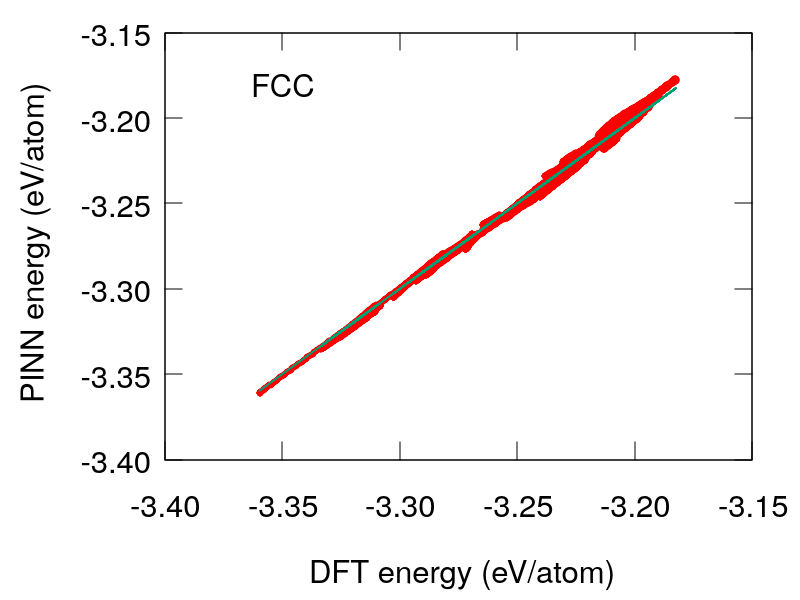}\enskip{}\textbf{(b)}\includegraphics[width=0.45\textwidth]{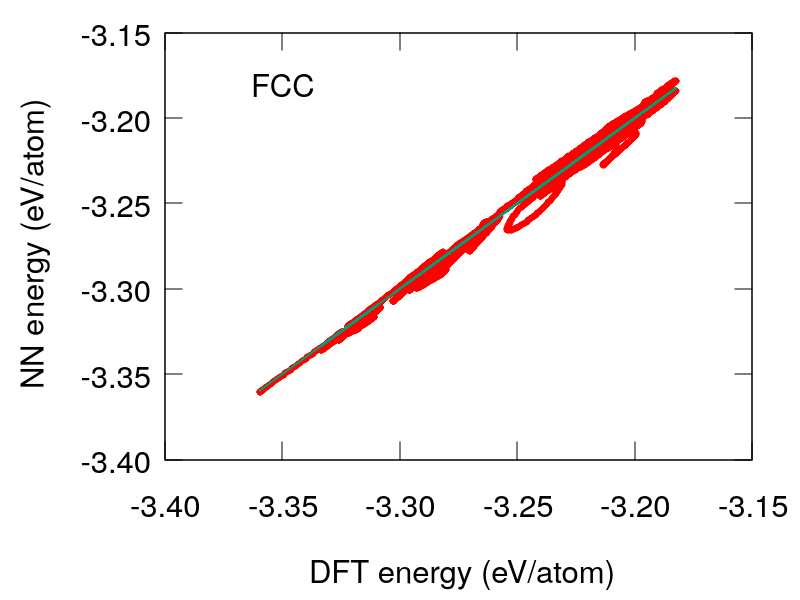}

\caption{Energy of FCC Al in NPT MD simulations at the temperatures of 300\,K
and 600\,K The energies predicted by the PINN (a) and NN (b) potentials
are compared with DFT calculations from \citep{Botu:2015aa,Botu:2015bb}.
The straight lines represent the perfect fit.\label{Fig:validation_2}}
\end{figure}

\begin{figure}
\textbf{(a)}\includegraphics[width=0.45\textwidth]{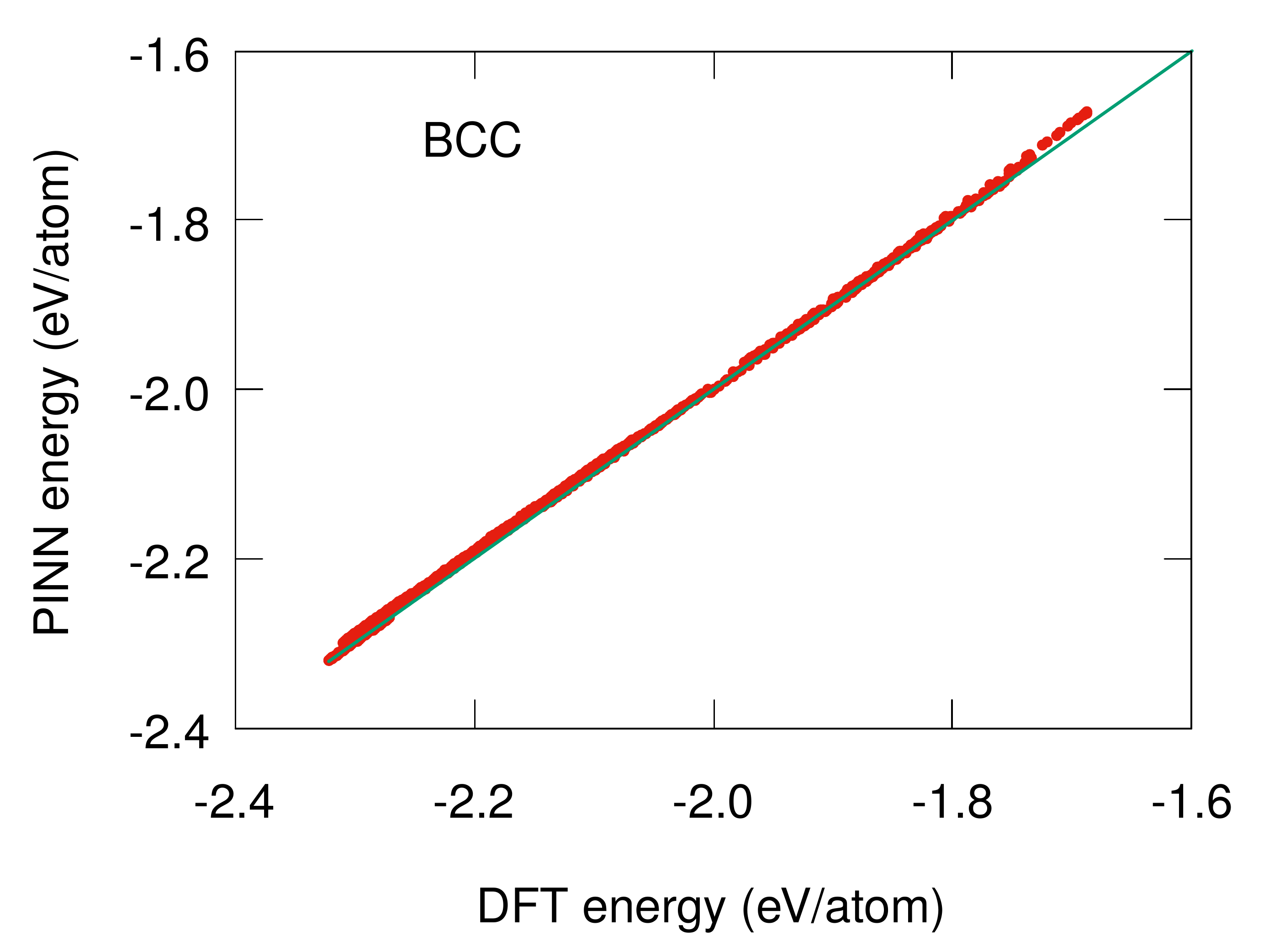}\enskip{}\textbf{(b)}\includegraphics[width=0.45\textwidth]{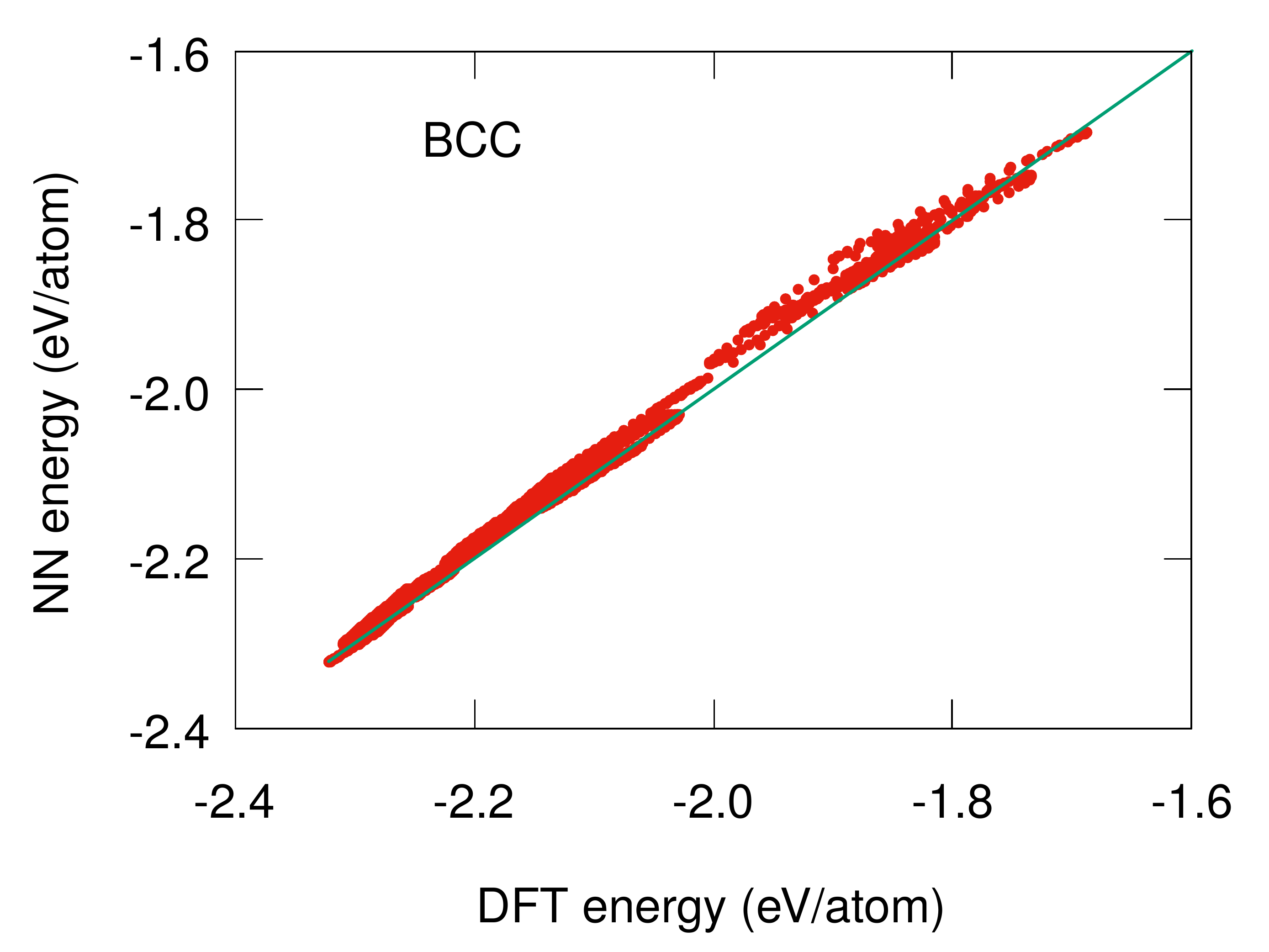}

\bigskip{}
\bigskip{}

\textbf{(c)}\includegraphics[width=0.45\textwidth]{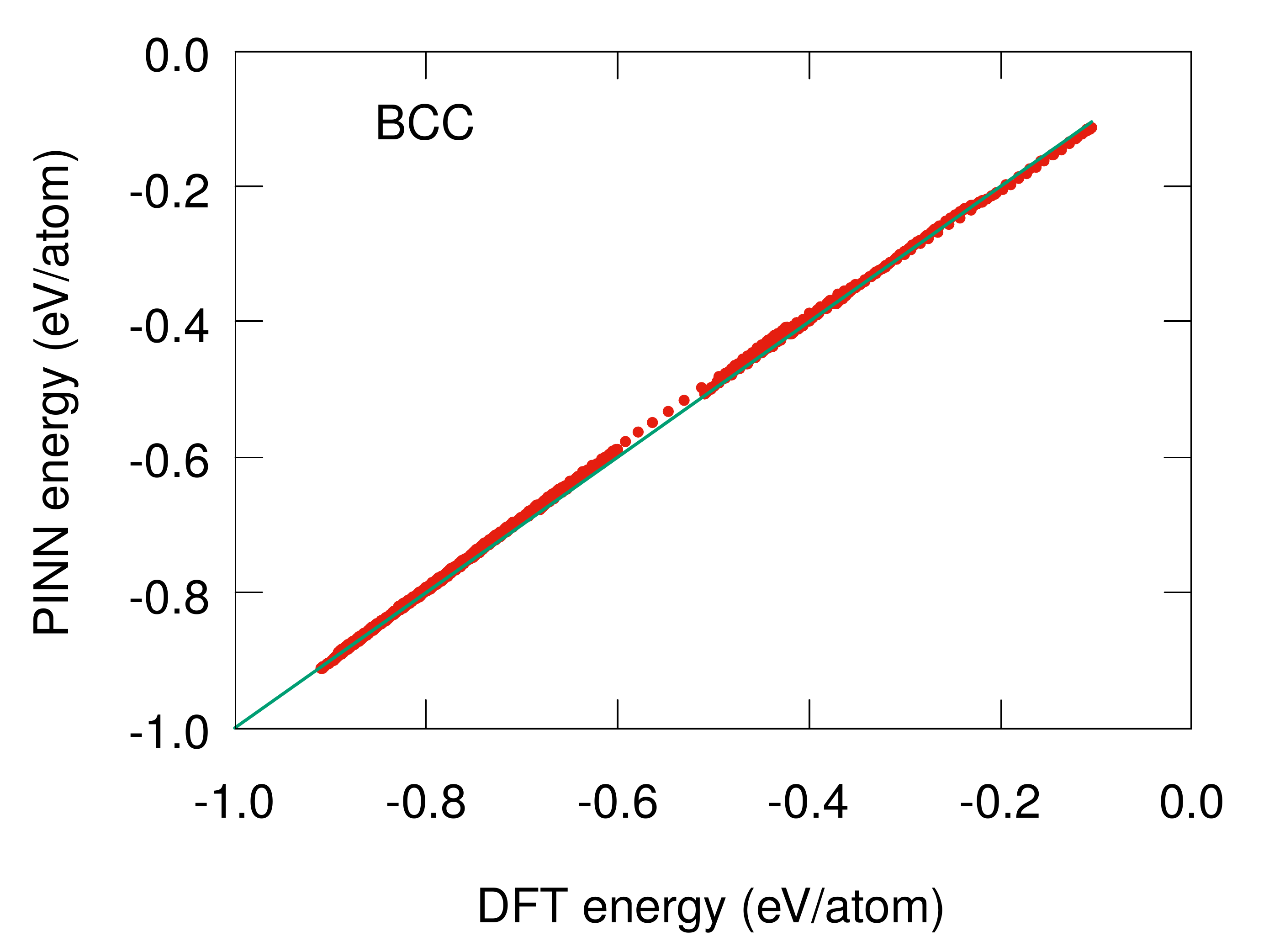}\enskip{}\textbf{(d)}\includegraphics[width=0.45\textwidth]{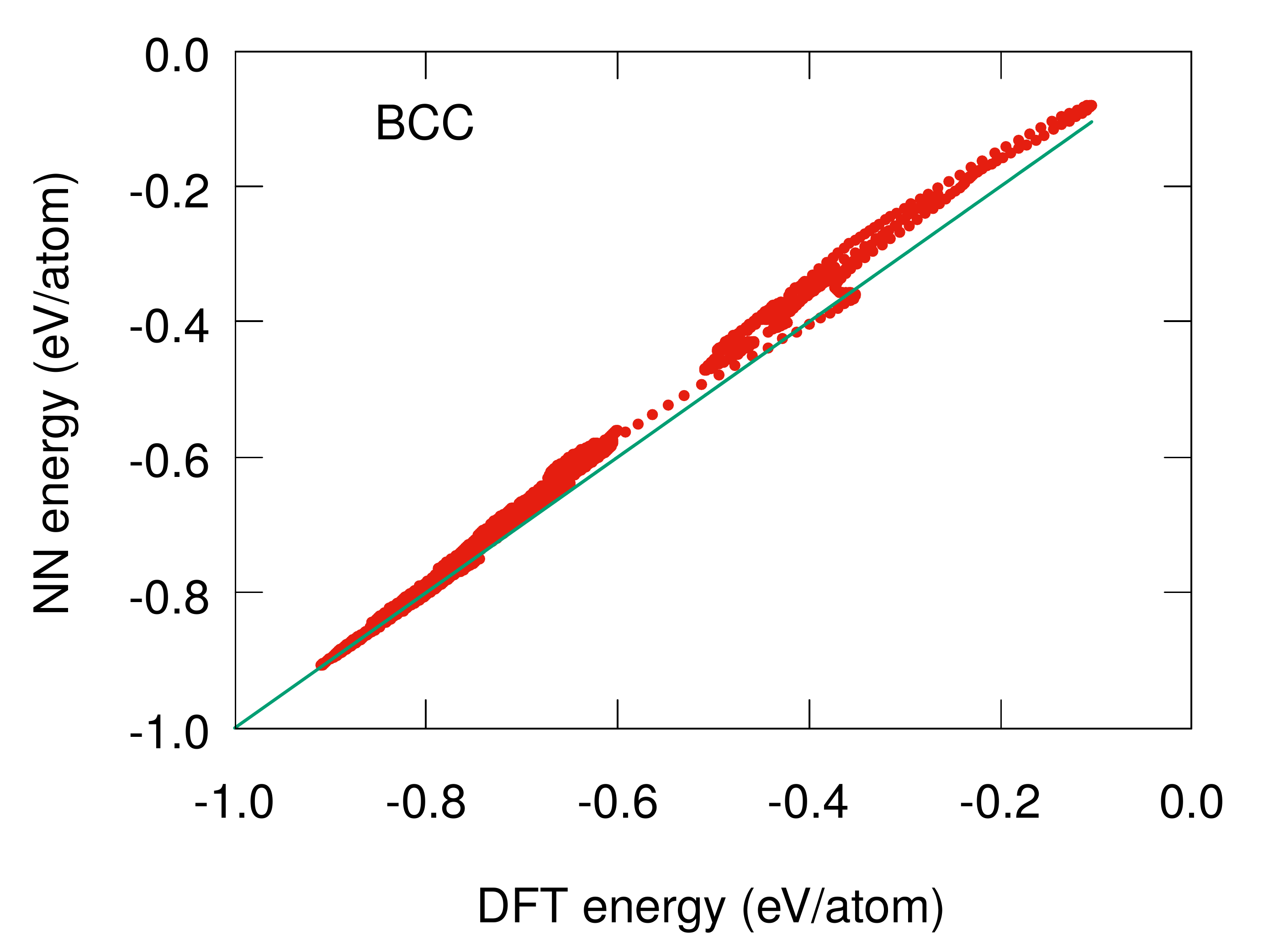}

\textbf{(e)}\includegraphics[width=0.45\textwidth]{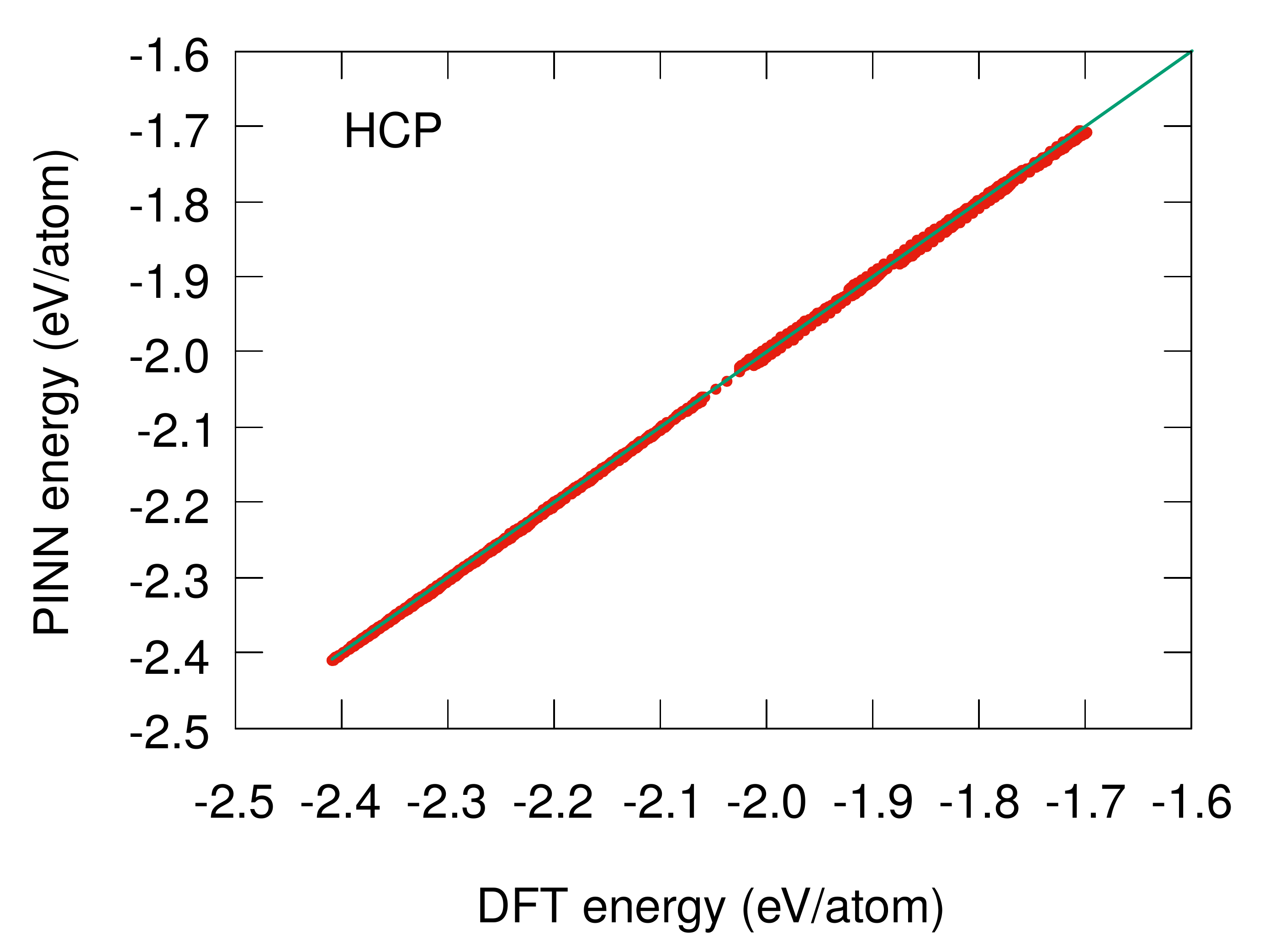}\enskip{}\textbf{(f)}\includegraphics[width=0.45\textwidth]{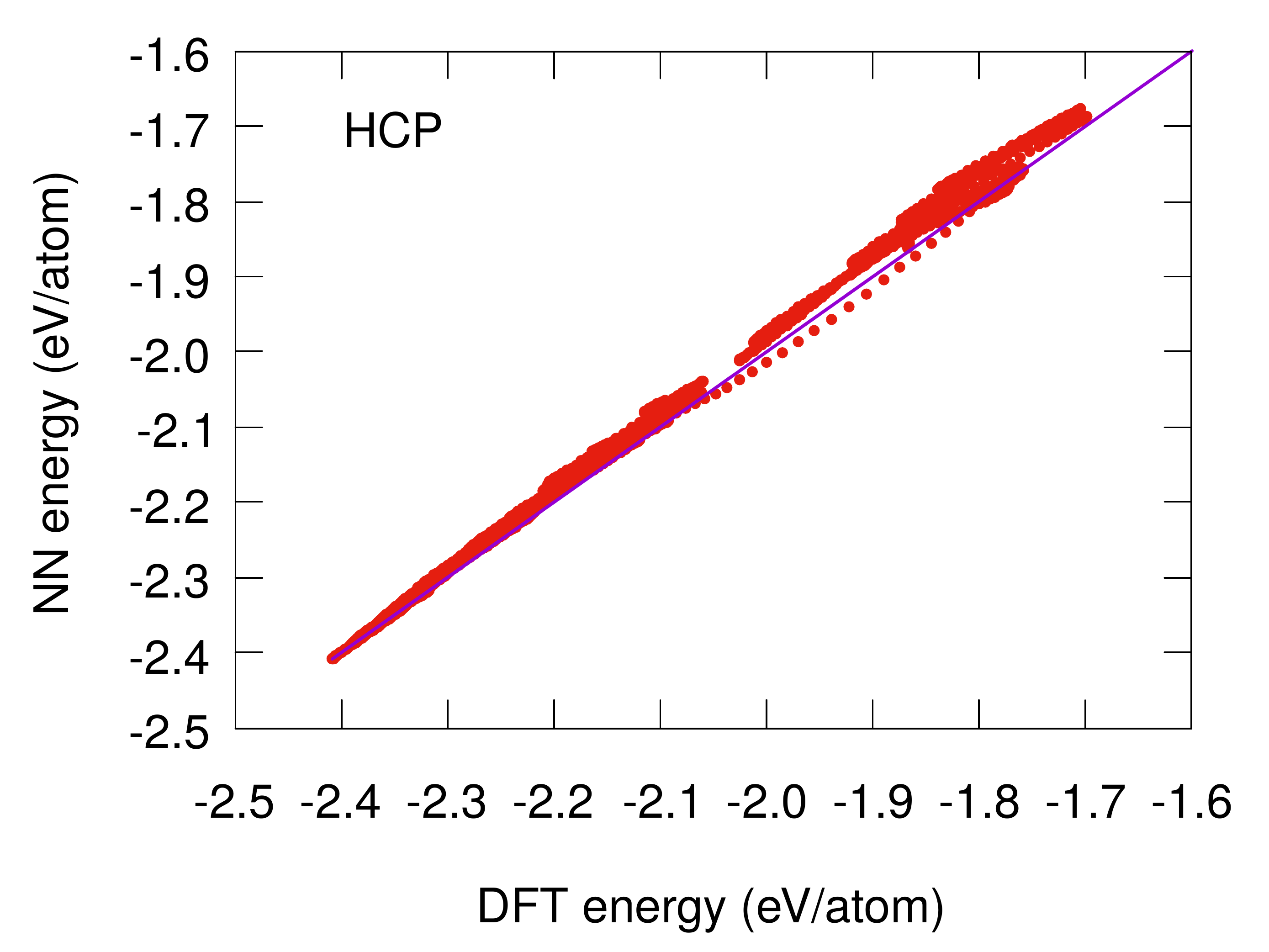}

\caption{Energy of BCC and HCP Al in NVT MD simulations at the temperatures
of (a,b,e,f) 300\,K and 600\,K and (c,d) 1000\,K, 1500\,K, 2000\,K
and 4000\,K. The energies predicted by the PINN (a,c,e) and NN (b,d,f)
potentials are compared with DFT calculations from \citep{Botu:2015aa,Botu:2015bb}.
The straight lines represent the perfect fit.\label{Fig:validation_3}}
\end{figure}

\begin{figure}
\textbf{(a)}\includegraphics[width=0.45\textwidth]{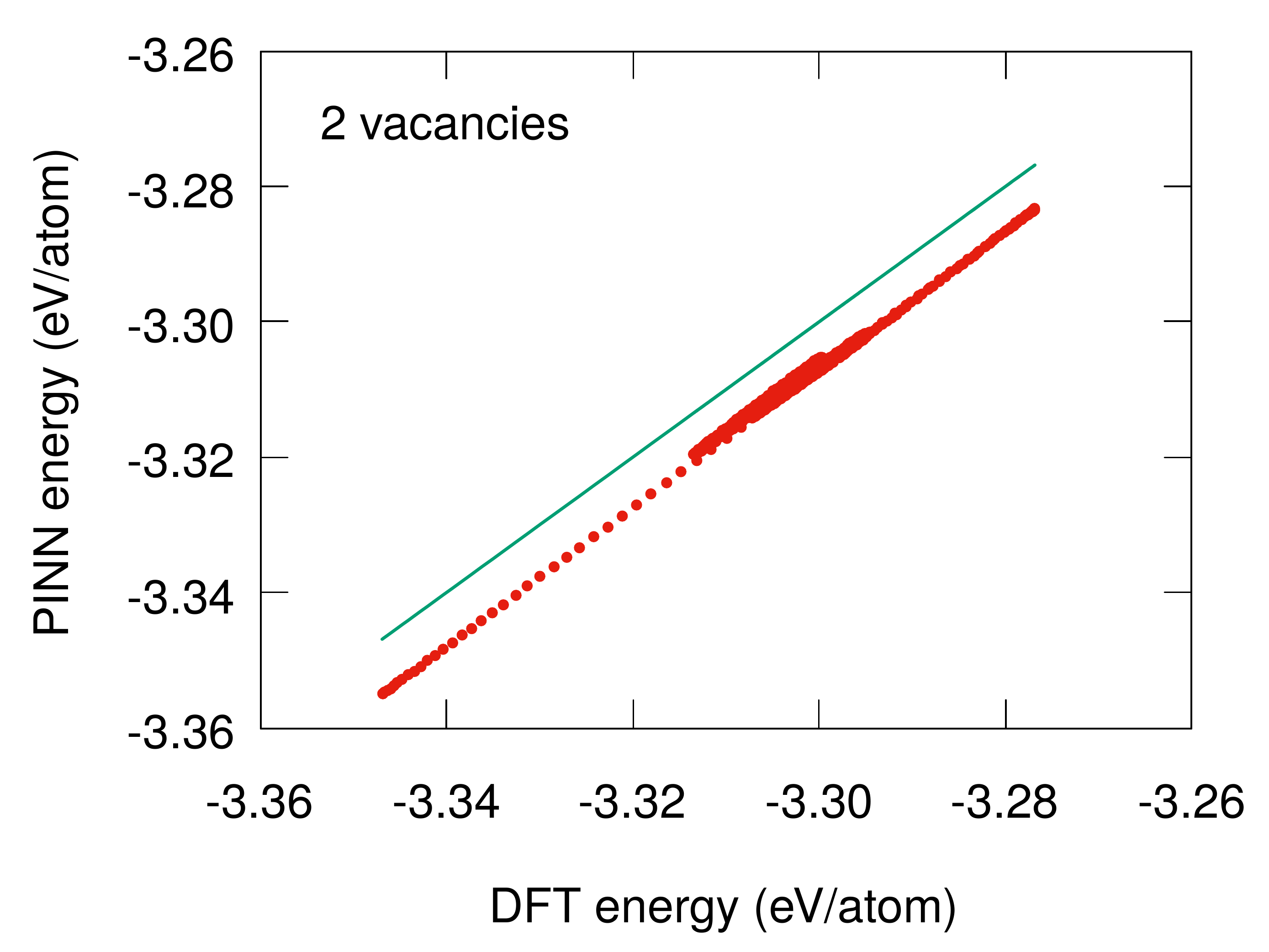}\enskip{}\textbf{(b)}\includegraphics[width=0.45\textwidth]{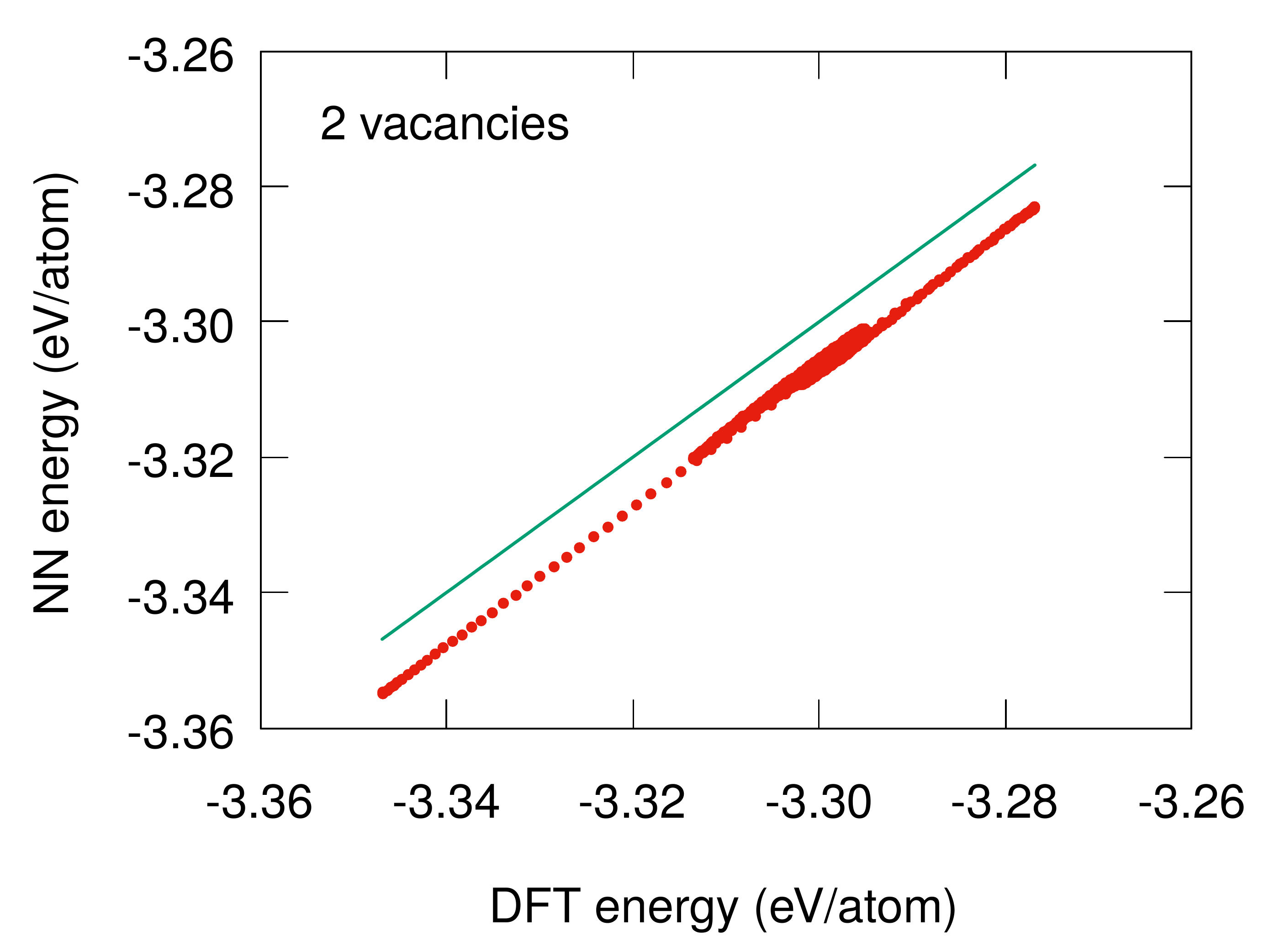}

\bigskip{}
\bigskip{}

\textbf{(c)}\includegraphics[width=0.45\textwidth]{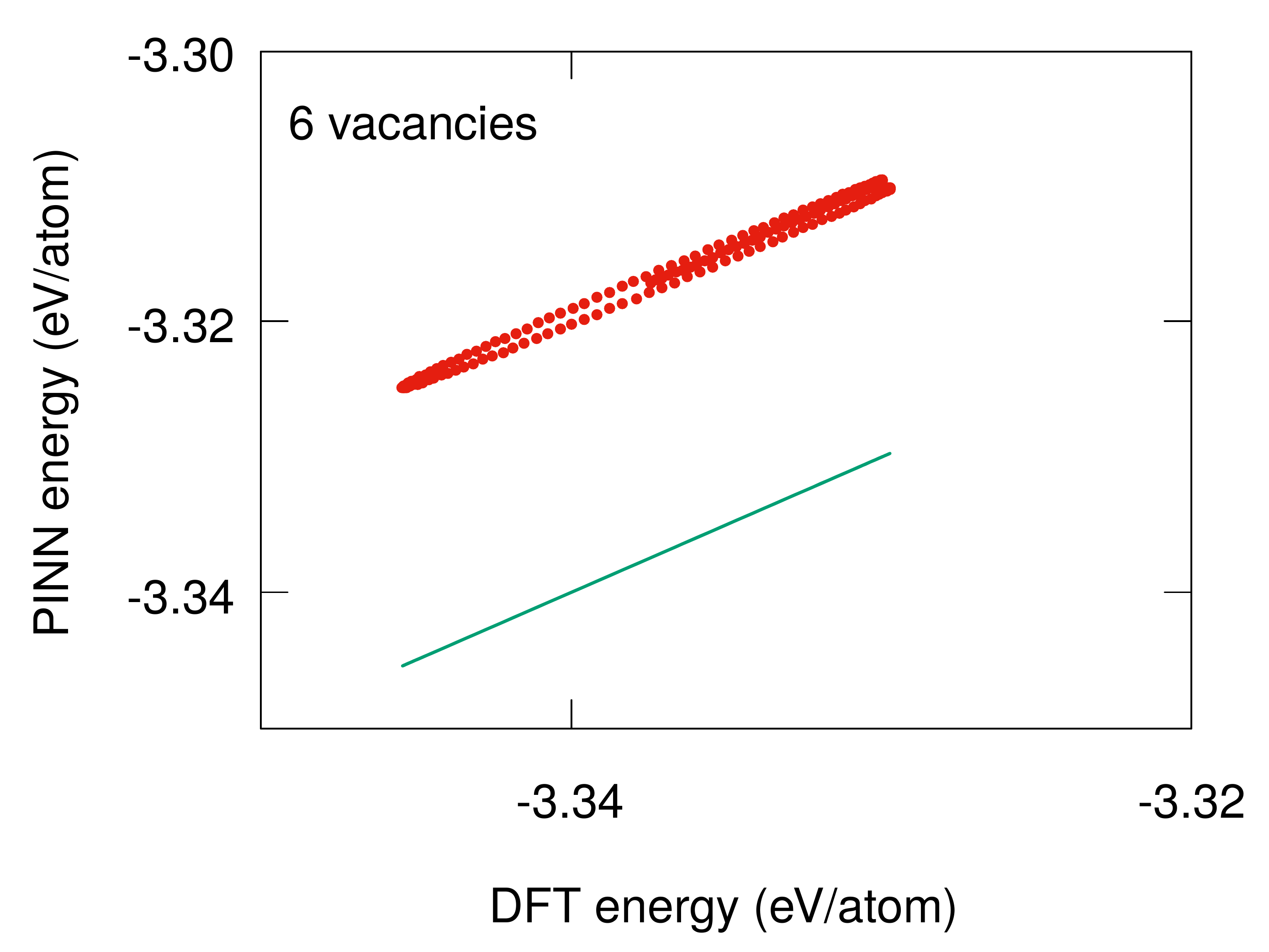}\enskip{}\textbf{(d)}\includegraphics[width=0.45\textwidth]{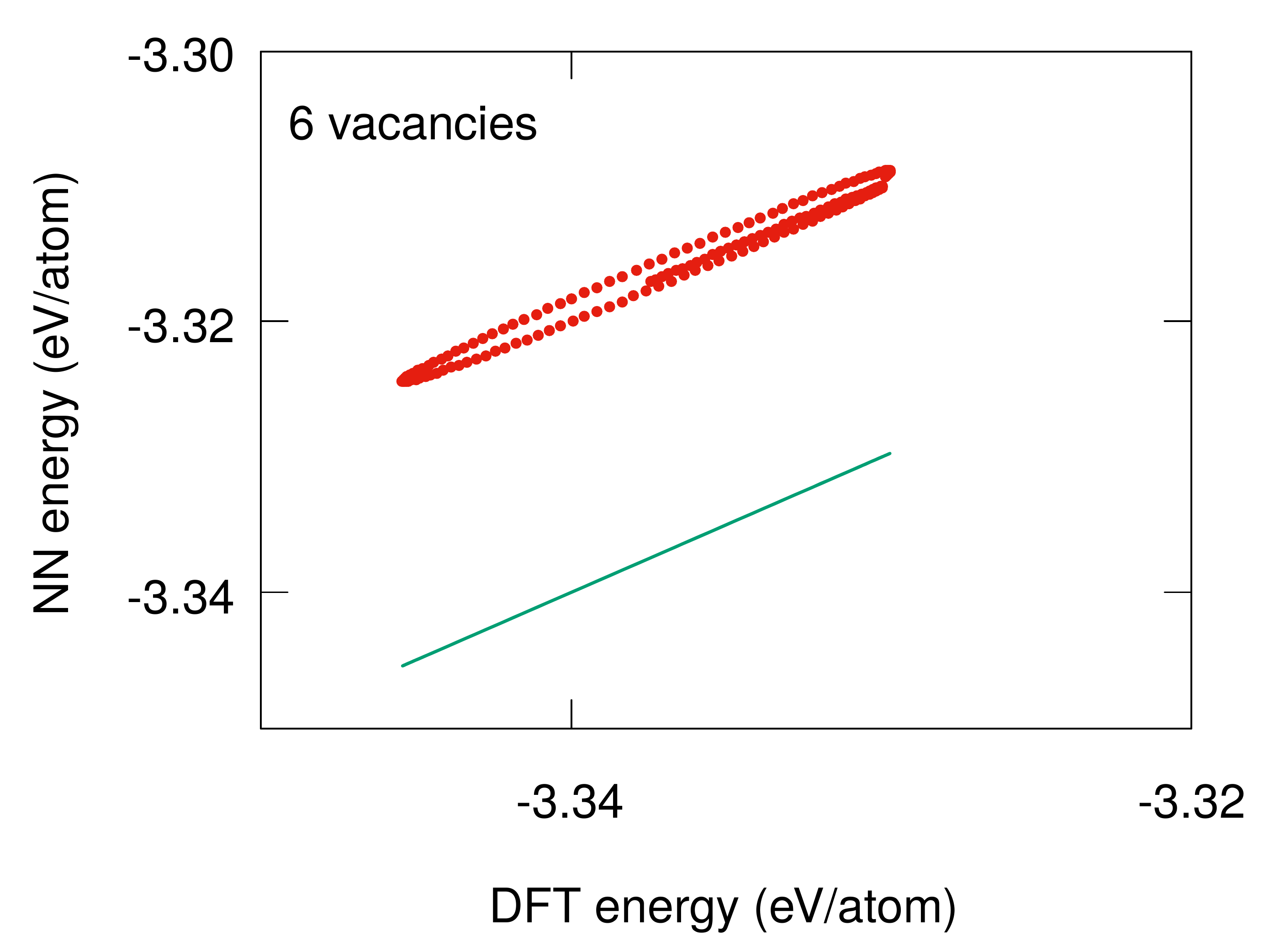}

\caption{Energy of Al supercells containing (a,b) 2 and (c,d) 6 vacancies in
NVE MD simulations starting at 700\,K. The energies predicted by
the PINN (a,c) and NN (b,d) potentials are compared with DFT calculations
from \citep{Botu:2015aa,Botu:2015bb}. The straight lines represent
the perfect fit.\label{Fig:validation_4}}
\end{figure}

\begin{figure}
\textbf{(a)}\includegraphics[width=0.45\textwidth]{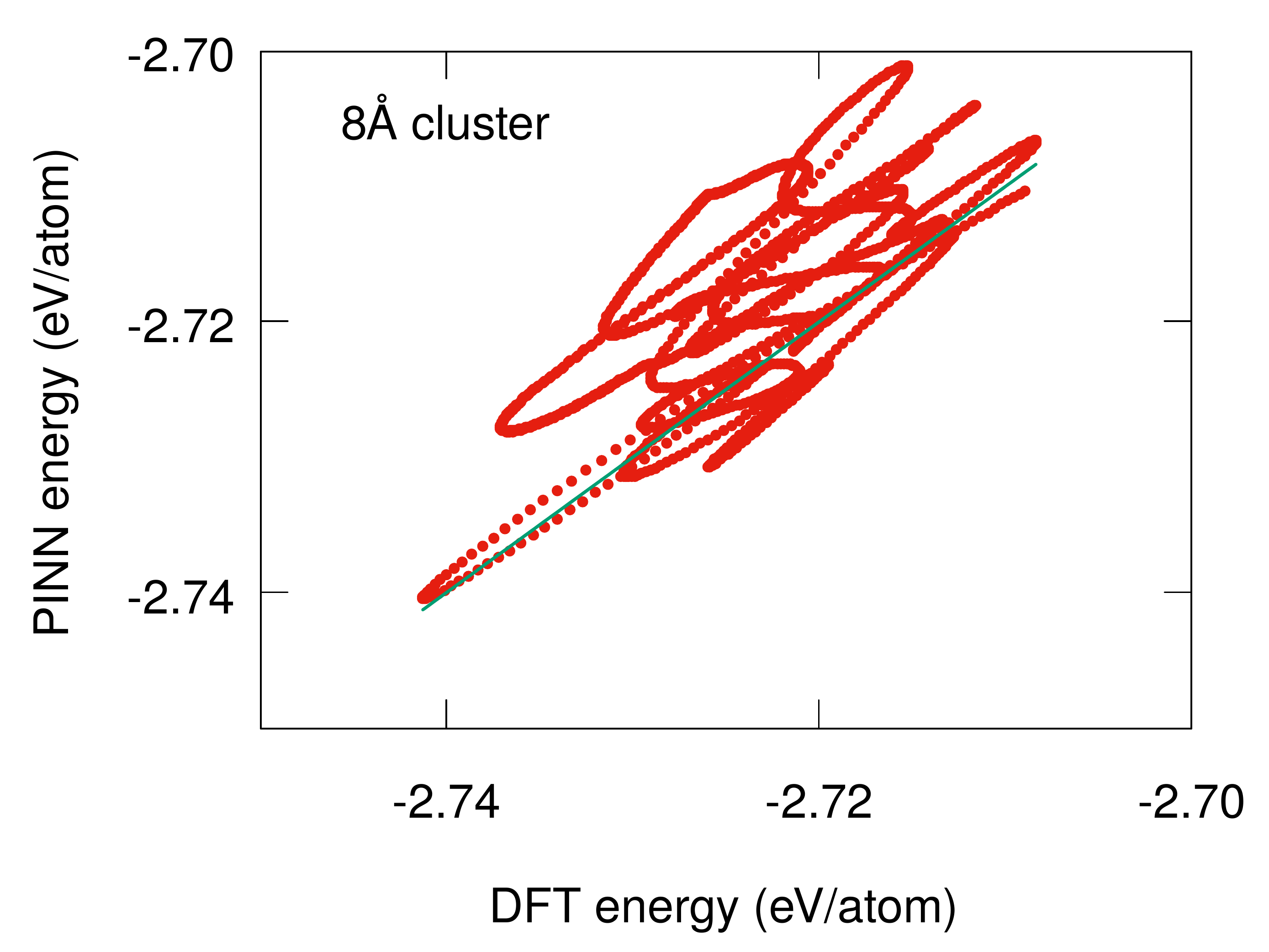}\enskip{}\textbf{(b)}\includegraphics[width=0.45\textwidth]{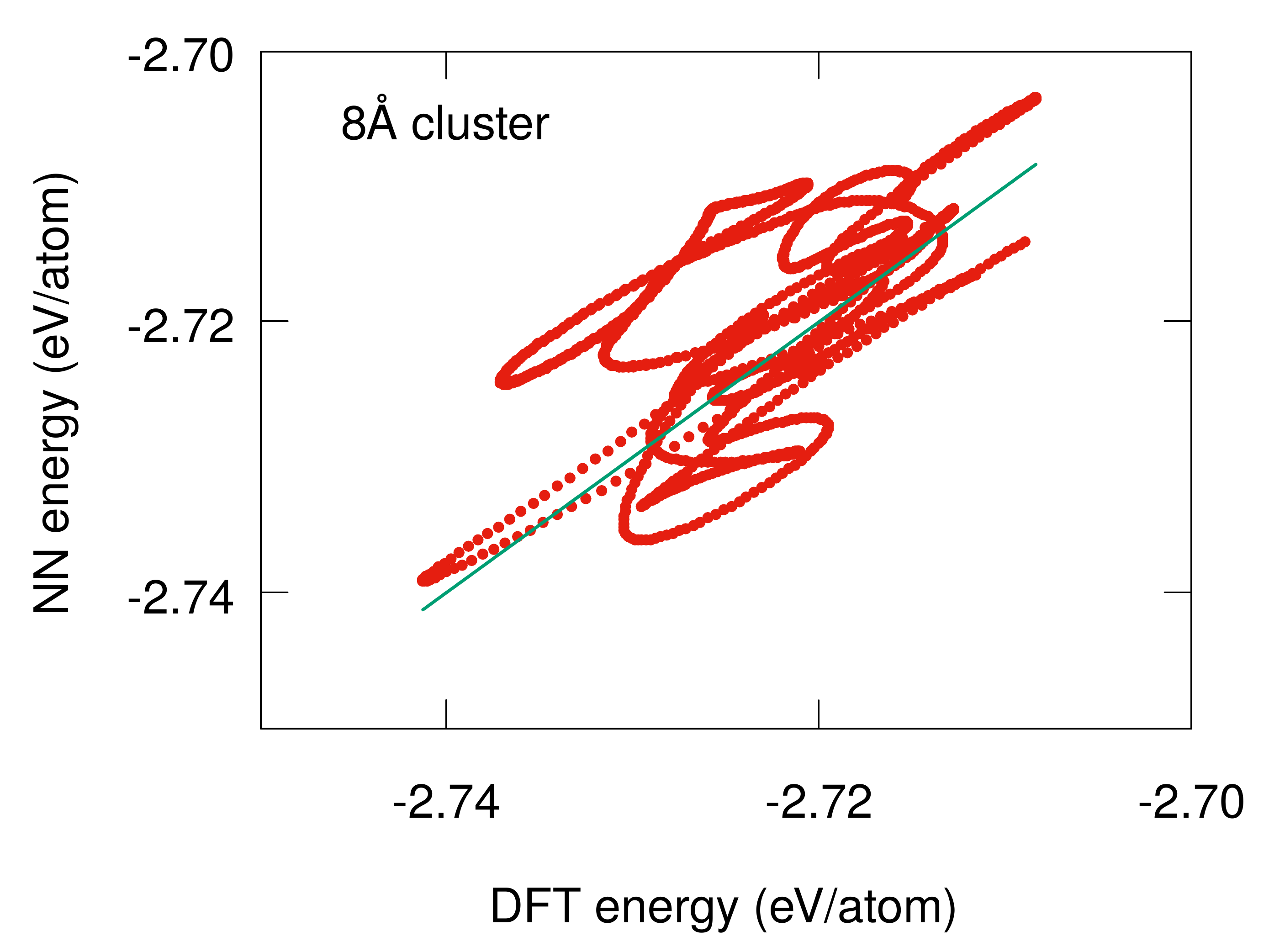}

\bigskip{}
\bigskip{}

\textbf{(c)}\includegraphics[width=0.45\textwidth]{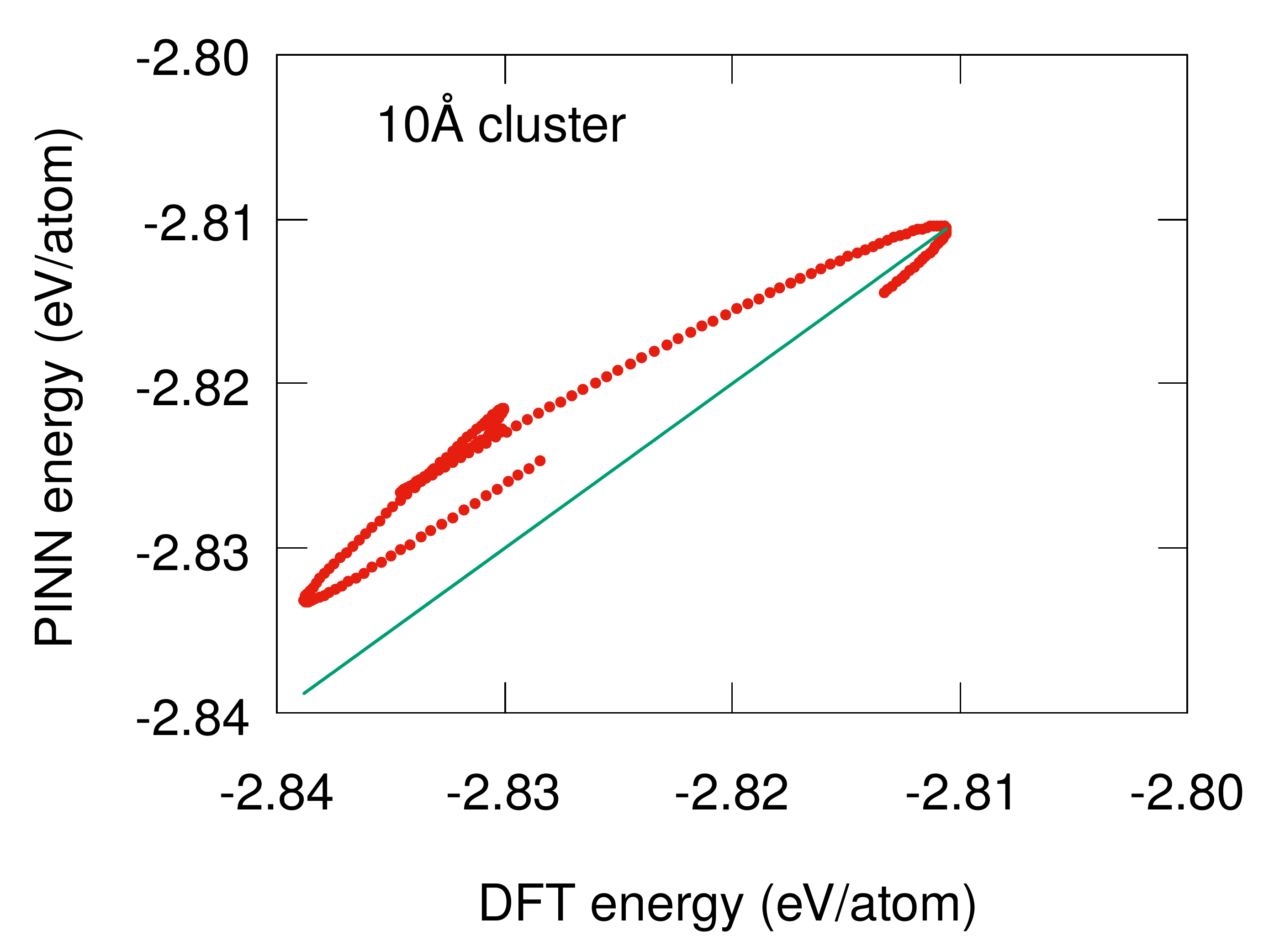}\enskip{}\textbf{(d)}\includegraphics[width=0.45\textwidth]{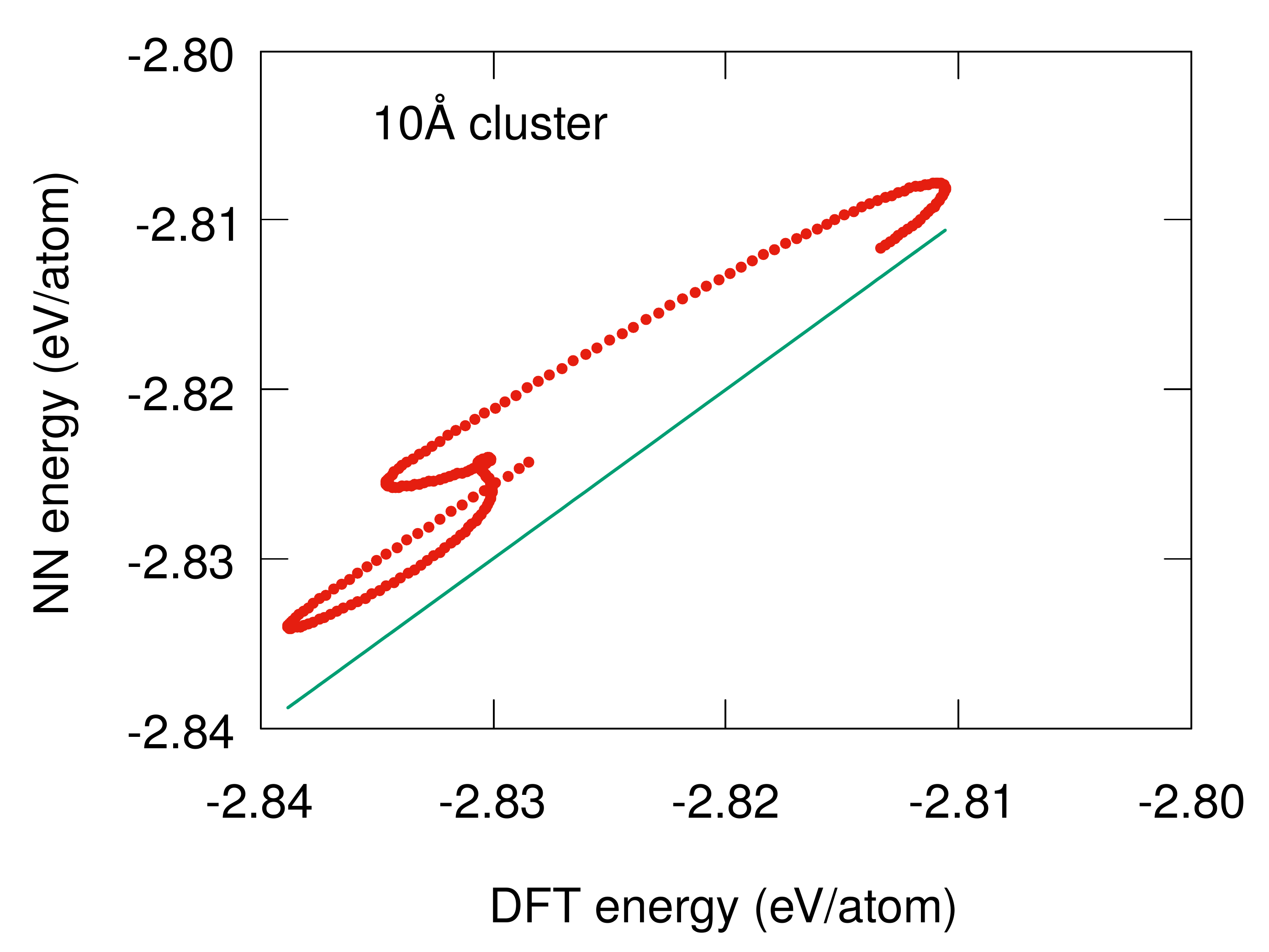}

\textbf{(e)}\includegraphics[width=0.45\textwidth]{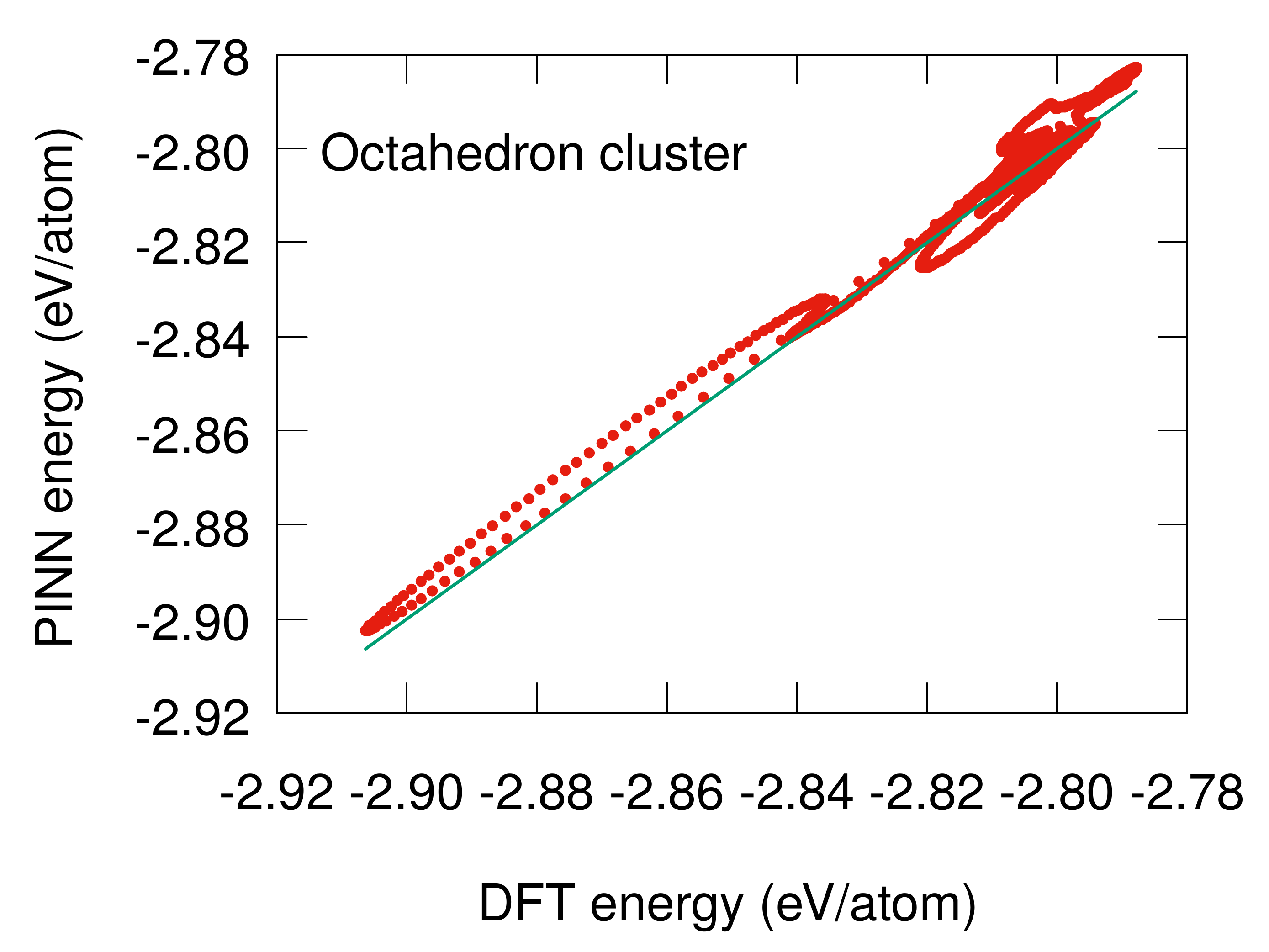}\enskip{}\textbf{(f)}\includegraphics[width=0.45\textwidth]{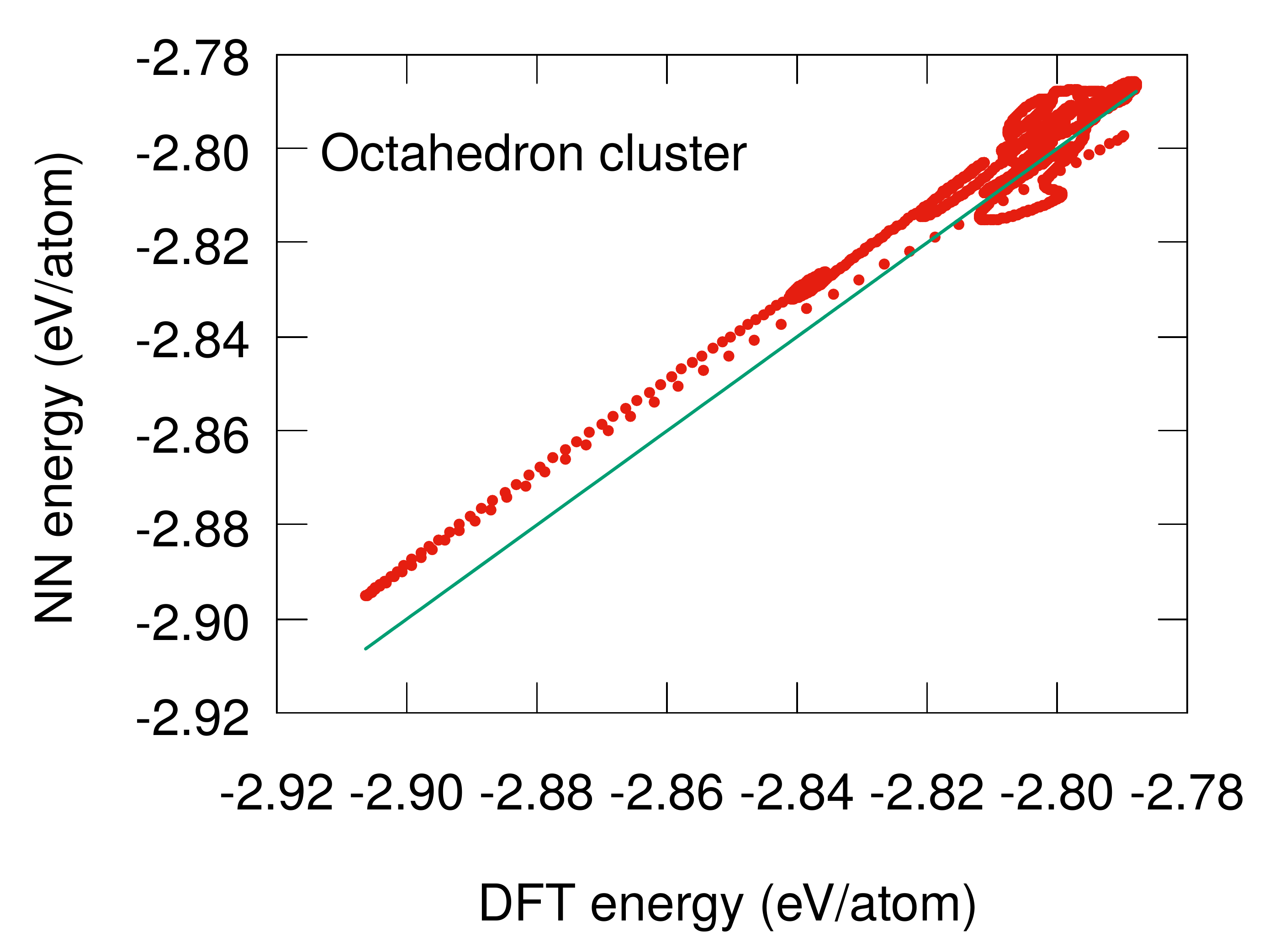}

\caption{Energy of the 8\,\AA\ (a,b), 10\,\AA\ (c,d) and octahedral Al clusters
in NVE MD simulations at the temperatures of (a-d) 1200\,K and (e,f)
1000\,K. The energies predicted by the PINN (a,c,e) and NN (b,d,f)
potentials are compared with DFT calculations from \citep{Botu:2015aa,Botu:2015bb}.
The straight lines represent the perfect fit.\label{Fig:validation_5}}
\end{figure}

\begin{figure}
\textbf{(a)}\includegraphics[width=0.45\textwidth]{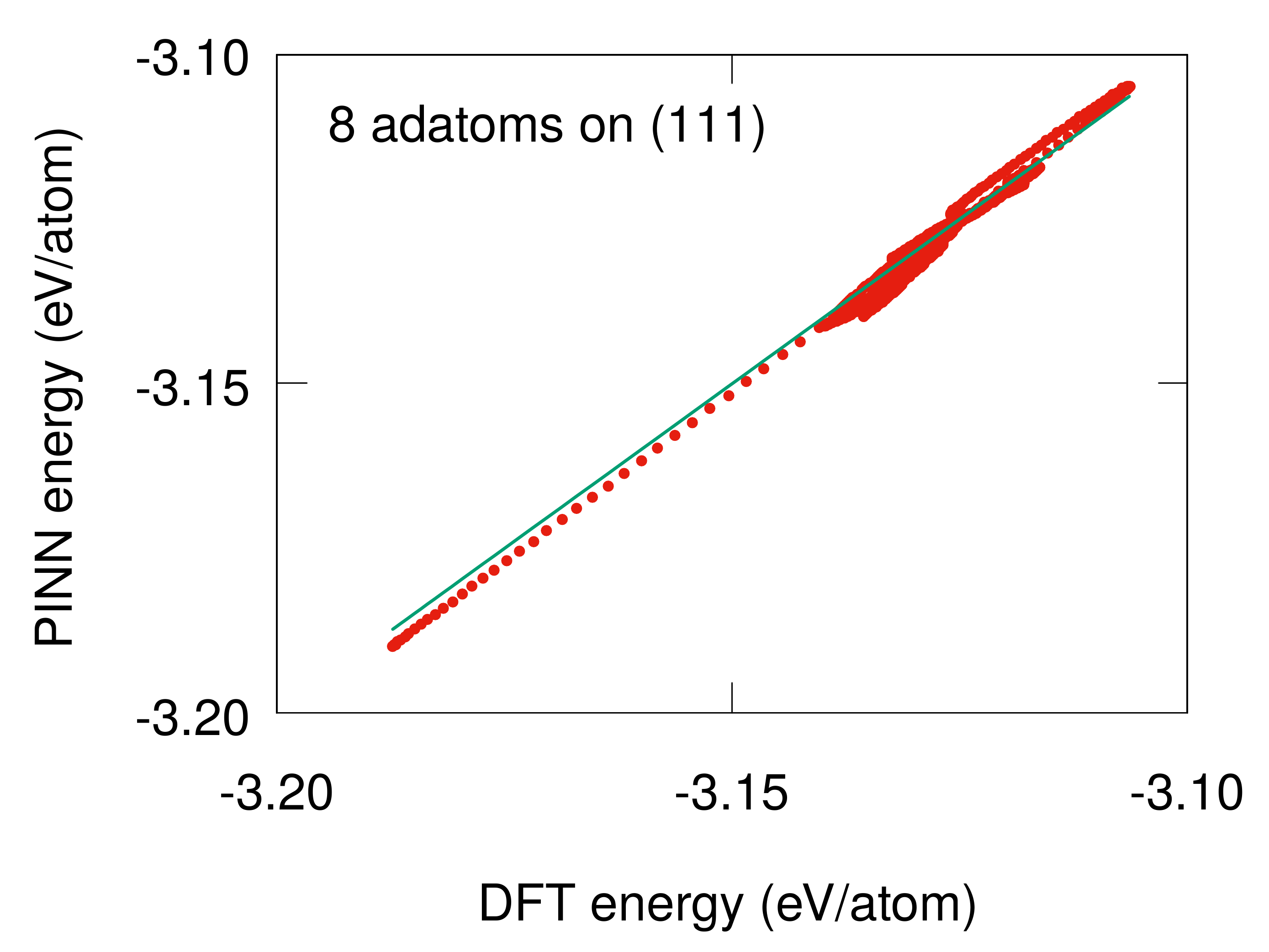}\enskip{}\textbf{(b)}\includegraphics[width=0.45\textwidth]{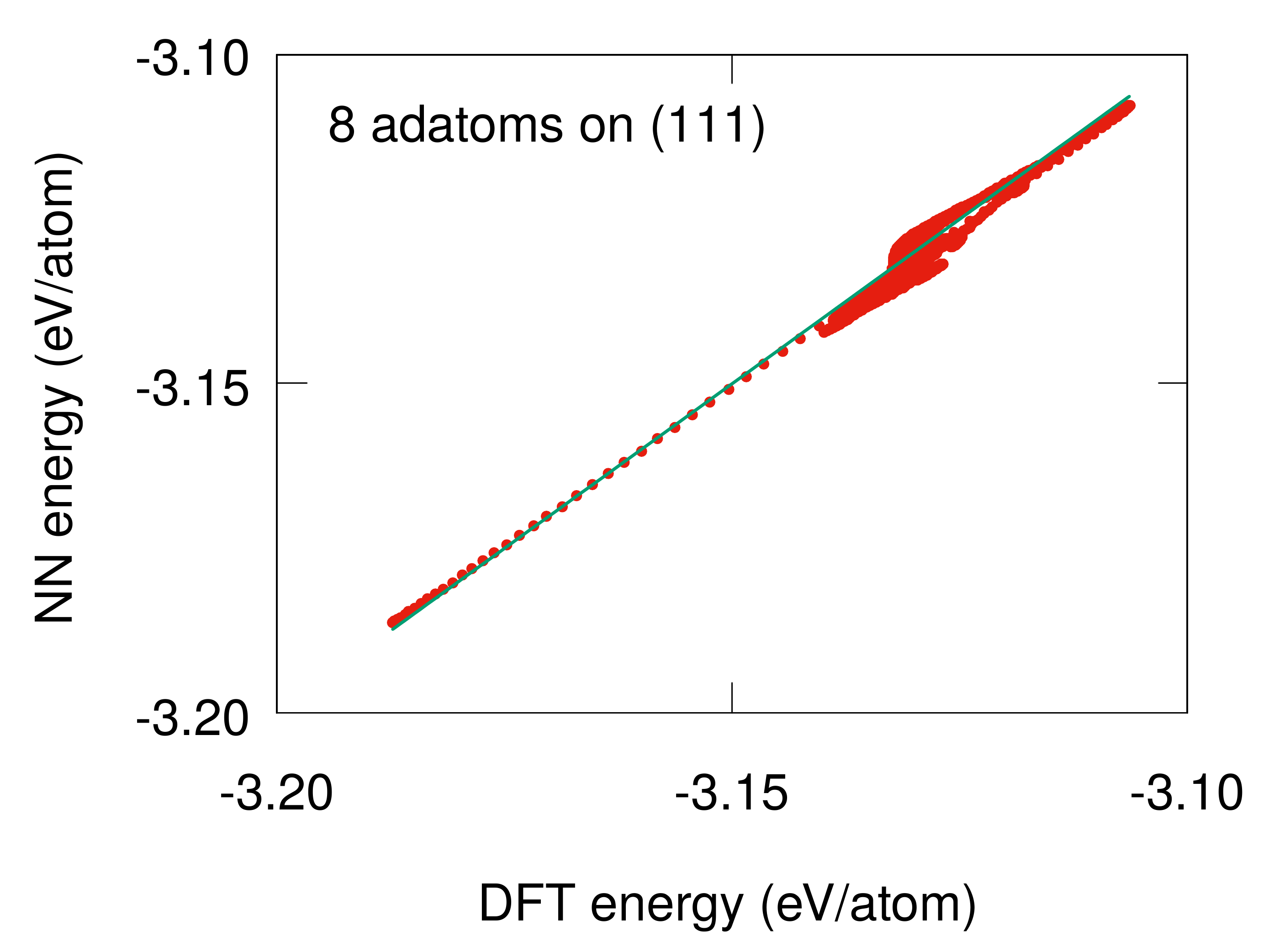}

\bigskip{}
\bigskip{}

\textbf{(c)}\includegraphics[width=0.45\textwidth]{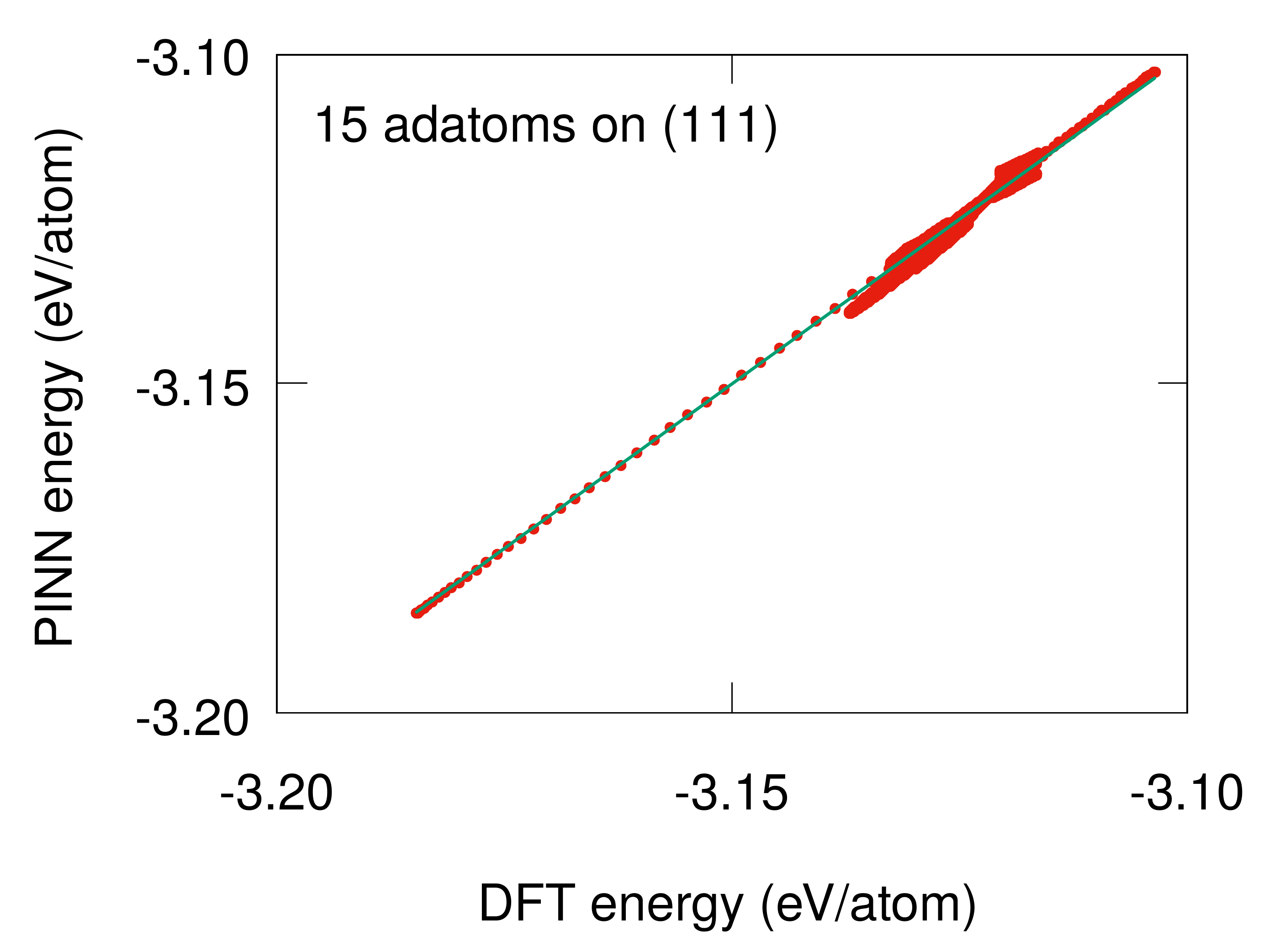}\enskip{}\textbf{(d)}\includegraphics[width=0.45\textwidth]{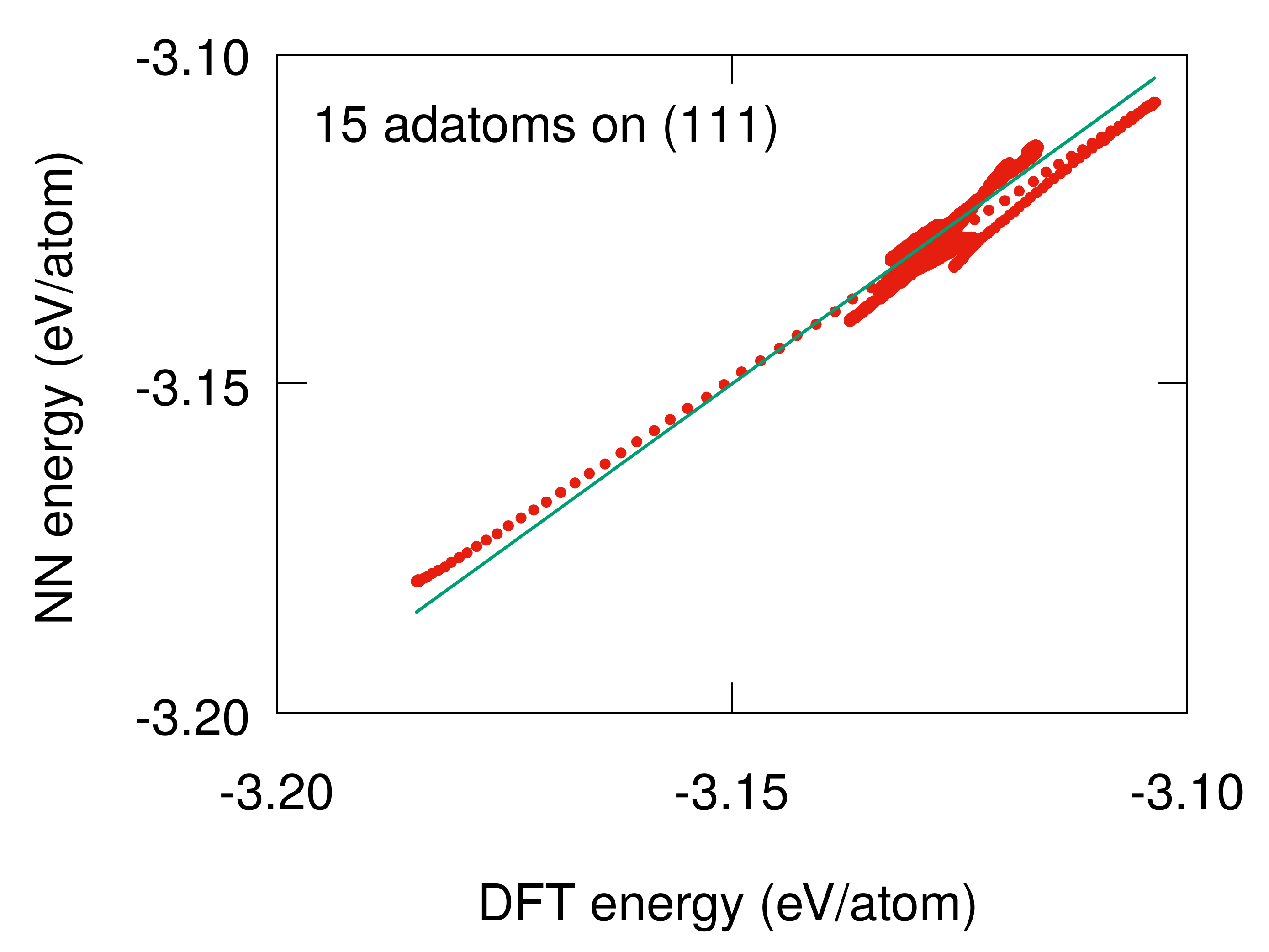}

\caption{Energy of Al supercells containing (a,b) 8 and (c,d) 15 adatoms on
the (111)FCC surface in NVE MD simulations starting at 1500\,K. The
energies predicted by the PINN (a,c) and NN (b,d) potentials are compared
with DFT calculations from \citep{Botu:2015aa,Botu:2015bb}. The straight
lines represent the perfect fit.\label{Fig:validation_6}}
\end{figure}

\begin{figure}
\textbf{(a)}\includegraphics[width=0.47\textwidth]{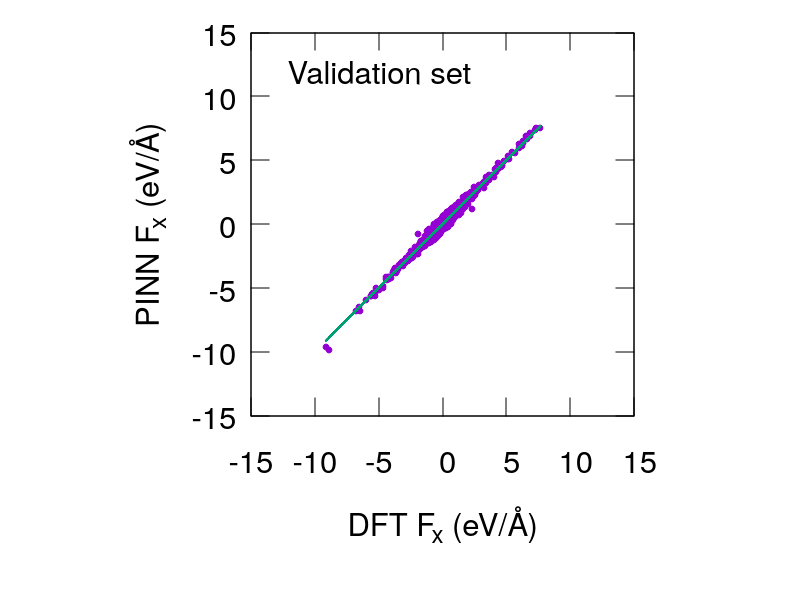}\textbf{(b)}\includegraphics[width=0.47\textwidth]{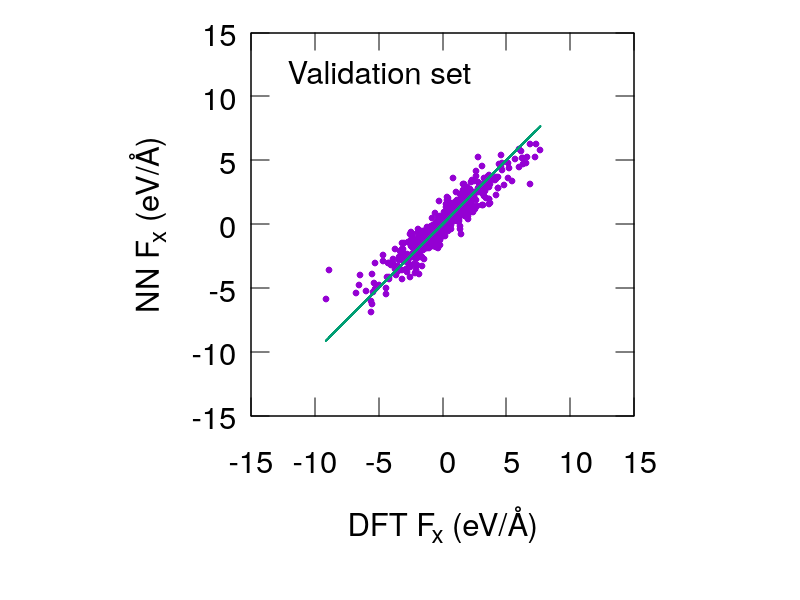}\bigskip{}
\bigskip{}
\textbf{(c)}\includegraphics[width=0.47\textwidth]{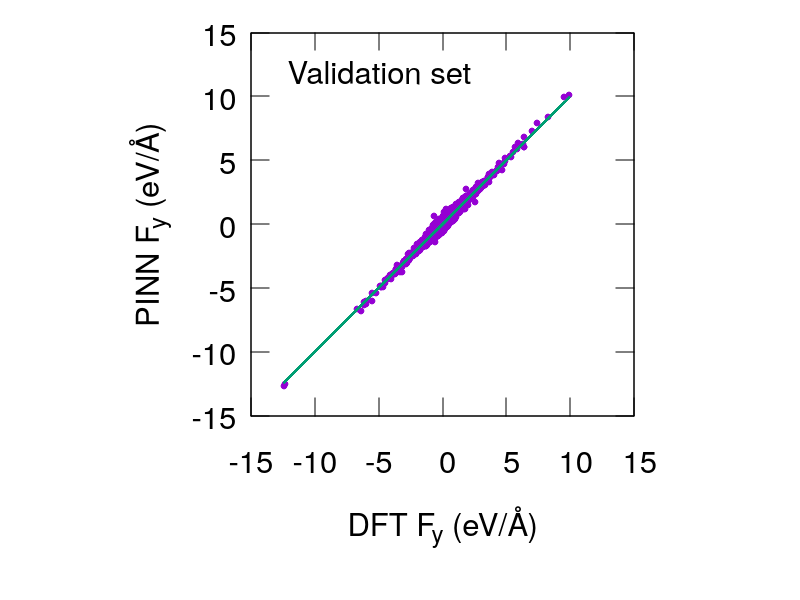}\textbf{(d)}\includegraphics[width=0.47\textwidth]{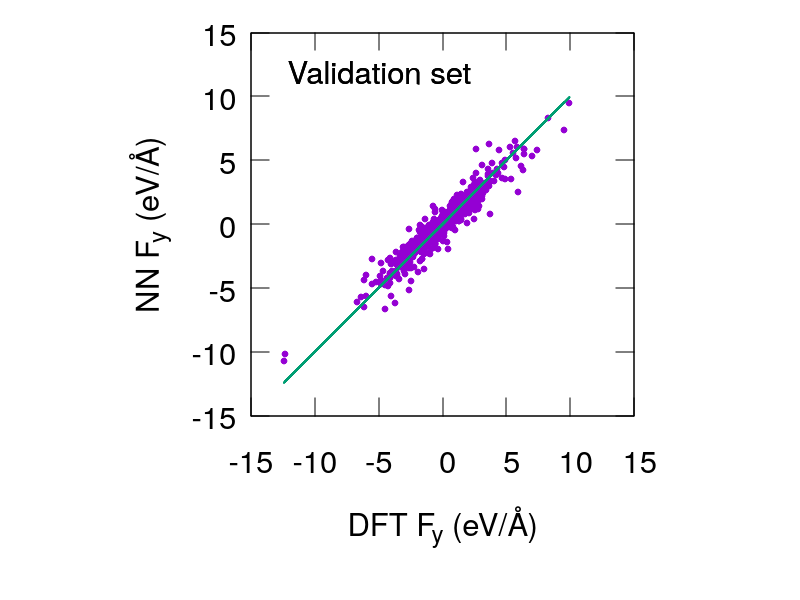}

\bigskip{}
\bigskip{}

\textbf{(e)}\includegraphics[width=0.47\textwidth]{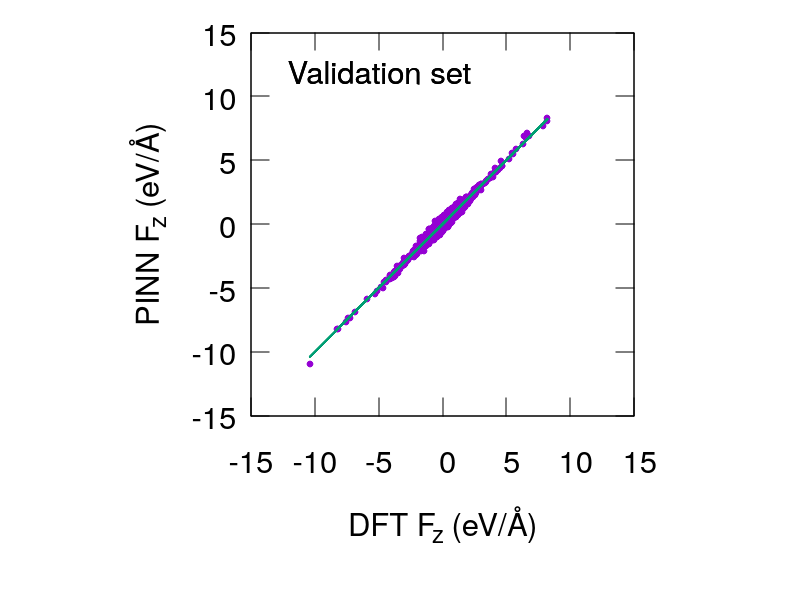}\textbf{(f)}\includegraphics[width=0.47\textwidth]{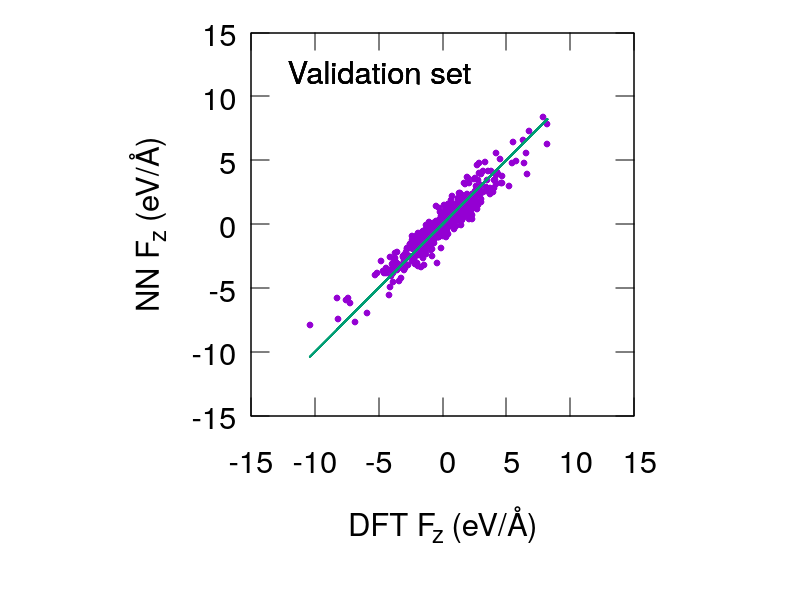}

\caption{Atomic force components in validation database predicted by the PINN
and NN potentials in comparison with with DFT calculations. The straight
lines represent the perfect fit. \label{Fig:validation_forces_S1}}
\end{figure}

\begin{figure}
\textbf{(a)}\includegraphics[width=0.47\textwidth]{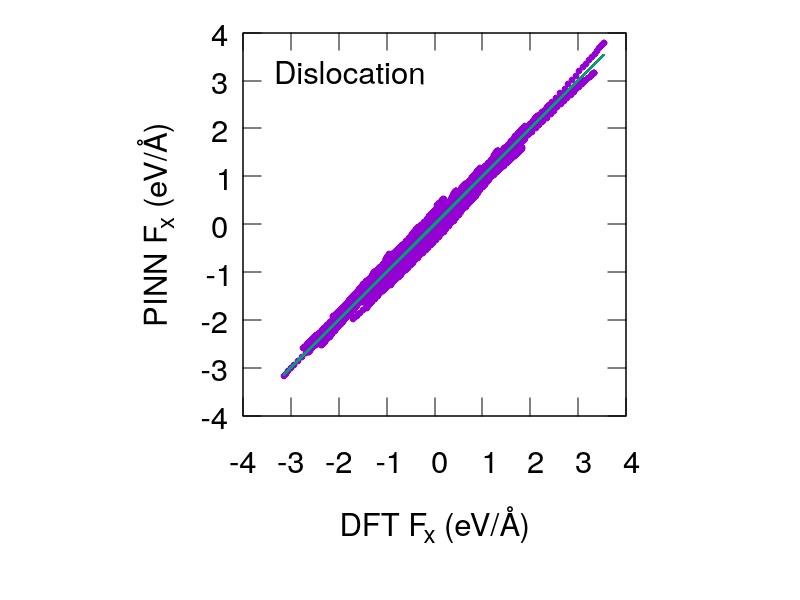}\textbf{(b)}\includegraphics[width=0.47\textwidth]{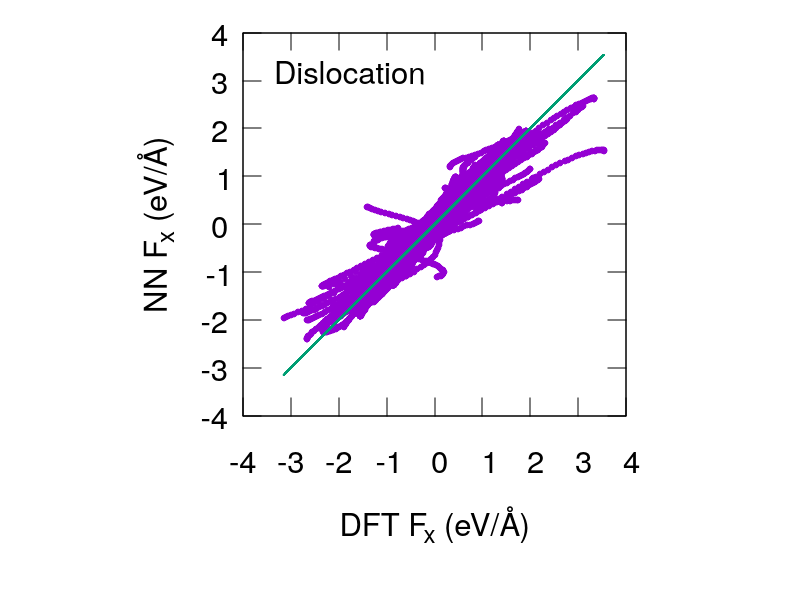}\bigskip{}
\bigskip{}
\textbf{(c)}\includegraphics[width=0.47\textwidth]{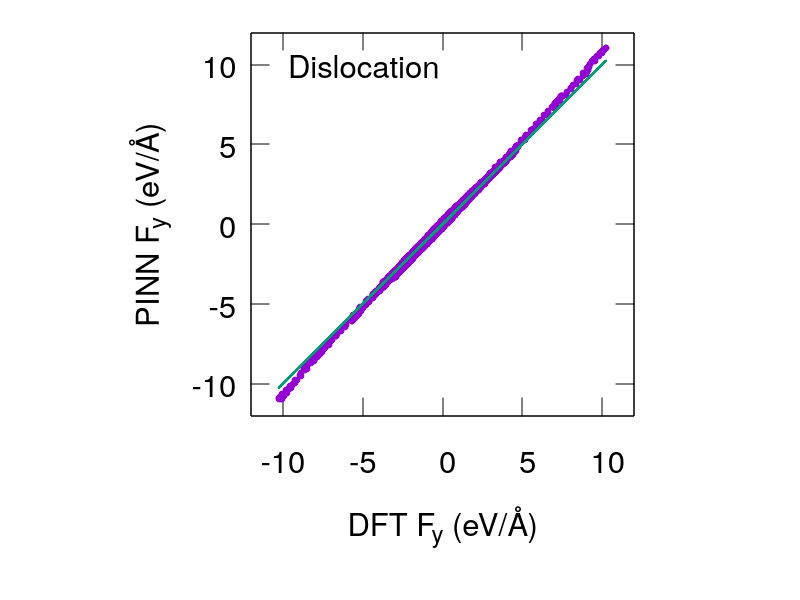}\textbf{(d)}\includegraphics[width=0.47\textwidth]{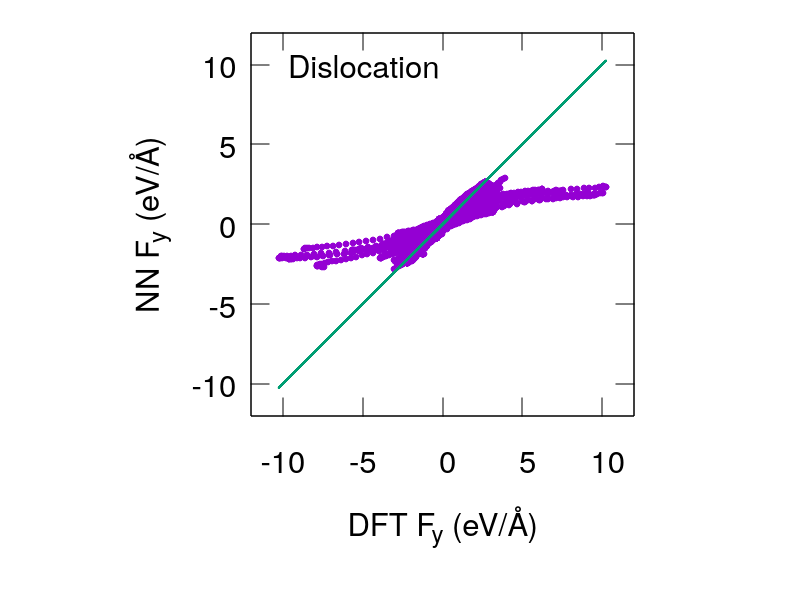}

\bigskip{}
\bigskip{}

\textbf{(e)}\includegraphics[width=0.47\textwidth]{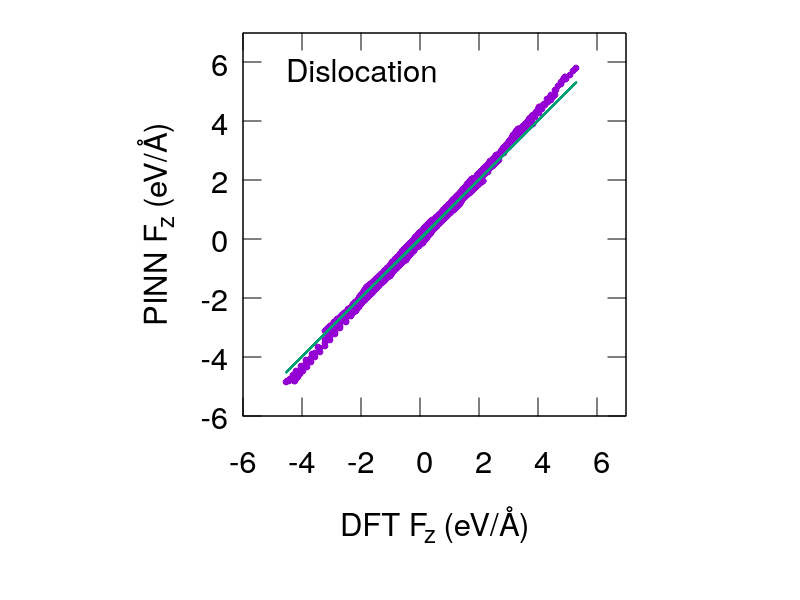}\textbf{(f)}\includegraphics[width=0.47\textwidth]{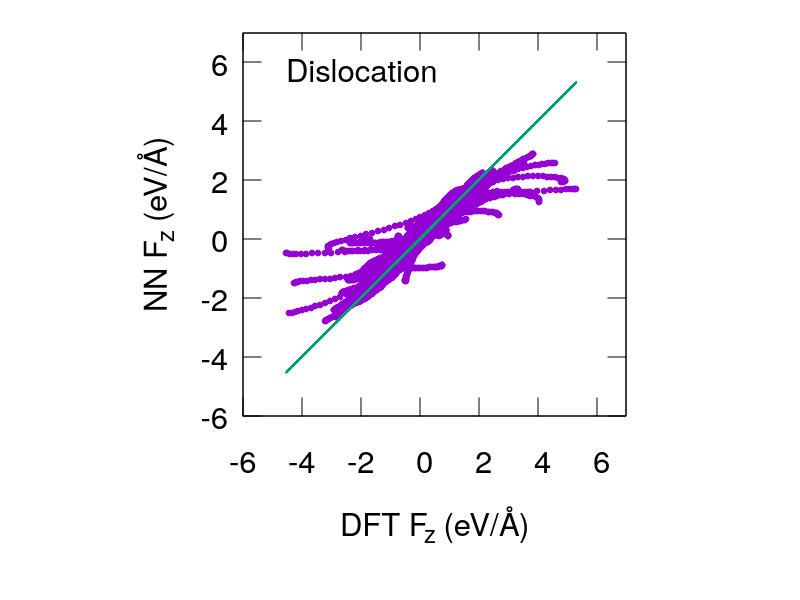}

\caption{Atomic forces for the edge dislocation in NVE MD simulations starting
at 700 K predicted by the PINN and NN potentials in comparison with
DFT calculations. The straight lines represent the perfect fit. \label{Fig:validation_forces_Disloc-1}}
\end{figure}

\begin{figure}
\textbf{(a)}\includegraphics[width=0.47\textwidth]{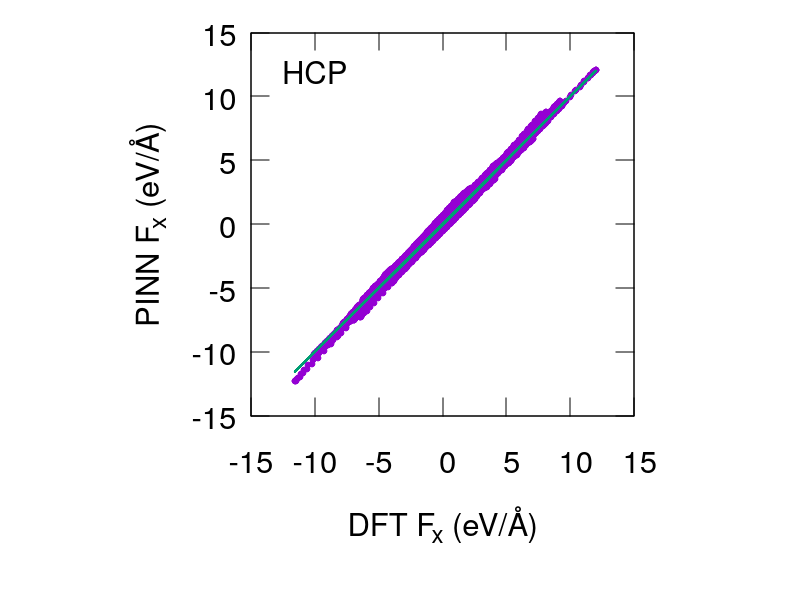}\textbf{(b)}\includegraphics[width=0.47\textwidth]{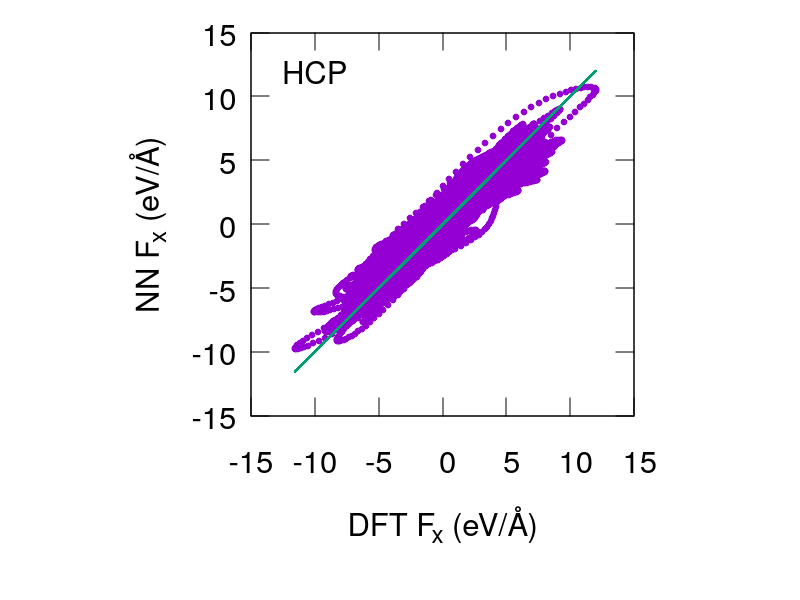}\bigskip{}
\bigskip{}
\textbf{(c)}\includegraphics[width=0.47\textwidth]{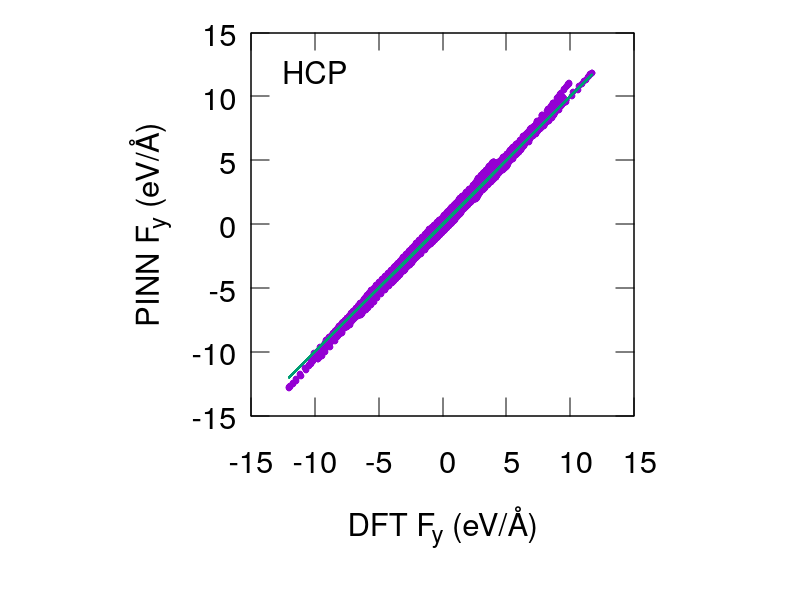}\textbf{(d)}\includegraphics[width=0.47\textwidth]{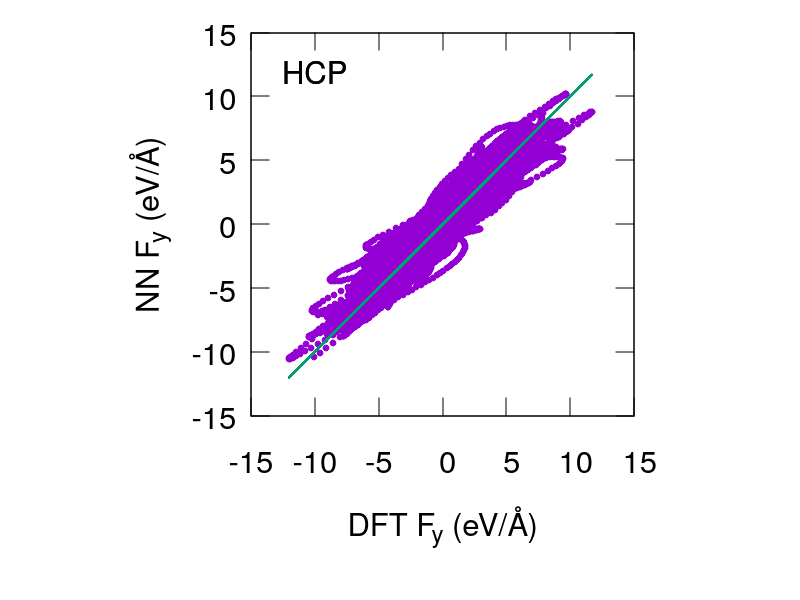}

\bigskip{}
\bigskip{}

\textbf{(e)}\includegraphics[width=0.47\textwidth]{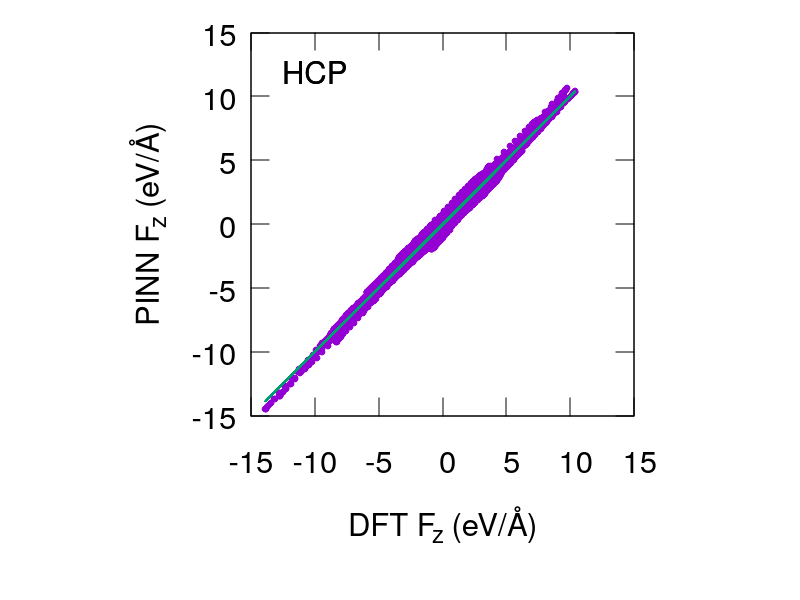}\textbf{(f)}\includegraphics[width=0.47\textwidth]{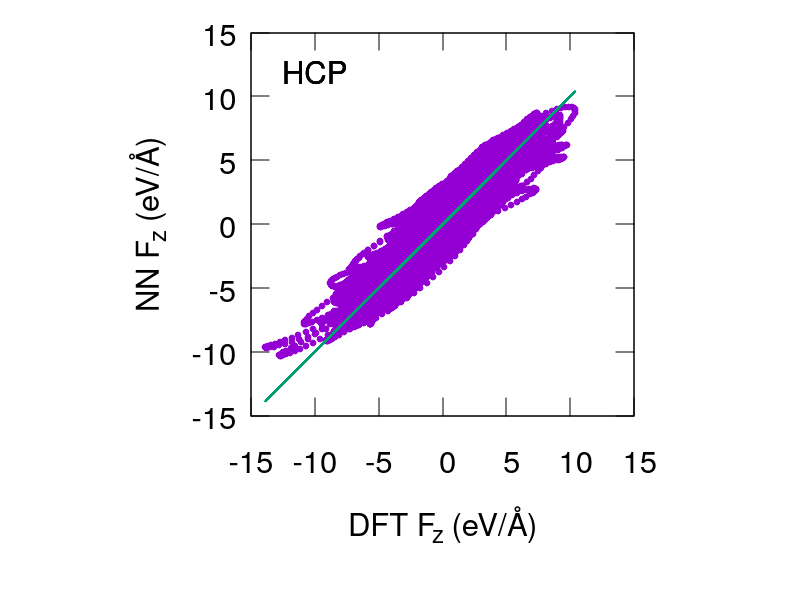}

\caption{Atomic forces in HCP Al during NVT MD simulations at 300\,K, 600\,K,
1000\,K, 1500\,K, 2000\,K and 4000\,K predicted by the PINN and
NN potentials in comparison with DFT calculations. The straight lines
represent the perfect fit. \label{Fig:validation_forces_HCP}}
\end{figure}

\begin{figure}
\noindent \begin{centering}
\textbf{(a)} \includegraphics[width=0.5\textwidth]{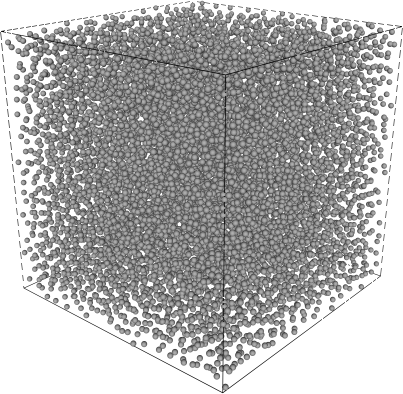}
\par\end{centering}
\bigskip{}
\bigskip{}
\bigskip{}

\noindent \begin{centering}
\textbf{(b)} \includegraphics[width=0.75\textwidth]{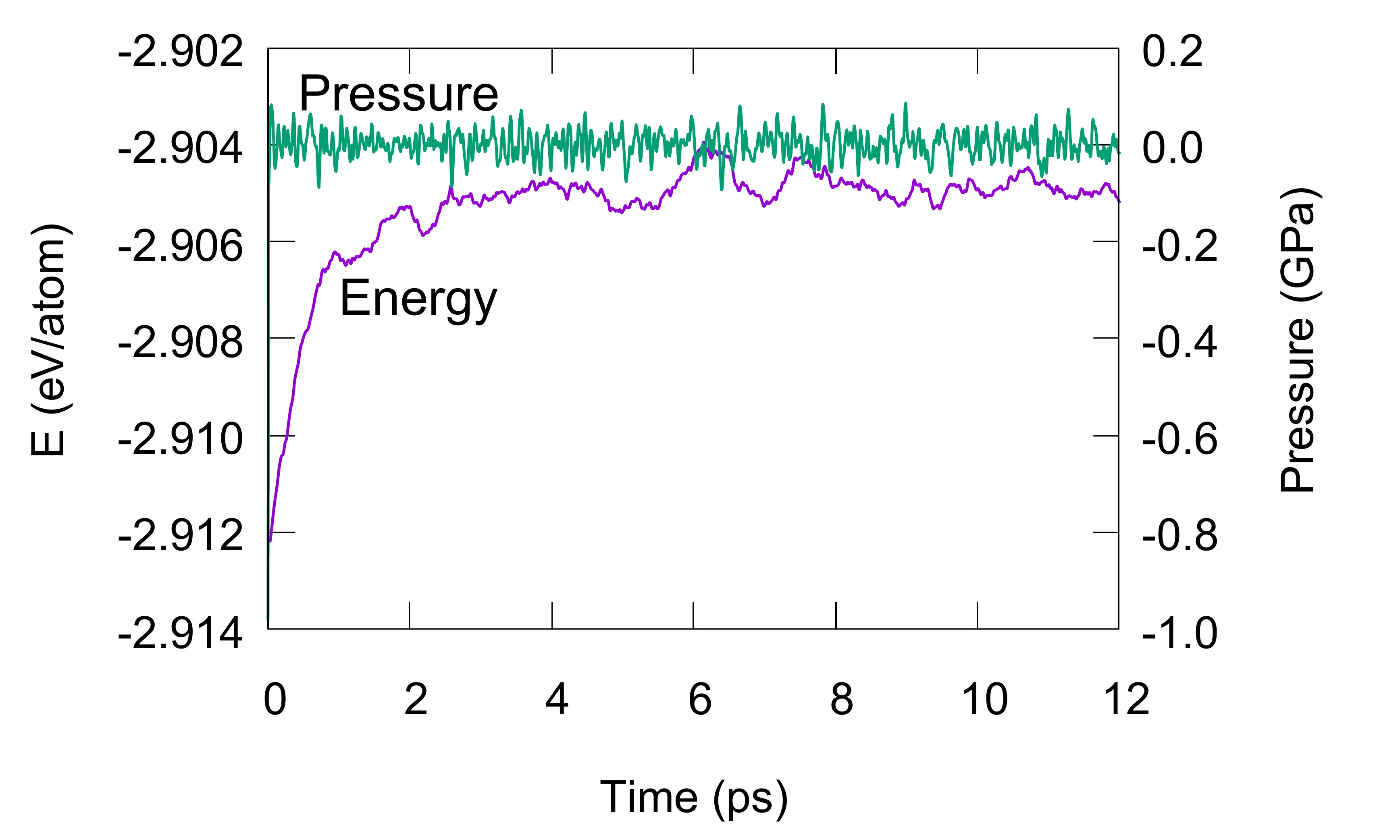}
\par\end{centering}
\caption{Demonstration of MD simulations for liquid Al with the PINN potential.
The simulation was conducted in the zero-pressure NPT ensemble at
the temperature of 1250 K using a beta-version of the ParaGrandMC
code (https://software.nasa.gov/software/LAR-18773-1). The system
contains 10,976 atoms. (a) Typical snapshot of the system. (b) Energy
and pressure as a function of time during initial stages of the simulation.
\label{fig:S_MD_example}}

\end{figure}

\end{document}